\documentclass[11pt,a4paper]{article}
\usepackage{jheppub}
\usepackage{amsmath,amssymb,amsfonts,amsthm}
\usepackage{graphicx}
\usepackage{epsfig} 
\usepackage{verbatim} 
\usepackage{enumerate}
\usepackage{slashed} 
\usepackage{booktabs}
\usepackage{natbib}
\usepackage[dvipsnames]{xcolor}

\definecolor{darkgray}{rgb}{0.35, 0.35, 0.35}
\definecolor{rossocorsa}{rgb}{0.83, 0.0, 0.0}
\definecolor{oceanboatblue}{rgb}{0.0, 0.47, 0.75}
\definecolor{greenmunsell}{rgb}{0.0, 0.66, 0.47}
\definecolor{fandango}{rgb}{0.71, 0.2, 0.54}
\definecolor{darkorange}{rgb}{1.0, 0.55, 0.0}

\usepackage{epsf}

\usepackage{todonotes}
\usepackage{multirow}
\def\stamp{--- {\bf \today} --- {\bf \jobname.tex}}

\renewcommand{\Re}{\operatorname{Re}}
\renewcommand{\Im}{\operatorname{Im}}
 
\def\stamp{--- {\bf \today} --- {\bf \jobname.tex}}

\def\sign(#1){\textrm{sign}(#1)}

\def\BE{\begin{equation}} 
\def\EE{\end{equation}}

 \def\<#1|#2){\left\langle#1|#2\right\rangle} 
 \def\<#1|#2|#3]{\left\langle#1|#2|#3\right ]} 
\def\(#1|#2|#3>{\left[#1|#2|#3\right\rangle} 
 \def\[#1|#2]{\left[#1|#2\right]}

\def\an[#1,#2]{\left\langle#1\,#2\right\rangle} 
\def\aq[#1,#2,#3]{\left\langle#1|#2|#3\right]} 
\def\qa[#1,#2,#3]{\left[#1|#2|#3\right\rangle} 
\def\sq[#1,#2]{\left[#1\,#2\right]} 
\def\spa#1.#2{\left\langle#1\,#2\right\rangle} 
\def\spab[#1,#2,#3]{\left\langle#1|#2|#3\right]} 
\def\spba[#1,#2,#3]{\left[#1|#2|#3\right\rangle} 
\def\spb#1.#2{\left[#1\,#2\right]} 
\def\lor#1.#2{\left(#1\,#2\right)}

\def\bfTheta{\mathbf{\Theta}}
\def\bfD{\mathbf{D}}
\def\Vbr{V_{\text{br}}}
\def\Vsym{V_{\text{sym}}}

\newcommand{\beq}{\begin{eqnarray}}
\newcommand{\eeq}{\end{eqnarray}}

\newcommand{\ba}{\begin{eqnarray}}
\newcommand{\ea}{\end{eqnarray}}

\newcommand{\baed}{\begin{equation}\begin{aligned}}
\newcommand{\eaed}{\end{aligned}\end{equation}}

\title{Disfavouring Electroweak Baryogenesis and a hidden Higgs in a $CP$-violating Two-Higgs-Doublet Model} 

\author[a]{Anders Haarr,}
\author[b]{Anders Kvellestad,}
\author[c]{Troels C.\ Petersen}

\affiliation[a]{Faculty of Science and Technology, University of Stavanger, \\4036 Stavanger, Norway}
\affiliation[b]{Nordita, KTH Royal Institute of Technology and Stockholm University,
Roslagstullsbacken 23, SE-106 91 Stockholm, Sweden}
\affiliation[c]{Discovery Center, Niels Bohr Institute, University of Copenhagen, Blegdamsvej 17, DK-2100 Copenhagen, Denmark}

\emailAdd{anders.haarr@uis.no}
\emailAdd{anders.kvellestad@nordita.org}
\emailAdd{petersen@nbi.ku.dk}

\preprint{
\begin{flushright}
Stavanger 2016/11\\
NORDITA-2016-121\\
\end{flushright}
}

\abstract{A strongly first-order electroweak phase transition is a necessary requirement for Electroweak Baryogenesis. We investigate the plausibility of obtaining a strong phase transition in a Two-Higgs-Doublet Model of type II with a minimal amount of $CP$ violation. By performing a Bayesian fit where we constrain the scalar sector with indirect and direct measurements, we find that current data disfavours a first-order phase transition in this model. This result is mainly driven by the interplay of three effects: Constraints from the LHC Higgs data on the magnitude of the quartic couplings, the requirement of a $H^\pm$ heavier than around $490$~GeV to avoid large contributions to $BR(b \rightarrow s\gamma)$ and the fact that a first-order phase transition requires relatively light scalar states in addition to the $125$~GeV Higgs. For similar reasons we find that a ``hidden-Higgs'' scenario, in which the $125$~GeV state is identified with the next-to-lightest scalar, is disfavoured by current data independent of any requirement on the phase transition strength.} 

 
\begin{document}
\maketitle

%
%
%
%%%%%%%%%%%%%%%%%%%%%%%%%%%%%%%%%%%%%%%%%%%%%%%%%%%%%%%%%%%%%%%%% 
\section{Introduction}\label{sec:introduction} 
%%%%%%%%%%%%%%%%%%%%%%%%%%%%%%%%%%%%%%%%%%%%%%%%%%%%%%%%%%%%%%
%
The observed baryon asymmetry of the Universe \cite{Kuzmin:1985mm} has, at present, no satisfying explanation. The mechanism of Electroweak Baryogenesis, where baryon number violating processes together with a strongly first-order phase transition (FOPT) and $CP$ violation produce the baryon asymmetry, provides a possible solution.\footnote{For a comprehensive review see \cite{Rubakov:1996vz}. For a more recent review see \cite{Morrissey:2012db}.} Electroweak Baryogenesis fails in the Standard Model for two reasons. Firstly, the Standard Model Higgs mass must be less than $\sim M_W$ to give a first-order phase transition \cite{Kajantie:1996mn}. Secondly, the magnitude of the  $CP$ violation in the Standard Model is not sufficient~\cite{Gavela:1993ts,Gavela:1994ds,Gavela:1994dt,TranbergCP}.
Models with new physics that modify the scalar sector of the SM can provide the necessary first-order phase transition \cite{Anderson:1991zb,Curtin:2014jma,Damgaard:2015con, Damgaard:2013kva,Profumo:2007wc}, but not all of these give additional sources of $CP$ violation, see for instance \cite{Damgaard:2015con}. Two-Higgs-Doublet Models (2HDM) can provide both \cite{Cline:2011mm}.  

However, 2HDMs are strongly constrained by experimental data. In particular, the
close agreement between the measured properties of the $125$~GeV Higgs and SM predictions puts severe constraints on the theory. In 2HDMs this leads to one of two scenarios: Kinematic decoupling, where a single SM-like scalar at $125$~GeV is obtained by pushing up the masses of the other scalars, or the alignment limit, where the mixing in the scalar sector is adjusted so that one of the scalars closely mimics the SM Higgs.

In the following, we consider a 2HDM of type II with a minimal amount of $CP$ violation \cite{Basso:2015dka,Basso:2013hs,Basso:2012st,ElKaffas:2007rq,WahabElKaffas:2007xd,ElKaffas:2006gdt,Khater:2003wq,Chen:2015gaa,Inoue:2014nva} and investigate how the available data from the LHC and elsewhere affect the possibility for a first-order phase transition. The question of phase transition strength in 2HDMs has been the focus of several studies \cite{Arnold:1992rz,Cline:1996mga,Fromme:2006cm,Dorsch:2014qja,Dorsch:2013wja,Blinov:2015vma,Cline:2011mm}. In this study we perform a Bayesian statistical fit and find that current data provides strong evidence against a first-order phase transition in the studied model.
%In this study we perform a Bayesian statistical fit to assess the plausibility of a strong phase transition. We find that current data provides strong evidence against a first-order phase transition in the studied model.

When exploring parameter regions with a strong phase transition we find a clear preference for light scalars. This draws us to perform a dedicated analysis of the ``hidden-Higgs'' scenario, where the 125 GeV resonance is identified as the next-to-lightest Higgs in the model. When calculating the mass spectrum of the model to tree-level accuracy, this region of parameter space avoids significant tension with experimental data. It is however strongly disfavoured when the spectrum calculation is performed at loop level. We emphasize that this conclusion is drawn independently of the strength of the phase transition.

The remainder of the paper is organized as follows: In Sec.\ \ref{sec:The model} we describe the 2HDM under consideration, while Sec.\ \ref{sec:Baryogenesis} is devoted to the one-loop effective potential. Here we also introduce the measure for the phase transition strength. The nuts and bolts of our Bayesian parameter scans are presented in Sec.\ \ref{sec:Parameter scan}, including our choice of free parameters, constraints from observables and the method used for calculating the strength of the phase transition. Our results are presented and discussed in Sec.\ \ref{sec:Results and discussion}. In this section, we perform five different scans. For clarity, the names and purposes of the different scans are summarized in Table \ref{tab:Scans}. Lastly, our conclusions are presented in Sec.\ \ref{sec:Conclusions}.
\begin{table*}[tb]
\centering
\begin{tabular}{llc}
\hline
Scan name & Purpose & Section \\
\hline
\textcolor{darkgray}{Prior} & \multirow{2}{0.56\linewidth}{Illustrate the effective priors for the main and direct-search scans}  & \ref{subsec:Input parameters} \\[1.4\normalbaselineskip]
\textcolor{rossocorsa}{Main} & \multirow{2}{0.56\linewidth}{Scan full parameter space including all experimental data}  & \ref{sec:Results and discussion} \\[1.4\normalbaselineskip]
\textcolor{oceanboatblue}{Direct search}  & \multirow{2}{0.56\linewidth}{Scan full parameter space using only direct Higgs searches} & \ref{sec:Results and discussion} \\[1.4\normalbaselineskip]
\textcolor{fandango}{Strong PT} & Explore region with strong phase transition  & \ref{subsec:Interplay} \\[0.4\normalbaselineskip] 
\textcolor{darkorange}{Hidden Higgs: tree level} & \multirow{1}{0.56\linewidth}{Compare to results in literature}  & \ref{subsec:Hidden Higgs} \\[0.4\normalbaselineskip]
\textcolor{Brown}{Hidden Higgs: loop level} & \multirow{1}{0.56\linewidth}{Extend results in literature}  & \ref{subsec:Hidden Higgs} \\
\hline
\end{tabular}
\caption{Names and purposes of the parameter scans in the paper. The colors refer to the plot colors used for the different scans. %When not referring to a specific scan, grey is used.
}
\label{tab:Scans}
\end{table*} 
%

%
%%%%%%%%%%%%%%%%%%%%
\section{The model}
\label{sec:The model}
%%%%%%%%%%%%%%%%%%%%
%
Two-Higgs-Doublet Models differs from the SM in the scalar part of the Lagrangian:
\ba
   \mathcal{L}_H = \sum\limits_{a=1}\limits^{2} [D_\mu \Phi_a]^{\dagger} [D^\mu \Phi_a] - V(\Phi_1,\Phi_2),
\ea
with the usual covariant derivative
\ba
D^\mu \Phi_a &= [\partial^\mu - ig \sigma_j W^{\mu}_j - \frac{1}{2}ig' B^\mu]\Phi_a \nonumber   
\ea
The most general renormalizable, gauge invariant scalar potential is given by \cite{Branco:2011iw}
\beq
\label{scalarpotential}
V(\Phi_1 ,\Phi_2) = &-&\frac{1}{2} \left[ m_{11}^2 \Phi_1^{\dagger} \Phi_1 + m_{22}^2 \Phi_2^{\dagger} \Phi_2
+ \left( m_{12}^2 \Phi_1^{\dagger} \Phi_2 + \text{h.c.} \right) \right] \nonumber \\
&+& \frac{1}{2} \lambda_1 \left( \Phi_1^{\dagger} \Phi_1 \right)^2 + \frac{1}{2} \lambda_2 \left( \Phi_2^{\dagger} \Phi_2 \right)^2 \nonumber \\
&+& \lambda_3 \left( \Phi_1^{\dagger} \Phi_1 \right) \left( \Phi_2^{\dagger} \Phi_2 \right) + \lambda_4 \left( \Phi_1^{\dagger} \Phi_2 \right) \left( \Phi_2^{\dagger} \Phi_1 \right) \\
&+& \frac{1}{2} \left[ \lambda_5 \left(\Phi_1^{\dagger} \Phi_2 \right)^2 + \text{h.c.}
\right] \nonumber
\nonumber \\
&+& \left[  \left(\lambda_6 \Phi_1^{\dagger} \Phi_1 + \lambda_7 \Phi_2^{\dagger} \Phi_2 \right)\Phi^{\dagger}_1 \Phi_2 + \text{h.c.}
\right] \nonumber,
\eeq
where the parameters $m^2_{11}$, $m^2_{22}$, $\lambda_{1-4}$ are real, while $m^2_{12}$ and $\lambda_{5-7}$ are complex. We constrain the model by taking $\lambda_{6,7}=0$ in Eq.\ (\ref{scalarpotential}). In addition we demand that the vacuum expectation values (VEVs) of the two doublets are real in this basis. This model is sometimes referred to as the $\text{2HDM}_5$ \cite{ElKaffas:2007rq}.

After setting $\lambda_{6,7}=0$ the scalar potential respects the $Z_2$ symmetry $\Phi_1 \rightarrow -\Phi_1$ in the quartic terms, which helps avoid the occurrence of flavour-changing neutral currents (FCNCs). This also limits the amount of $CP$ violation. On the other hand, the term proportional to $m^2_{12}$ in Eq.\ (\ref{scalarpotential}) breaks the $Z_2$ symmetry softly and allows for some $CP$ violation. The term ``soft'' refers to the fact that the $Z_2$ symmetry is respected at small distances to all orders of the perturbative series \cite{Ginzburg:2004vp}.

As thoroughly described in \cite{Grzadkowski:2014ada,Grzadkowski:2013rza}, the $\text{2HDM}_5$ can exhibit $CP$ conservation, explicit $CP$ violation or spontaneous $CP$ violation depending on the choice of parameter values. The conditions for the three different $CP$ scenarios are expressed through the values of three Jarlskog-like invariants. Since we find that the data does not support a FOPT we do not investigate the values of these invariants in our analysis. Previous studies seem to suggest that the phase transition is not sensitive to a $CP$-violating phase \cite{Fromme:2006cm,Dorsch:2013wja}.

With the introduction of $\langle \Phi_1 \rangle \equiv v \text{c}_{\beta}$ and $\langle \Phi_2 \rangle \equiv v \text{s}_{\beta}$,\footnote{$c_\beta = \cos \beta$ and $s_\beta = \sin \beta$} we parametrize the doublets as
\ba
\Phi_1 &=&\left(\begin{array}{c}
\text{G}^+ \text{c}_{\beta}-\text{H}^+ \text{s}_{\beta}\\
\frac{1}{\sqrt{2}}\left[v \text{c}_{\beta} + \eta_1 + i \left(\text{G}^0 \text{c}_{\beta} - \eta_3 \text{s}_{\beta}\right)\right]
\end{array}\right) \label{Phi1_parametrization} \\
\Phi_2 &=&\left(\begin{array}{c}
\text{G}^+ \text{s}_{\beta}+ \text{H}^+ \text{c}_{\beta}\\
\frac{1}{\sqrt{2}}\left[v \text{s}_{\beta} + \eta_2 + i \left(\text{G}^0 \text{s}_{\beta} + \eta_3 \text{c}_{\beta}\right)\right]
\end{array}\right) \label{Phi2_parametrization},
\ea
where the fields $\eta_{1-3}$ and $\text{G}^0$ are real while $\text{H}^+$ and $\text{G}^+$ are complex fields. The fields $\text{G}^0$, $\text{G}^+$ and $\text{H}^+$ are the neutral Goldstone, charged Goldstone and charged Higgs, respectively. The $\eta_{1-3}$ are not mass eigenstates, but will mix among themselves to form the mass eigenstates $\text{H}_{1-3}$.

%
%
%%%%%%%%%%%%%%%%%%%%%%%%%%%%%%%%%%%%%%%%%%%%%%%%%%%%%%%%%%%%%%
\subsection{Extremization, mass matrices and parametrizations}
\label{subsec:Extrem,massm,inputpar}
%%%%%%%%%%%%%%%%%%%%%%%%%%%%%%%%%%%%%%%%%%%%%%%%%%%%%%%%%%%%%%
%
Extremizing the potential through the constraints\footnote{The subscript `vac' denotes the vacuum, where all fields are zero.}
\ba
\frac{\partial V}{\Phi_1}\bigg|_{\text{vac}} = \frac{\partial V}{\Phi^{\dagger}_1}\bigg|_{\text{vac}} = \frac{\partial V}{\Phi_2}\bigg|_{\text{vac}}= \frac{\partial V}{\Phi^{\dagger}_2}\bigg|_{\text{vac}} = 0  
\ea
yields three independent constraints, which can be used to replace the three parameters
$m^2_{11}$, $m^2_{22}$ and $\Im(m^2_{12})$:
\ba
\label{extremization}
m^2_{11} &=& \left[-2\mu^2 + v^2 \lambda_{345}\right]\text{s}^2_{\beta} + v^2 \lambda_1 \text{c}^2_{\beta} \\
m^2_{22} &=& \left[-2\mu^2 + v^2 \lambda_{345}\right]\text{c}^2_{\beta} + v^2 \lambda_2 \text{s}^2_{\beta} \\
\Im(m^2_{12}) &=& v^2 \text{c}_{\beta} \text{s}_{\beta} \lambda_5^I, 
\ea
where $\lambda_5^R \equiv \Re(\lambda_5)$, $\lambda_5^I \equiv \Im(\lambda_5)$, $\mu^2 \equiv \frac{\Re(m^2_{12})}{2 \text{c}_{\beta} \text{s}_{\beta}}$ and $\lambda_{345} \equiv \lambda_3 + \lambda_4 + \lambda_5^R$.

The neutral mass matrix is given by the second derivatives of $V$ with respect to the $\eta_i$:\footnote{The neutral mass matrix is really 4 by 4, but the neutral Goldstone contributes two lines of zeroes and an extra zero eigenvalue.}
\ba
\label{massmatrixvacuum}
\mathcal{M}^2=\left(\begin{array}{ccc}
v^2 \lambda_1 \text{c}^2_{\beta}+ \mu^2 \text{s}^2_{\beta}&(-\mu^2+v^2 \lambda_{345})\text{c}_{\beta}\text{s}_{\beta}&-\frac{1}{2}v^2 \text{s}_{\beta} \lambda_5^I \\

(-\mu^2+v^2 \lambda_{345})\text{c}_{\beta}\text{s}_{\beta}&\mu^2\text{c}^2_{\beta} + v^2 \lambda_2\text{s}^2_{\beta} &-\frac{1}{2}v^2 \text{c}_{\beta} \lambda_5^I\\

-\frac{1}{2}v^2 \text{s}_{\beta} \lambda_5^I&-\frac{1}{2}v^2 \text{c}_{\beta} \lambda_5^I&\mu^2-v^2 \lambda_5^R
\end{array}\right).
\label{eq:neutral_mass_matrix}
\ea
This mass matrix is diagonalized by a matrix $R$, which rotates the $\eta_i$ into the mass eigenstates $H_i$:
\ba
\label{mixingrelations}
\left(\begin{array}{c}
\text{H}_1\\
\text{H}_2\\
\text{H}_3
\end{array}\right) &=& 
R \left(\begin{array}{c}
\eta_1\\
\eta_2\\
\eta_3
\end{array}\right),\\\nonumber \\
R \mathcal{M}^2 R^T = \mathcal{M}^2_{\text{diag}}&=& \text{diag}(m_{H_1}^2,m_{H_2}^2,m_{H_3}^2).
\ea
The neutral mass matrix has six independent components. Diagonalizing the charged mass matrix yields the mass of the charged Higgs
\ba
\label{ChargedHiggsMass}
m^2_{H^\pm} = \frac{1}{2} \left(2 \mu^2 - v^2\left[\lambda_4 + \lambda_5^R \right] \right)
\ea
and a zero eigenvalue corresponding to the charged Goldstone mass. This gives six equations from the neutral sector and one from the charged sector that can be used to exchange the quartic couplings $\lambda_i$ with the physical masses $m_{H_{1-3}},m_{H^\pm}$ and three mixing angles $\alpha_{1-3}$. Since the $\text{2HDM}_5$ only has six $\lambda_i$ there are different ways of substituting parameters. In \cite{Basso:2015dka,Basso:2013hs,Basso:2012st,ElKaffas:2007rq,WahabElKaffas:2007xd,ElKaffas:2006gdt,Khater:2003wq} the convention is to solve for $m_{H_3}$ and keep $\tan\beta$ as an input parameter, while \cite{Chen:2015gaa,Inoue:2014nva} keeps $m_{H_3}$ as an input parameter and solves for $\tan\beta$. 

Here we follow a different approach: We implement the model in the \texttt{Mathematica} package \texttt{SARAH  v.4.8.6} \cite{Staub:2015kfa} and use \texttt{SARAH} to generate a one-loop spectrum generator based on \texttt{SPheno} \cite{SPheno1,SPheno2}. As input parameters we take the seven Lagrangian parameters $\lambda_{1-4}$, $\lambda_5^R$, $\lambda_5^I$, $\Re(m_{12}^2)$ and the symmetry breaking parameter $\tan\beta$. From this input \texttt{SARAH}/\texttt{SPheno} is used to solve the loop-corrected tadpole equations to obtain the remaining Lagrangian parameters $m^2_{11},m^2_{22}$ and $\Im(m^2_{12})$, and calculate the scalar masses at one loop.

%%%%%%%%%%%%%%%%%%%%%%%%%%%%%%%%%%%%%%%%%%%%%%%%%%%%%%%%%%%%%%
\subsection{Yukawa couplings}
\label{subsec:Yukawa couplings}
%%%%%%%%%%%%%%%%%%%%%%%%%%%%%%%%%%%%%%%%%%%%%%%%%%%%%%%%%%%%%%
The Yukawa couplings are the type-II version, where charged leptons and down-type quarks couple to $\Phi_1$ and up-type quarks couple to $\Phi_2$. The couplings of the neutral mass eigenstates $H_j$ to the third-generation quarks are then given by \cite{Basso:2012st}
\begin{align}
H_j b \bar{b}:&\quad \frac{-i g m_b}{2 m_W} \left(\frac{1}{\cos \beta} \left[R_{j1} - i \gamma_5 \sin \beta R_{j3} \right]\right) \\
H_j t \bar{t}:&\quad \frac{-i g m_t}{2 m_W} \left(\frac{1}{\sin \beta} \left[R_{j2} - i \gamma_5 \cos \beta R_{j3} \right]\right),
\end{align}
where the $R_{ji}$ are elements of the rotation matrix $R$ that rotates the $\eta_i$ to the physical fields $H_j$. The factors in the large brackets represent the change relative to the SM couplings. For the charged sector we have the couplings
\begin{align}
H^+ b \bar{t}:&\quad \frac{ig}{2\sqrt{2} m_W} V_{tb} \left[m_b(1+ \gamma_5)\tan \beta + m_t(1- \gamma_5)\cot \beta \right] \\
H^- t \bar{b}:&\quad \frac{ig}{2\sqrt{2} m_W} V_{tb}^{*} \left[m_b(1- \gamma_5)\tan \beta + m_t(1+ \gamma_5)\cot \beta \right].
\end{align}

%%%%%%%%%%%%%%%%%%%%%
\section{The effective potential and strength of the phase transition}
\label{sec:Baryogenesis}
%%%%%%%%%%%%%%%%%%%%
%
In this section we detail the construction of the one-loop effective potential at finite temperature. We further define notions used in our minimization procedure in Sec.\ \ref{subsec:determining PT strength} and our criterion for determining that the electroweak phase transition be first order.

%
%%%%%%%%%%%%%%%%%%%%%%%%%%%%%%%%%%%%%%%%%%%%%%%%
\subsection{One-loop effective potential}
\label{subsec:One-loop effective potential}
%%%%%%%%%%%%%%%%%%%%%%%%%%%%%%%%%%%%%%%%%%%%%%%%
%
We follow \cite{Cline:2011mm} and utilize an $SU(2)_L \times U(1)$ gauge transformation to make the final effective potential a function of $\eta_1$, $\eta_2$, and $\eta_3$. 
Further, we write the effective potential in Landau gauge, in which ghosts decouple but Goldstone bosons are included in the sum over species.

The one-loop effective potential is constructed in the usual way
\beq
V(T)= V_{\rm tree} + V_{\rm ct} + V_{\rm CW}+ V_{T}(T),
\eeq
where $V_{\rm tree}$ is the tree-level potential from Eq.\ (\ref{scalarpotential}), with $\lambda_{6,7}=0$ and $\Phi_1,\Phi_2$ parametrized as in Eqs.\ (\ref{Phi1_parametrization}) and (\ref{Phi2_parametrization}). Below we describe each of the terms contributing to the effective potential in detail. 

%
%%%%%%%%%%%%%%%%%%%%%%%%%%%%%%%%%%%%%%%%%%%%%%%%
\subsubsection{The Coleman-Weinberg contribution $V_{\text{CW}}$}
\label{subsubsec:CW and ct}
%%%%%%%%%%%%%%%%%%%%%%%%%%%%%%%%%%%%%%%%%%%%%%%%
%
When divergences have been subtracted, the one-loop contribution to the effective potential at zero temperature takes the form\footnote{Note that finite-temperature masses enter $V_{\text{CW}}$ after including thermal masses in Sec.\ \ref{subsubsec:finiteT}.} 
\beq
\label{CW-weinberg contribution}
V_{\rm CW}=\sum_{i}\frac{1}{64\pi^2} N_i M_i^4(\eta_j)\left[\log\frac{M_i^2(\eta_j)}{Q^2}-C_i\right],
\eeq
where $M^2_i(\eta_j)$ are the field-dependent masses given in Appendix \ref{app: Field dependent masses}, $i$ sums over particle species and
\beq
\label{particle degeneracy numbers}
N_{t,b}&=&-12,\quad N_{W}=6,\quad N_Z=3,\\
\quad N_{\text{H}_i,\text{G}^0}&=&1,\quad N_{\text{H}_{\pm}}=2 \nonumber.
\eeq
The photon and gluons do not couple directly to the Higgs fields and so give a constant contribution to the potential as a function of $\eta_j$. They therefore play no role in locating the minimum of the potential and are ignored. In the $\overline{\text{MS}}$ scheme, $C_i$ is 5/6 for gauge bosons and 3/2 for the rest. $Q$ is a renormalization scale which we take to be $m_t$. 

%
%%%%%%%%%%%%%%%%%%%%%%%%%%%%%%%%%%%%%%%%%%%%%%%%%%
\subsubsection{Finite-temperature contribution $V_T$}
\label{subsubsec:finiteT}
%%%%%%%%%%%%%%%%%%%%%%%%%%%%%%%%%%%%%%%%%%%%%%%%%%
%
The one-loop correction at finite temperature is given by
\beq
\label{oneloopT}
V_T(\eta_j,T) = \sum_i \frac{T^4}{2\pi^2} N_i \int_0^{\infty} dx\, x^2 \text{log}\Bigg[ 1 \pm e^{-\sqrt{x^2 + \frac{M_i^2(\eta_j)}{T^2}}} \Bigg],
\eeq
where the $N_i$ are defined in (\ref{particle degeneracy numbers}) and $M_i(\eta_j)$ are the field-dependent masses at zero temperature. The plus sign refers to fermions and the minus sign to bosons. This can be expanded around small $\frac{M_i}{T}$ giving \cite{Arnold:1992rz}:
\begin{align}
&\text{const.} + \frac{1}{24} N_i M_i^2(\eta_j) T^2 - N_i \frac{T}{12 \pi} M_i^3(\eta_j) + O(M_i^4), &\text{(bosons)\;}
\label{expansion-bosons} \\
&\text{const.} + \frac{1}{48} N_i M_i^2(\eta_j)T^2 + O(M_i^4), &\text{(fermions)}.
\label{expansion-fermions}
\end{align}
Thermal corrections to the effective masses can now be included in two ways: In the ``Parwani method'' \cite{Cline:1996mga,Parwani:1991gq} the replacement
\ba
M_i^2(\eta_j)\rightarrow M_i^2(\eta_j,T),
\ea
is made in Eqs.\ (\ref{CW-weinberg contribution}) and (\ref{oneloopT}). Alternatively, the ``Arnold-Espinosa method'' \cite{Arnold:1992rz,Carrington} exchanges
\ba
M_i^3(\eta_j)\rightarrow M_i^3(\eta_j,T),
\ea
in Eq.\ (\ref{expansion-bosons}). In practice one adds the term
\beq
\label{eq:ring1}
V_T^{\text{ring}}(\eta_j,T) = \sum_i \frac{T}{12\pi} N_i \text{Tr}\bigg[M_i^3(\eta_j) - M_i^3(\eta_j,T) \bigg],
\eeq
to the potential. The cubic term from the expansion Eq.\ (\ref{expansion-bosons}) is then effectively swapped for one that is also thermally dependent. To implement either of the methods, the thermally corrected masses, $M_i(\eta_j,T)$, must be calculated for each bosonic degree of freedom. The two methods differ by two-loop terms \cite{Cline:1996mga}.

We follow the Parwani method. The integral in Eq.\ (\ref{oneloopT}) is evaluated numerically and tabulated as a function of $\alpha_i(\eta_j,T) = M_i(\eta_j,T)^2/T^2$ for use in our numerical studies. The calculation of the thermal masses is given in Appendix \ref{app: Field dependent masses}.

%
%%%%%%%%%%%%%%%%%%%%%%%%%%%%%%%%%%%%%%%%%%%%%%%%%%
\subsubsection{The tree-level potential $V_{\rm tree}$ and counterterms $V_{\rm ct}$}
\label{subsubsec:Vtree and Vct}
%%%%%%%%%%%%%%%%%%%%%%%%%%%%%%%%%%%%%%%%%%%%%%%%%%
%
The renormalization scheme implemented in \texttt{SARAH} and \texttt{Spheno} is the $\overline{\text{MS}}$ scheme. The specific renormalization procedure is described in \cite{Staub:2015kfa}. In constructing $V_{\rm tree} + V_{\rm ct}$ we follow \cite{Camargo-Molina:2013qva}. $\texttt{Spheno}$ outputs both tree-level and one-loop parameters. The one-loop parameters inserted into the tree-level potential gives $V_{\rm tree} + V_{\rm ct}$, while the tree-level parameters go into the thermal masses. The thermal masses are used in the one-loop corrections to $V$ described in Sections \ref{subsubsec:CW and ct} and \ref{subsubsec:finiteT}. We note that the calculations in \texttt{Spheno} are performed in the Feynman-'t Hooft gauge, which is the $\xi \rightarrow 1$ limit of the class of $R_{\xi}$ gauges, while the effective potential is written in the Landau gauge, which is the limit $\xi \rightarrow 0$. As noted in \cite{Camargo-Molina:2013qva}, the VEVs are slightly different in these gauges. Numerically, we find this effect to be small in the physically interesting regions of parameter space.  

%
%%%%%%%%%%%%%%%%%%%%%%%%%%%%%%%%%%%%%%%% 
\subsection{The phase transition}
\label{subsec:Phase transition}
%%%%%%%%%%%%%%%%%%%%%%%%%%%%%%%%%%%%%%%%
%
At very high temperatures we have symmetry restoration, which in our parametrization means that the global minimum of the effective potential tends towards the point $(\eta_1,\eta_2,\eta_3) = (-v \cos \beta,-v \sin \beta,0)$. We refer to the minimum that resides at this point at very high temperatures as the \textit{symmetric minimum}. 

At zero temperature the symmetry of the theory has been spontaneously broken and each of the Higgs doublets acquire a VEV, defined by the global minimum of the effective potential.\footnote{For simplicity we disregard the possibility that the broken minimum is only metastable, with a lifetime longer than the age of the universe.} After \texttt{SARAH}/\texttt{SPheno} solves the tadpole equations, the point $(\eta_1,\eta_2,\eta_3)=(0,0,0)$ is guaranteed to be an extremal point of the effective potential, but not necessarily a minimum. At this point the doublets satisfy $\Phi_1^{\dagger}\Phi_1 + \Phi_2^{\dagger}\Phi_2 = v^2 \approx (246\;\text{GeV})^2$. For a given parameter point it is therefore necessary to check that $(\eta_1,\eta_2,\eta_3)=(0,0,0)$ is indeed the global minimum of the zero-temperature potential. In Sec.\ \ref{subsec:determining PT strength} we bundle this criterion into our stability condition. We refer to the minimum residing at the point $(\eta_1,\eta_2,\eta_3)=(0,0,0)$ when $T=0$ as the \textit{broken minimum}.

At intermediate temperatures the broken minimum will typically have shifted to a new position in field space. The same is true for the symmetric minimum, although the displacement in field space is generally small. We still refer to these minima as the broken and symmetric minimum at intermediate temperatures because we can track their movement in field space numerically as $T$ varies.

The \textit{critical temperature} $T_c$ of the phase transition is defined as the temperature at which the global minimum shifts from the symmetric to the broken minimum, as $T$ decreases. In a first-order phase transition, the two degenerate minima $\Vbr(T_c)$ and $\Vsym(T_c)$ will be at different points in field space, typically with a potential barrier in between. For a second order (cross-over) transition, the broken and symmetric minimum are not degenerate until they are at the same point in field space. For such transitions, the minima melt together so that the global minimum moves continuously (smoothly) from the symmetric minimum to the broken minimum as the temperature is lowered.

The strength of the electroweak phase transition is quantified by the ratio\footnote{The criterion has some issues related to gauge-invariance \cite{Dolan:1973qd}. Still, the numerical difference is believed to be small for appropriate choices of gauge, such as the Landau gauge \cite{Garny:2012cg,Wainwright:2011qy}. See also \cite{Patel:2011th}.} 
\begin{align}
  \xi_c = \frac{v_c}{T_c}.
  \label{eq:PT_strenth}
\end{align}
The parameter $v_c$ is the VEV of the broken minimum at this temperature. In the literature a common demand for a strong enough phase transition is $\xi_c > 1$. This criterion stems from the requirement that sphalerons should be inactive in the broken phase and not equilibriate the baryon asymmetry \cite{Rubakov:1996vz}. The criterion can be relaxed to $\xi \gtrsim 0.5$ by encorporating uncertainties related to the sphaleron rate in the broken phase \cite{sphaleroncriterion}. In Sec.\ \ref{subsec:determining PT strength} we detail our method for determining $\xi_c$.

%
%%%%%%%%%%%%%%%%%%%%
\section{Parameter scan}
\label{sec:Parameter scan}
%%%%%%%%%%%%%%%%%%%%
%
In order to determine the most probable $\text{2HDM}_5$ parameter regions in light of current data we perform a Bayesian fit of the model. The goal of such a fit is to obtain the posterior probability density $p(\bfTheta|\bfD)$ for the set of model parameters $\bfTheta$, given the experimental data $\bfD$. From Bayes' theorem the posterior distribution $p(\bfTheta|\bfD)$ is given by the prior distribution $p(\bfTheta) \equiv \pi(\bfTheta)$ and the likelihood $p(\bfD|\bfTheta) \equiv \mathcal{L}(\bfTheta)$, 
\begin{align}
  p(\bfTheta|\bfD) = \frac{\mathcal{L}(\bfTheta) \pi(\bfTheta)}{P(\bfD)}.
  \label{eq:Bayes}
\end{align}
The denominator $P(\bfD) = \int\mathcal{L}(\bfTheta) \pi(\bfTheta) d\bfTheta$ is the Bayesian evidence, which in our analysis only serves to normalize the posterior distribution. Our choice of prior distribution and the construction of the likelihood function will be detailed in Sections \ref{subsec:Input parameters} and \ref{subsec:likelihood}.

To numerically estimate $p(\bfTheta|\bfD)$ we scan the parameter space with \texttt{MultiNest v3.10} \cite{Multinest1,Multinest2,Multinest3}, via the Python interface \texttt{PyMultiNest} \cite{Pymulti}. In short, \texttt{MultiNest} explores the parameter space based on a nested sampling algorithm in which a fixed number of ``live points'' is gradually moved towards regions of increasing likelihood \cite{Skilling}. This is achieved by repeatedly sampling the parameter space until a point is found with higher likelihood than the current lowest-likelihood live point. The lowest-likelihood point is then replaced by the new point in the set of live points and the process is repeated. The main purpose of \texttt{MultiNest} is to estimate the Bayesian evidence $P(\bfD)$, but once the evidence has been estimated the set of current and previous live points can be re-weighted according to Bayes' theorem to produce a representative set of samples from $p(\bfTheta|\bfD)$. The performance of \texttt{MultiNest} is governed by three main settings: The number of live points (\texttt{nlive}), where a larger number ensures better coverage of the parameter space; the convergence criteria (\texttt{tol}) on \texttt{MultiNest}'s estimate of $\ln\left[P(\bfD)\right]$; and an approximate sampling efficiency (\texttt{efr}) describing how strictly the sampling region should be narrowed down as the scan progresses towards higher-likelihood regions of parameter space.  

The posterior predictive density for a derived quantity $f(\bfTheta)$, 
\begin{align}
  p(f|\bfD) = \int p(f,\bfTheta|\bfD) d\bfTheta,
  \label{eq:PT_post_pred}
\end{align}
is obtained from the posterior samples by calculating $f(\bfTheta)$ for each sample and then marginalize over the model parameters by histogramming the samples in $f$. In particular, we will be interested in the posterior density $p(\xi_c|\bfD)$ for the phase transition strength $\xi_c$.

The prior and posterior distributions in a Bayesian analysis are probability densities. This has two important implications for how the results should be interpreted. First, only the shapes of the distributions are relevant. For simplicity we therefore normalize the height of the maximum probability point to unity in all plotted distributions. The total number of samples in a distribution is only a measure of the accuracy with which the probability density has been determined. Second, there is a dependence on parameter space volume. For instance, given two parameter regions with similar likelihood values, the fit will prefer the largest parameter region as measured by the integrated prior probability. In this way, a Bayesian fit automatically and consistently penalizes fine tuning in the parameters. However, this also means that if the model is defined with a large parameter space that is weakly constrained by data, this effect can dominate the posterior results.

%
%
%
%%%%%%%%%%%%%%%%%%%%%%%%%%%%%%%%%%%%%%%%%%%%%%%%%%%%%%%%%%%%%%
\subsection{Input parameters and prior distribution}
\label{subsec:Input parameters}
%%%%%%%%%%%%%%%%%%%%%%%%%%%%%%%%%%%%%%%%%%%%%%%%%%%%%%%%%%%%%% 
%
Here we describe the input parameters and joint prior distribution used for the main scan in Sec.\ \ref{sec:Results and discussion}. For the other scans we specify any changes relative to the main setup when discussing the respective scan results.

As our input parameters we take
\begin{align}
  \lambda_1,\, \lambda_2,\, \lambda_3,\, \lambda_4,\, \lambda_5^R,\, \lambda_5^I,\, \Re(m_{12}^{2}),\, \tan\beta.
  \label{eq:parameters}
\end{align}
The remaining mass-squared Lagrangian parameters $m_{11}^{2}$, $m_{12}^{2}$ and $\Im(m_{12}^{2})$ are determined by \texttt{SPheno} by solving the loop-corrected tadpole equations.

Our joint prior distribution is based on independent priors for the input parameters. For the quartic couplings we choose flat prior distributions cut off at the perturbativity bound of $|4\pi|$. Also $\tan\beta$ is assigned a flat prior, with the range $[0.1,60]$. $\Re(m_{12}^{2})$ is the only dimensionful input parameter and sets the overall mass scale for the scalar sector. For this parameter we choose a log prior, i.e.\ a prior distribution flat in $\log[\Re(m_{12}^{2})]$, as this is the least informative prior for a scale parameter.\footnote{Most other 2HDMs scans in the literature sample the parameter space using flat distributions for the mass scale parameter or the physical masses. Looking at the results in Sec.\ \ref{sec:Results and discussion}, a flat prior for $\Re(m_{12}^2)$ would only serve to weaken the probability of a first-order phase transition in the $\text{2HDM}_5$, thus strengthening the conclusion of our paper.} Since $\Re(m_{12}^{2})$ can be negative, we formulate a ``two-sided'' log prior that combines a log prior for the magnitude $|\Re(m_{12}^{2})|$ over the range $[10^2,1500^2]$~GeV$^2$ with a $50/50$ prior on the sign. The two sides of the distribution are joined by a flat distribution on the range $[-10^2,10^2]$~GeV$^2$. Our choice of individual parameter priors for the main scan is summarized in the left-hand side of Table~\ref{tab:priorsMain}.
\begin{table*}[tb]
\centering
\textbf{Priors and settings for main scan} \\
\begin{tabular}{lll|ll}
\hline
 Parameter & Range & Type & \texttt{MultiNest} setting & Value  \\
\hline
$\lambda_1$  & $[0,\,4\pi]$  & flat &\texttt{nlive} & 4000  \\ 
$\lambda_2$  & $[0,\,4\pi]$  & flat & \texttt{tol} & 0.5  \\ 
$\lambda_3$  & $[-4\pi,\,4\pi]$  & flat & \texttt{efr} & 0.8 \\ \cline{4-5}  
$\lambda_4$  & $[-4\pi,\,4\pi]$  & flat & Likelihood & Included \\ \cline{4-5}  
$\lambda_5^R$  & $[-4\pi,\,4\pi]$  & flat & Stability & Yes \\ 
$\lambda_5^I$  & $[-4\pi,\,4\pi]$  & flat & Direct searches & Yes \\ 
$\tan\beta$  & $[0.1,\,60]$  & flat & $\Delta \rho$ + B physics & Yes  \\  
$\Re(m_{12}^2)$  & $[-1500^2,\, 1500^2]$ $\text{GeV}^2$  & two-sided log & Strong PT & No \\ \cline{4-5}
 & & & Sampled points & Posterior \\ \cline{4-5}
 & & & $1.6 \times 10^7$ & $3.3 \times 10^4$ \\
\hline
\end{tabular}
\caption{Prior distributions, likelihood contributions and \texttt{MultiNest} settings for the main scan. The ``two-sided'' log prior for $\Re(m_{12}^2)$ is a log prior for the magnitude $|\Re(m_{12}^2)|$ combined with a $50/50$ prior for the sign of $\Re(m_{12}^2)$. Since the log prior diverges when $\Re(m_{12}^{2}) \rightarrow 0$ this prior distribution is joined by a flat prior on the range $[-10^2,10^2]$~GeV$^2$.}
\label{tab:priorsMain}
\end{table*}

While $\Re(m_{12}^2)$ sets the overall mass scale, the physical scalar masses can still deviate significantly from $\sqrt{|\Re(m_{12}^2)|}$ due to $\tan\beta$ and the quartic couplings. Earlier investigations of the phase transition strength in 2HDMs indicate that a not too heavy scalar mass spectrum is required for a first-order phase transition, see e.g.\ \cite{Dorsch:2013wja,Dorsch:2014qja,Cline:2011mm}.\footnote{However, a significant mass splitting between the heavy scalars seems to improve the possibility for a first-order phase transition.} We have confirmed this in preliminary investigations of the high-mass region ($> 1$~TeV) of our model. To focus our study on parameter regions that may provide a first-order phase transition, we impose the requirement that all scalars should be lighter than $1$~TeV. 

\begin{figure*}[t]
\centering
\includegraphics[width=0.24\textwidth]{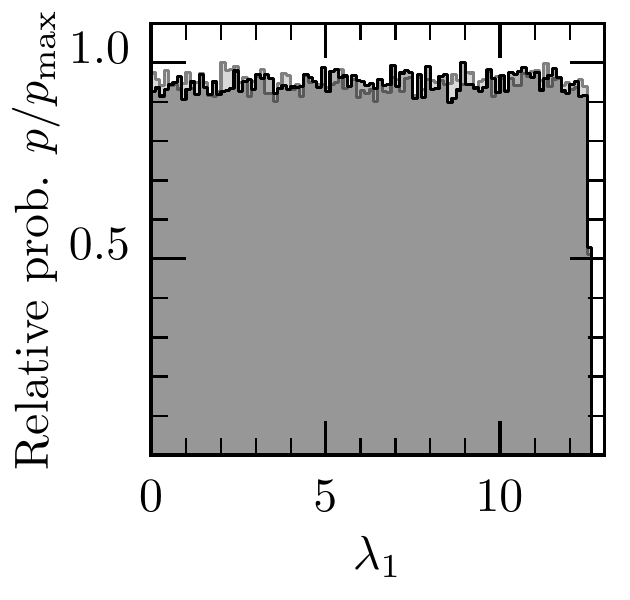}
\includegraphics[width=0.24\textwidth]{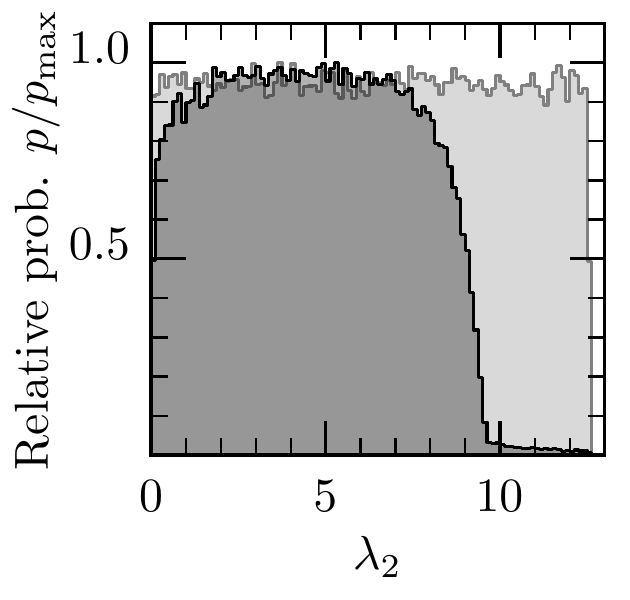}
\includegraphics[width=0.24\textwidth]{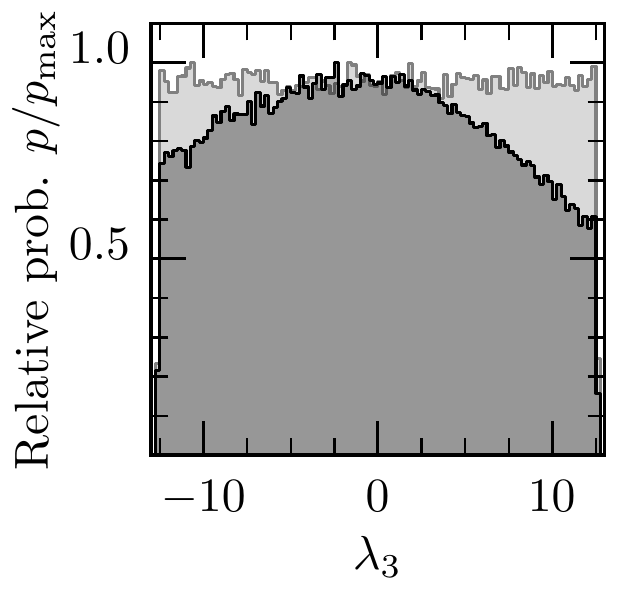}
\includegraphics[width=0.24\textwidth]{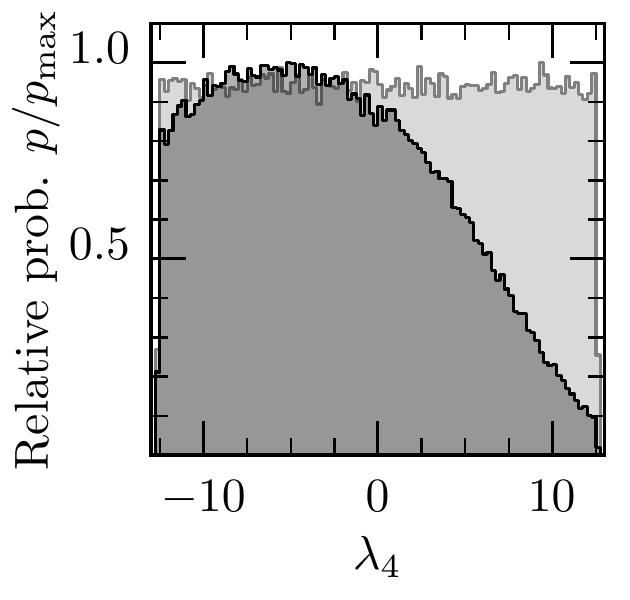}\\
\includegraphics[width=0.24\textwidth]{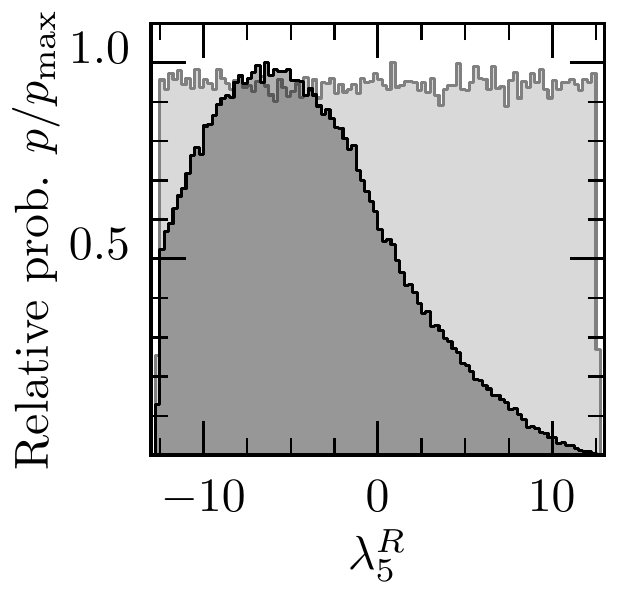}
\includegraphics[width=0.24\textwidth]{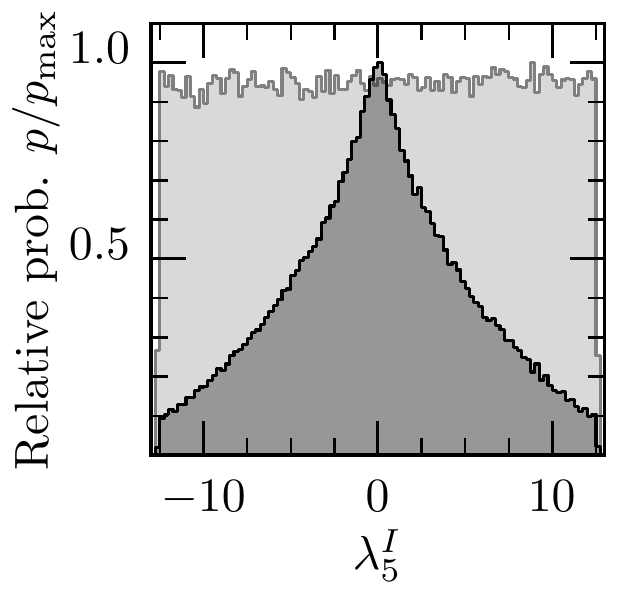}
\includegraphics[width=0.24\textwidth]{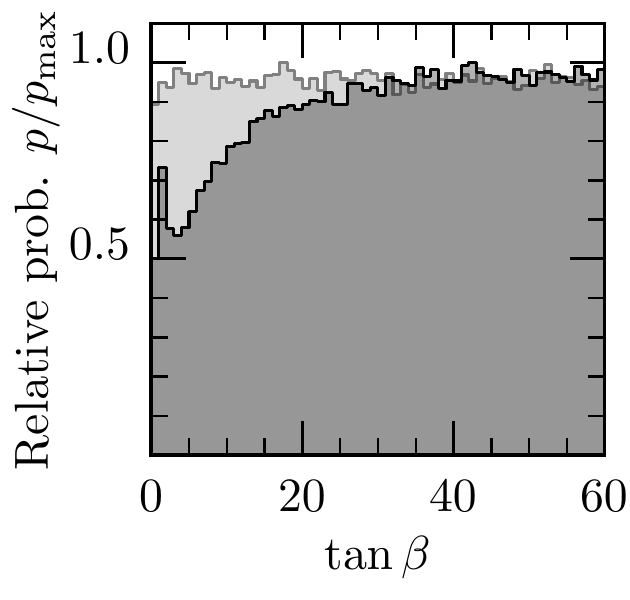}
\includegraphics[width=0.24\textwidth]{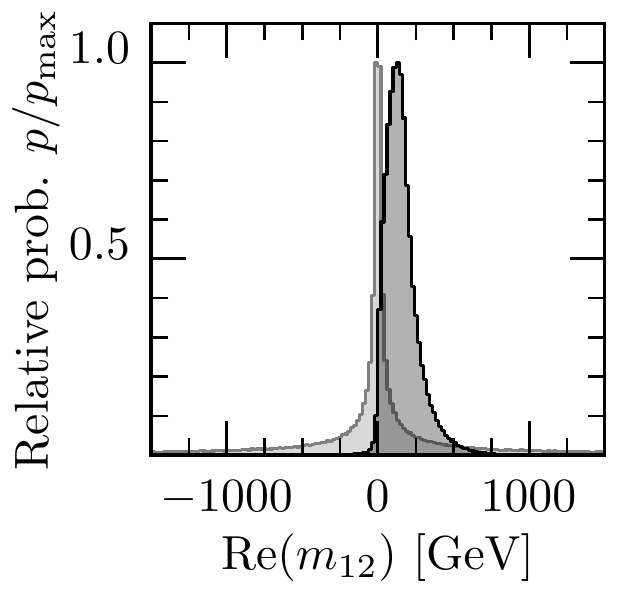}
\caption{\textit{Light grey:} Prior distributions for the input parameters.  \textit{Dark grey:} Effective prior distributions after requiring a physical spectrum with all scalar masses below $1$~TeV. Here and in subsequent figures we use the simplified notation $\Re(m_{12}) \equiv \text{sgn}[\Re(m_{12}^2)] \sqrt{|\Re(m_{12}^2)|}$.}
\label{fig:priors}
\end{figure*}

With this additional constraint on the scalar masses the effective prior for our model is given by  
\ba
  \pi(\bfTheta) = \prod\limits_i H\left(1 - \dfrac{m_{H_i}(\bfTheta)}{\text{TeV}}\right) \prod\limits_j \pi_j(\theta_j),
  \label{eq:priorMain}
\ea
where $H$ is the Heaviside step function, $m_{H_i}$ are the scalar masses and $\pi_j$ are the individual parameter priors, appropriately normalized. Implicit in the constraint from $H$ is of course also the requirement that the parameter point $\bfTheta$ gives a physically valid spectrum. In Fig.~\ref{fig:priors} we show two sets of marginalized one-dimensional priors for the input parameters.  The light grey distributions are the input priors from Tables~\ref{tab:priorsMain} and \ref{tab:priorsDirect}, while the dark grey distributions are the effective priors resulting from from Eq.\ (\ref{eq:priorMain}). We note that the joint requirement of obtaining a physical spectrum and having $m_{H_i} < 1$~TeV affects the parameter distributions significantly. In particular, the probability for $\Re(m_{12}^2) < (0\;\text{GeV})^2$ or $\Re(m_{12}^2) > (750\;\text{GeV})^2$ is greatly reduced.    

%
%%%%%%%%%%%%%%%%%%%%%%%%%%%%%%%%%%%%%%%%%%%%%%%%%%%%%%%%%%%%%%
\subsection{Likelihood function}
\label{subsec:likelihood}
%%%%%%%%%%%%%%%%%%%%%%%%%%%%%%%%%%%%%%%%%%%%%%%%%%%%%%%%%%%%%% 
%
The likelihood function $\mathcal{L}(\bfTheta)$ is obtained by taking the predicted joint probability distribution $p(\bfD|\bfTheta)$ for the data $\bfD$ and regard this as a function of the input parameters $\bfTheta$. When the data consists of independent observations $D_i$ the likelihood function factorizes as  
\ba
  \mathcal{L}(\bfTheta) = \prod\limits_i \mathcal{L}_i(\bfTheta).
  \label{eq:likelihood}
\ea

We construct our likelihood function with contributions from the electroweak precision observable $\Delta \rho$; the $B$-physics observables $\Delta M_{B_s}$, $BR(b \rightarrow s \gamma)$ and $BR(B_u \rightarrow \tau \nu_{\tau})$; 95\% CL exclusion limits on $\sigma \times BR$ from LEP, Tevatron and LHC Higgs searches, as derived by \texttt{HiggsBounds v4.3.1} \cite{HB-1,HB-2,HB-3,HB-4,HB-5}; and a set of 85 signal rate and  4 mass measurements for the observed $125$~GeV Higgs, checked using \texttt{HiggsSignals v1.4.0} \cite{HS-1,HS-2,HS-3}. 

For $\Delta \rho$, $\Delta M_{B_s}$, $BR(b \rightarrow s \gamma)$ and $BR(B_u \rightarrow \tau \nu_{\tau})$ we use likelihoods that are gaussian in the predicted observable, centred on the observed value and with a width determined by adding experimental and theoretical errors in quadrature. When only 95\% CL exclusion limits are available, as for the $\sigma \times BR$ limits checked with \texttt{HiggsBounds}, we follow the approach of Ref.\ \cite{deAustri:2006jwj} and formulate a likelihood that is a step function at the observed limit, smeared by a 5\% gaussian uncertainty. For the 125 GeV Higgs, \texttt{HiggsSignals} calculates a combined $\chi^2$-value for the included set of mass and signal rate measurements, taking available correlations into account. We then take the likelihood contribution to be\footnote{This likelihood contribution is missing the usual gaussian normalization factor, $(\sqrt{2\pi}\sigma)^{-1}$. A constant factor in a likelihood does not affect the posterior distribution as it cancels between the likelihood and evidence in Eq.\ (\ref{eq:Bayes}). With a parameter-dependent theory error, $\sigma=\sigma(\bfTheta)$, this is no longer true. However, when theory errors are estimated to be a fraction of the theory prediction and the data is strongly constraining, as is the case for the $125$~GeV Higgs, this effect remains small across the parameter regions that give a reasonable fit to the data.}
\ba
  \mathcal{L}_{\text{HS}} = \exp\left(-\frac{\chi^2}{2}\right).
  \label{eq:L_HS}
\ea
Below we give some further details on the individual observables.

The new contribution to $\Delta \rho$ coming from the additional scalars in the $\text{2HDM}_5$ is calculated using Eqs.\ (4.10) and (4.11) in Ref.\ \cite{ElKaffas:2006gdt}. From fits to the SM \cite{PDG}, the allowed BSM contribution is constrained to   
% http://pdg.lbl.gov/2014/reviews/rpp2014-rev-standard-model.pdf,
%
\ba
  \Delta \rho = (4.0 \pm 2.4) \times 10^{-4}.
  \label{eq:delta rho}
\ea
Phenomenologically, the effect of this constraint is to force the mass splitting between $H^\pm$ and either $H_2$ or $H_3$ to be small, typically within a few tens of GeV.  

The $B_s$-$\bar{B}_s$ mixing observable $\Delta M_{B_s}$ is sensitive to box diagrams involving $H^\pm$ and quarks, with the top quark contribution generally dominating. We use the theory calculation provided by the \texttt{SARAH}-generated version of \texttt{SPheno}. To this calculation we assign a conservative 20\% theory error to account for uncertainties in phenomenological parameters in the hadronic matrix element, and the fact that the calculation only includes the top quark contributions. For the observed value we take the LHCb and CDF average from Ref.\ \cite{Amhis:2014hma},  
\ba
  \Delta M_{B_s} = 17.757 \pm 0.021\;\text{ps}^{-1}.
  \label{eq:delta M_B_s}
\ea
Due to the dependence on the top Yukawa coupling, the main effect of this observable is to disfavour very low values of $\tan\beta$.

The decay $B_u \rightarrow \tau \nu$ can be mediated by a charged Higgs at tree level. In the type-II 2HDM this alters the predicted $BR(B_u \rightarrow \tau \nu)$ by a factor $[1-(m_B^2/m_{H^\pm}^2)\tan^2\beta]^2$ relative to the SM prediction \cite{Akeroyd:2007eh,Akeroyd:2003zr}. Hence, for small $m_{H^\pm}$ and large $\tan\beta$, the predicted branching ratio can deviate significantly from the SM prediction. We calculate the predicted branching ratio using \texttt{SuperIso v3.4} \cite{Mahmoudi:2007vz,Mahmoudi:2008tp} and take the experimental average
\ba
  BR(B_u \rightarrow \tau \nu) = (1.14 \pm 0.22) \times 10^{-4} 
  \label{eq:BR_B_tau_nu}
\ea
as the observed value~\cite{Amhis:2014hma}.

At leading order in the SM, the FCNC process $b \rightarrow s \gamma$ proceeds through a $W$-quark loop. In the 2HDM there are additional contributions from loops where the $W$ is replaced by a charged Higgs. We calculate the theory prediction with \texttt{SuperIso} and use the average from Ref.\ \cite{Amhis:2014hma} as the observed value, 
\ba
  BR(b \rightarrow s \gamma) = (3.43 \pm 0.22) \times 10^{-4}, 
  \label{eq:BR_b_s_gamma}
\ea
with statistical and systematic uncertainties added in quadrature. We find that, at large $\tan\beta$, $m_H^{\pm} > 493$~GeV is required to avoid $>2\sigma$ tension with the observed value, in agreement with Refs.\ \cite{Misiak:2015xwa,Enomoto:2015wbn}. For low $\tan\beta$ the constraint on $m_{H^{\pm}}$ gets even stronger.

We use the effective coupling approach in \texttt{HiggsBounds} to test the extended Higgs sector of the 2HDM against exclusion limits from Higgs searches at LEP, Tevatron and LHC. Here we give a brief summary of the method: First, for each Higgs boson $H_i$ the relevant theory prediction, typically $(\sigma\times BR)_{H_i}^{\text{th}}$, is calculated for each search, using masses, decay rates and effective couplings provided by \texttt{SPheno}. This is then compared to the \textit{expected} limits for the different searches, in order to determine which search is expected to give the strongest constraint. Finally, for the search channel found to have the strongest expected sensitivity, the theory prediction is compared to the observed limit through the ratio
\ba
  R_i = \frac{(\sigma \times BR)_{H_i}^{\text{th}}}{(\sigma \times BR)_{H_i}^{\text{lim}}},
\ea 
so that $R_i > 1$ signifies exclusion at the 95\% CL. The overall result is then taken to be 
\ba
  R = \max\limits_i R_i, 
\ea 
which we use to formulate our smeared upper-limit likelihood at $R=1$. For further details on the methodology of \texttt{HiggsBounds} and which searches are included, we refer the reader to Refs.\ \cite{HB-1,HB-2,HB-3,HB-4,HB-5}.

To test the model against the measurements of the observed $125$ GeV signal, we run \texttt{HiggsSignals} in the ``peak-centered'' mode. In this mode, \texttt{HiggsSignals} calculates a $\chi^2$ for the hypothesis that a signal observed at a specific mass is explained by the model. The total $\chi^2$ is a sum of contributions from both signal rate and mass measurements. For the signal rates, the basis for comparison with experimental results is the signal rate modifier,
\ba
  \mu = \sum\limits_i \omega_i \frac{(\sigma \times BR)_i^{\text{th}}}{(\sigma \times BR)_i^{\text{SM}}}. 
\ea
Here $i$ denotes the specific channels considered in the experimental analysis and the $\omega_i$ are the SM channel weights, defined as
\ba
  \omega_i = \frac{\epsilon_i (\sigma \times BR)_i^{\text{SM}}}{\sum\limits_j \epsilon_j (\sigma \times BR)_j^{\text{SM}}}, 
\ea
where $\epsilon_i$ is the relative experimental efficiency for channel $i$. We run \texttt{HiggsSignals} with an assigned 1\% theory error on the predicted Higgs boson masses. Further details on the $\chi^2$ calculation in \texttt{HiggsSignals} can be found in Refs.\ \cite{HS-1,HS-2,HS-3}. 

We use the most up-to-date Higgs data sets included in \texttt{HiggsBounds v4.3.1} and \texttt{HiggsSignals v1.4.0}. For the $125$ GeV Higgs data in \texttt{HiggsSignals} this includes analyses published prior to July 2015,\footnote{While this does not include the final combination of ATLAS and CMS Higgs measurements at 7/8 TeV \cite{Khachatryan:2016vau}, almost all the individual measurements contributing to that combination are included.} while for \texttt{HiggsBounds} analyses published before October 2015 are included. No Higgs searches at $13$~TeV are included in our analysis. As we will see, the level of agreement with SM predictions seen in the 7/8 TeV data is already enough to strongly favor the regions of $\text{2HDM}_5$ parameter space with SM-like predictions. Since the ATLAS+CMS Higgs combination \cite{Khachatryan:2016vau} and $13$~TeV Higgs data so far \cite{ATLAS_Hgg_13TeV, ATLAS_HZZ_13TeV, CMS_Hgg_13TeV, CMS_HZZ_13TeV} show no evidence of deviations from the SM, we expect that including this data would only strengthen the conclusions of our analysis.

In the next subsection we detail our check of positivity and stability for the zero-temperature effective potential. Parameter points that fail this check are assigned zero likelihood. We do not impose the commonly used constraint on tree-level perturbative unitarity \cite{Ginzburg:2005dt}, which places upper bounds on combinations of the quartic couplings.\footnote{Perturbative unitarity bounds have recently been computed at the one-loop level for $CP$-conserving 2HDMs \cite{Grinstein:2015rtl}. The impact of this on a Bayesian global fit is studied in \cite{Cacchio:2016qyh}.} However, we have investigated the expected impact by checking how the posterior distributions from our scans change when discarding samples that fail this constraint. Relative to the results presented in Sec.\ \ref{sec:Results and discussion}, the most important difference is that $\lambda_1$ is constrained to $\lambda_1 < 8.4$. For the main and direct-search scans in Sec.\ \ref{sec:Results and discussion} the only other difference is a slightly increased preference for lower values of $|\lambda_3|$, $|\lambda_4|$ and $|\lambda_5^R|$, as well as for lower values of $\tan\beta$, due to a correlation with $\lambda_1$. This has no impact on the conclusions we draw from these scans. For the strong-PT scan presented in Sec.\ \ref{subsec:Interplay} the perturbative unitarity constraint has virtually no impact on the posterior distributions. In the case of the hidden-Higgs scans the impact is more significant and we give a dedicated discussion of this in Sec.\ \ref{subsec:Hidden Higgs}. We will find that enforcing tree-level perturbative unitarity strengthens the conclusion that the hidden-Higgs scenario is disfavoured by current data.

%
%%%%%%%%%%%%%%%%%%%%%%%%%%%%%%%%%%%%%%%% 
\subsection{Stability condition and calculation of $\xi_c$ }
\label{subsec:determining PT strength}
%%%%%%%%%%%%%%%%%%%%%%%%%%%%%%%%%%%%%%%% 
%
Before calculating the phase transition strength $\xi_c$ for a given parameter point, we need to establish the positivity of $V(T=0)$ and the stability of the broken minimum. Using \texttt{MINUIT2}~\cite{James:1975dr} we first minimize $V(T=0)$ in the vicinity of $(\eta_1,\eta_2,\eta_3) = (0,0,0)$ to establish that the vacuum corresponds to a minimum of the effective potential. Next, we minimize $V(T=0)$ multiple times, starting from different points at large field values. If a minimum deeper than the SM minimum is found, or if the minimization algorithm tends towards too large field values, $|\eta_i| > 20 m_t$, we regard the potential as unstable. Similarly, we also check that for very high temperatures, the only minimum of $V(T)$ is the symmetric minimum, $V_{\text{sym}}$, in $(\eta_1,\eta_2,\eta_3) = (-v \cos \beta,-v \sin \beta,0)$. In what follows we will refer to these constraints collectively as the ``stability condition''. 

The complete stability check requires repeated numerical minimization of the effective potential. Due to this computational expense it is impractical to include this check directly in the scan likelihood for a large scan. A workaround is then to run the scan with the likelihood constructed from the observables described in the previous section, and afterwards perform the stability check on all points in the resulting posterior distribution. By discarding posterior points that fail the stability condition we effectively re-weight the posterior to incorporate the stability condition. As long as the scan sample the posterior distribution with high enough accuracy, this re-weighting is equivalent to including the stability check directly in the scan likelihood. This is the approach we use for the main and direct-search scans in Sec.\ \ref{sec:Results and discussion}, which are the largest scans in our study. For the medium-sized hidden-Higgs scans in Sec.\ \ref{subsec:Hidden Higgs} the check that the vacuum is a minimum is included in the scan likelihood, but the complete stability check is again only done on the resulting posterior samples. Finally, for the smaller strong-PT scan in Sec.\ \ref{subsec:Interplay} the complete stability check is included directly in the scan likelihood. 

For points that pass the above checks we can go on to calculate $\xi_c$. We need to determine the critical temperature $T_c$ and the field space distance $v_c$ between $\Vbr(T_c)$ and $\Vsym(T_c)$. A simple algorithm is used to estimate $T_c$: Starting from $T=0$~GeV, we increase the temperature in small steps $\Delta T$, for each step tracking the position of the broken minimum by minimizing the potential with \texttt{MINUIT2}. When a temperature is reached for which $\Vsym(T) < \Vbr(T)$, the step size $\Delta T$ is reduced and the algorithm is repeated, starting from the previous temperature, for which $\Vsym(T) > \Vbr(T)$. We use the step sizes $\Delta T = 20$, $2$ and $0.2$~GeV. As pointed out in \cite{Cline:2011mm}, the symmetric minimum can develop a non-zero VEV at lower temperatures. To take this into account we track the position of $\Vsym$ by minimizing the potential in the region around $(\eta_1,\eta_2,\eta_3) = (-v \cos \beta,-v \sin \beta,0)$ each time $\Delta T$ is reduced. We note that there is a possibility of having several minima at intermediate temperatures, which may alternate between being the global minimum. Since we track the global minimum from its position at zero temperature we might miss multi-step transitions occurring at intermediate temperatures. 

%
%%%%%%%%%%%%%%%%%%%%
\section{Results and discussion}
\label{sec:Results and discussion}
%%%%%%%%%%%%%%%%%%%%
%

For the result of our main scan we combine the posterior samples from four identical \texttt{MultiNest} scans. The scan likelihood function include all the observables discussed in Sec.\ \ref{subsec:likelihood}. The choice of priors, \texttt{MultiNest} settings and included likelihood contributions are summarized in Table \ref{tab:priorsMain}. In total, the scans visited around $1.6 \times 10^7$ parameter points, producing a posterior distribution of $6 \times 10^4$ points. After discarding posterior samples that fail the stability condition, we end up with a posterior distribution of $3.3 \times 10^4$ samples.   

\begin{table*}[tb]
\centering
\textbf{Priors and settings for direct-search scan} \\
\begin{tabular}{lll|ll}
\hline
 Parameter & Range & Type & \texttt{MultiNest} setting & Value  \\
\hline
$\lambda_1$  & $[0,\,4\pi]$  & flat &\texttt{nlive} & 4000  \\ 
$\lambda_2$  & $[0,\,4\pi]$  & flat & \texttt{tol} & 0.5  \\ 
$\lambda_3$  & $[-4\pi,\,4\pi]$  & flat & \texttt{efr} & 0.8 \\ \cline{4-5}  
$\lambda_4$  & $[-4\pi,\,4\pi]$  & flat & Likelihood & Included \\ \cline{4-5}  
$\lambda_5^R$  & $[-4\pi,\,4\pi]$  & flat & Stability & No \\ 
$\lambda_5^I$  & $[-4\pi,\,4\pi]$  & flat & Direct searches & Yes \\ 
$\tan\beta$  & $[0.1,\,60]$  & flat & $\Delta \rho$ + B physics & No  \\  
$\Re(m_{12}^2)$  & $[-1500^2,\, 1500^2]$ $\text{GeV}^2$  & two-sided log & Strong PT & No \\ \cline{4-5}
 & & & Sampled points & Posterior  \\ \cline{4-5}
 & & & $5.5 \times 10^6$ & $2.4 \times 10^4$ \\
\hline
\end{tabular}
\caption{Prior distributions, likelihood contributions and \texttt{MultiNest} settings for the direct-search scan. Compared to the main scan, only direct Higgs search and measurement results are included as constraints, while $\Delta \rho$ + B physics and stability complience has not been considered.}
\label{tab:priorsDirect}
\end{table*}
To illustrate the impact of the direct Higgs searches and measurements alone we also perform a ``direct-search scan'' where we only include the likelihood contributions calculated with \texttt{HiggsBounds} and \texttt{HiggsSignals}, leaving out the stability condition, $\Delta\rho$ and the B-physics observables. In this case the result consists of the combined posterior samples from two identical scans, with the scan details summarized in Table \ref{tab:priorsDirect}.

\begin{figure*}[t]
\centering
\includegraphics[width=0.24\textwidth]{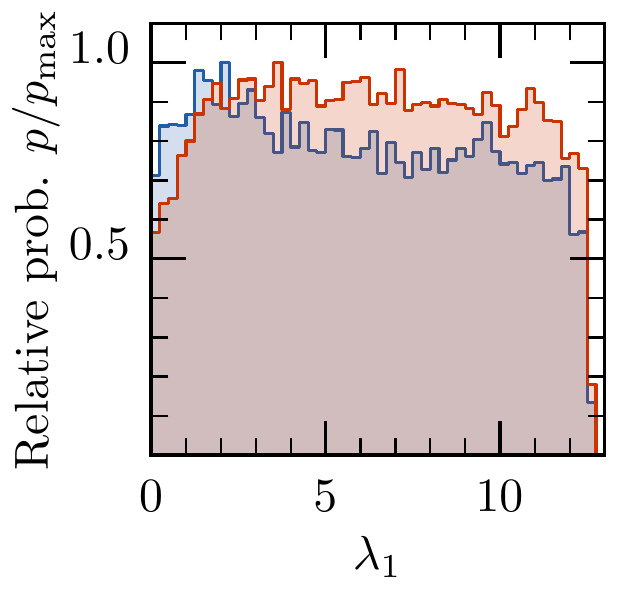}
\includegraphics[width=0.24\textwidth]{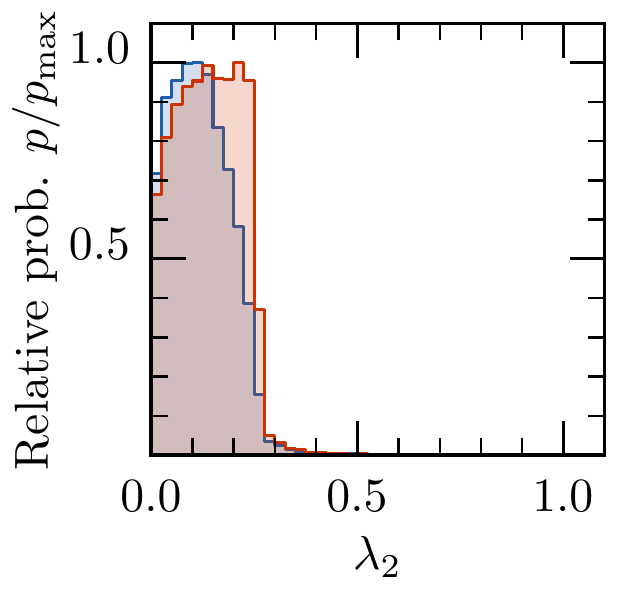}
\includegraphics[width=0.24\textwidth]{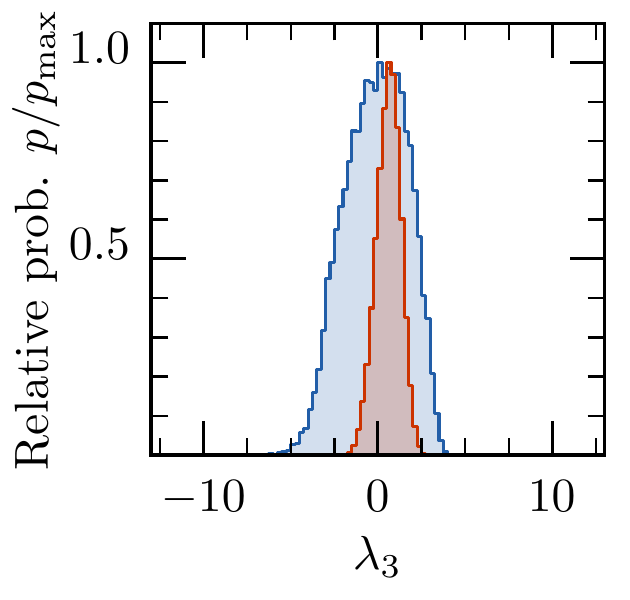}
\includegraphics[width=0.24\textwidth]{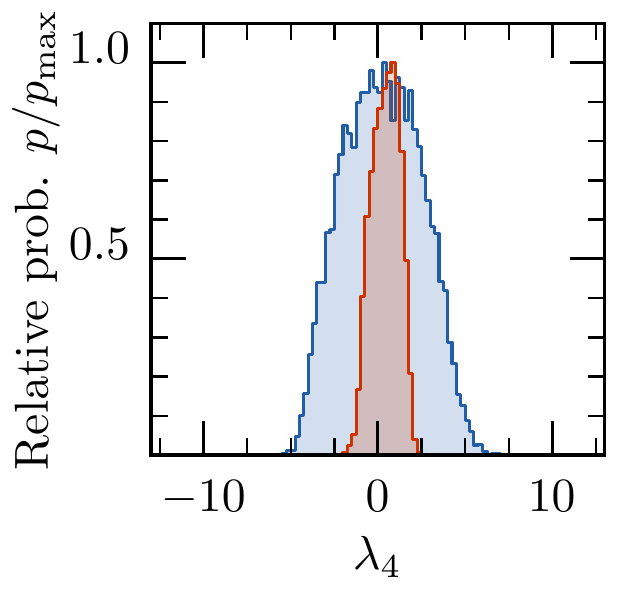}\\
\includegraphics[width=0.24\textwidth]{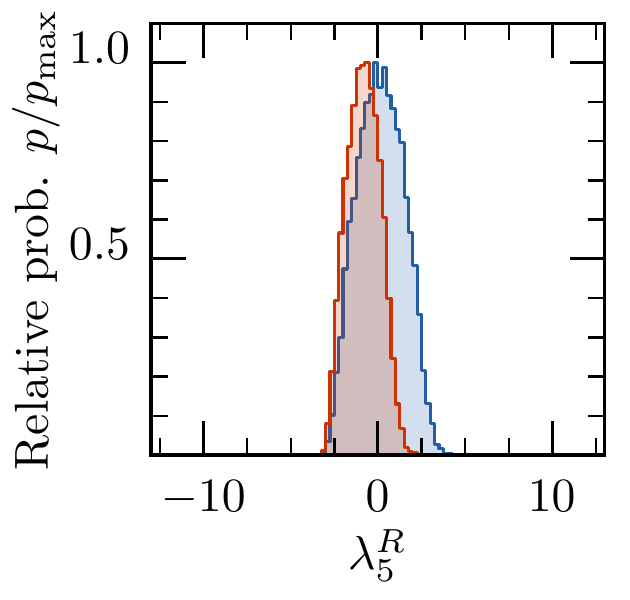}
\includegraphics[width=0.24\textwidth]{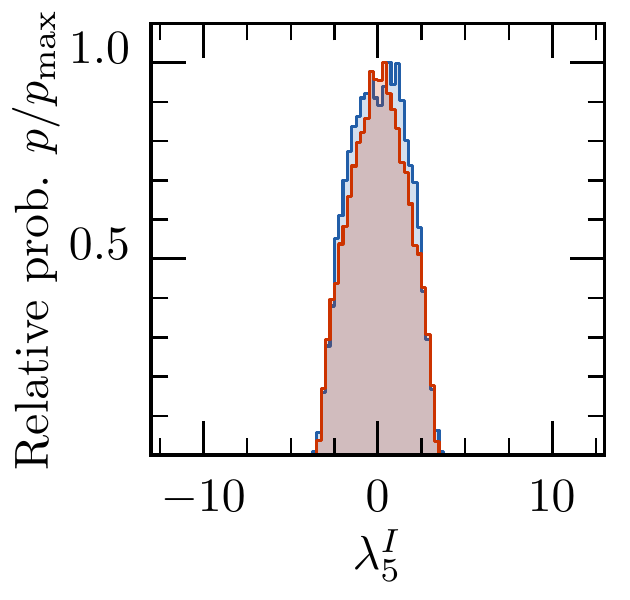}
\includegraphics[width=0.24\textwidth]{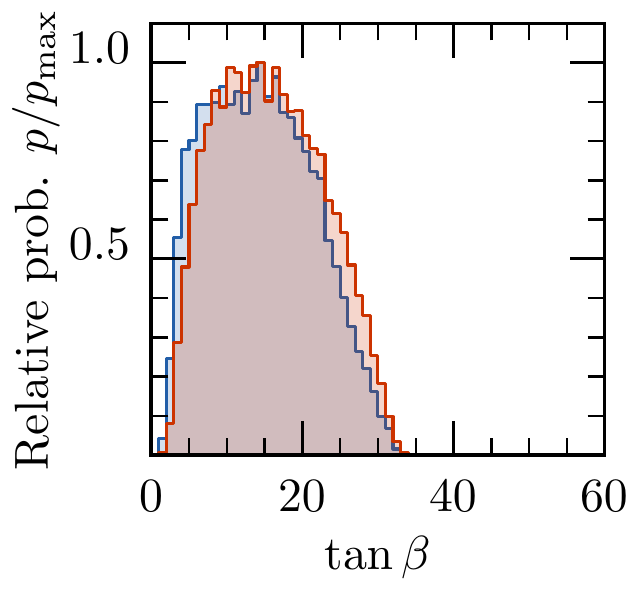}
\includegraphics[width=0.24\textwidth]{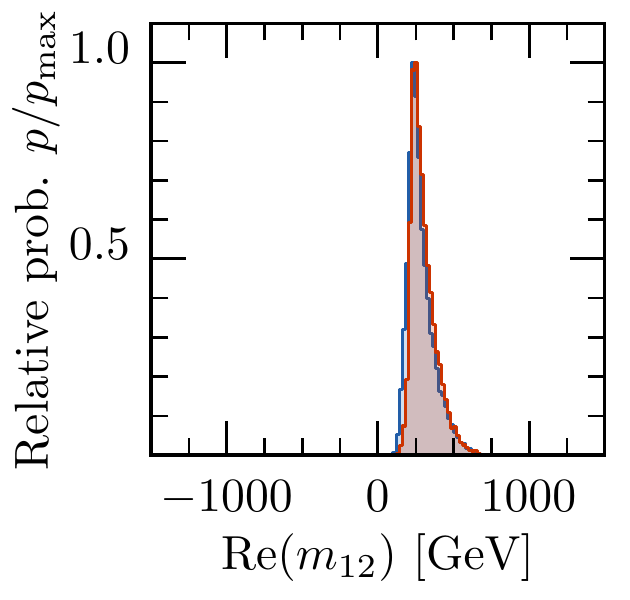}
\caption{Marginalized posterior distributions for the input parameters. \textit{Red:} The result of the main scan with all likelihood contributions included. \textit{Blue:} The result of the direct-search scan, which includes likelihood contributions from direct Higgs searches and measurements. Note that the plotted range for $\lambda_2$ is $[0,\,1]$, in contrast to Fig.\ \ref{fig:priors}, where it is $[0,\,13]$.}
\label{fig:posteriors}
\end{figure*}
\begin{figure*}
\centering
\includegraphics[width=0.32\textwidth]{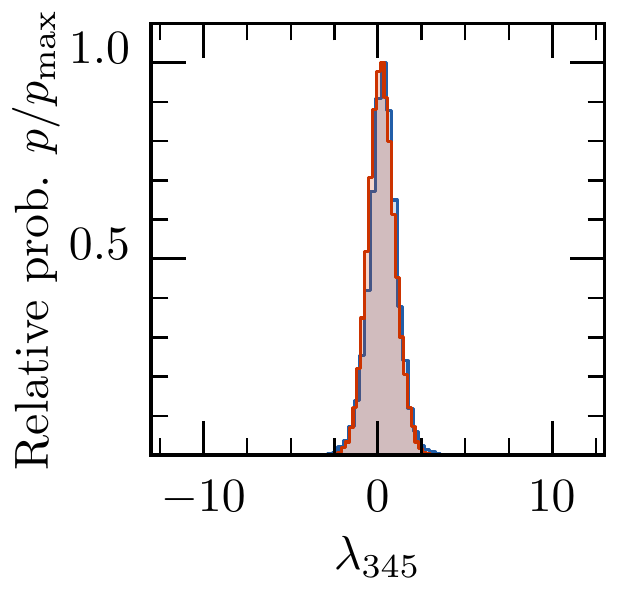}
\caption{Marginalized posterior distributions for $\lambda_{345} \equiv \lambda_3 + \lambda_4 + \lambda_5^R$ from the main scan (red) and the direct-search scan (blue).}
\label{fig:posteriors_la345}
\end{figure*}
In Fig.\ref{fig:posteriors} we show the marginalized posterior distributions for the input parameters. The result of the direct-search scan is shown in blue, while the main scan results are shown in red. Comparing first the posteriors from the direct-search scan in blue with the effective priors in dark grey from Fig.\ref{fig:priors}, we see that the direct-search observables constrain the probable ranges of all quartic couplings except $\lambda_1$. Most notably, there is a clear preference for $\lambda_2 < 0.3$. This is due to the Higgs mass measurement and the fact that in the alignment limit $m_{H_1}^2$ is at tree level dominated by the term $v^2 \lambda_2 s_{\beta}^2$, appearing in element $(\mathcal{M}^2)_{22}$ in Eq.\ (\ref{eq:neutral_mass_matrix}). For $\tan\beta \gg 1$ the observed value $m_{H_1} \approx 125$~GeV then implies $\lambda_2 \approx (125~\mbox{GeV})^2/(246~\mbox{GeV})^2 \approx 0.26$. Except for $\lambda_1$, the least constrained quartic couplings are $\lambda_3$ and $\lambda_4$. At tree level these couplings only appear in the neutral-scalar mass matrix through the combination $\lambda_{345} \equiv \lambda_3 + \lambda_4 + \lambda_5^R$, see Eq.\ (\ref{eq:neutral_mass_matrix}). As a consequence, $\lambda_{345}$ is more strongly constrained by the Higgs data than the individual couplings. The posterior distribution for $\lambda_{345}$ is shown in Fig.\ \ref{fig:posteriors_la345}, where we see that $|\lambda_{345}| < 2.5$ is favoured.

Probable $\tan\beta$ values are limited to the range $(1.5, 33)$. This is mainly due to the type-II Yukawa couplings. At very low $\tan\beta$ we have enhanced rates for processes depending on the $t t H_i$ vertex, e.g.\ gluon fusion production, while at large $\tan\beta$ the vertices $b b H_i$ and $\tau \tau H_i$ become important. Finally, we see that the Higgs searches and measurements are most easily fit when the free mass scale parameter $\Re(m_{12}^2)$ is greater than about $(100\;\text{GeV})^2$.

\begin{figure*}[t]
\centering
\includegraphics[width=0.32\textwidth]{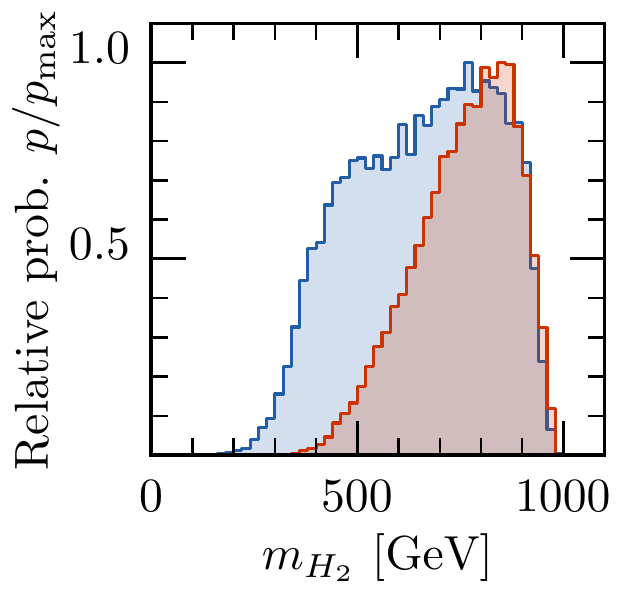}
\includegraphics[width=0.32\textwidth]{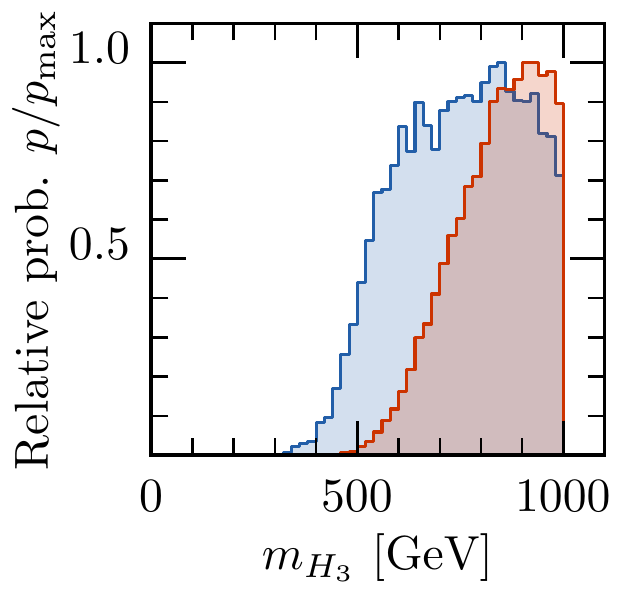}
\includegraphics[width=0.32\textwidth]{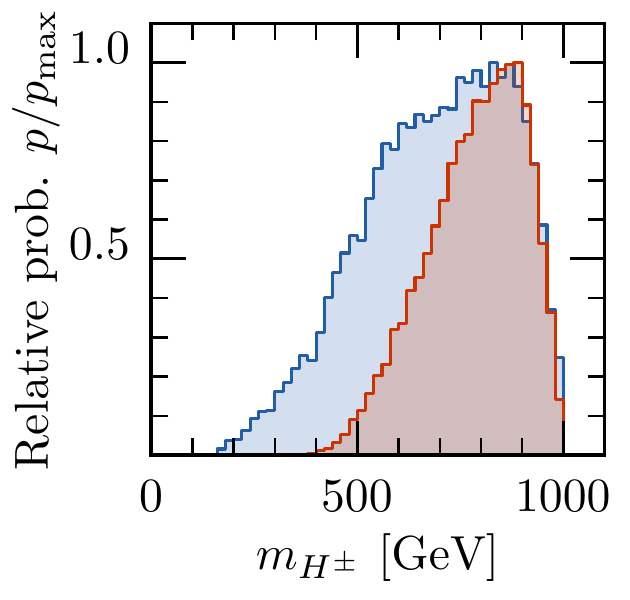}
\caption{Marginalized posterior distributions for the heavy scalar masses. \textit{Red:} The result of the main scan with all likelihood contributions included. \textit{Blue:} The result of the direct-search scan, which includes likelihood contributions from direct Higgs searches and measurements.}
\label{fig:posteriors_scalar_masses}
\end{figure*}    
\begin{figure*}[t]
\centering
\includegraphics[width=0.48\textwidth]{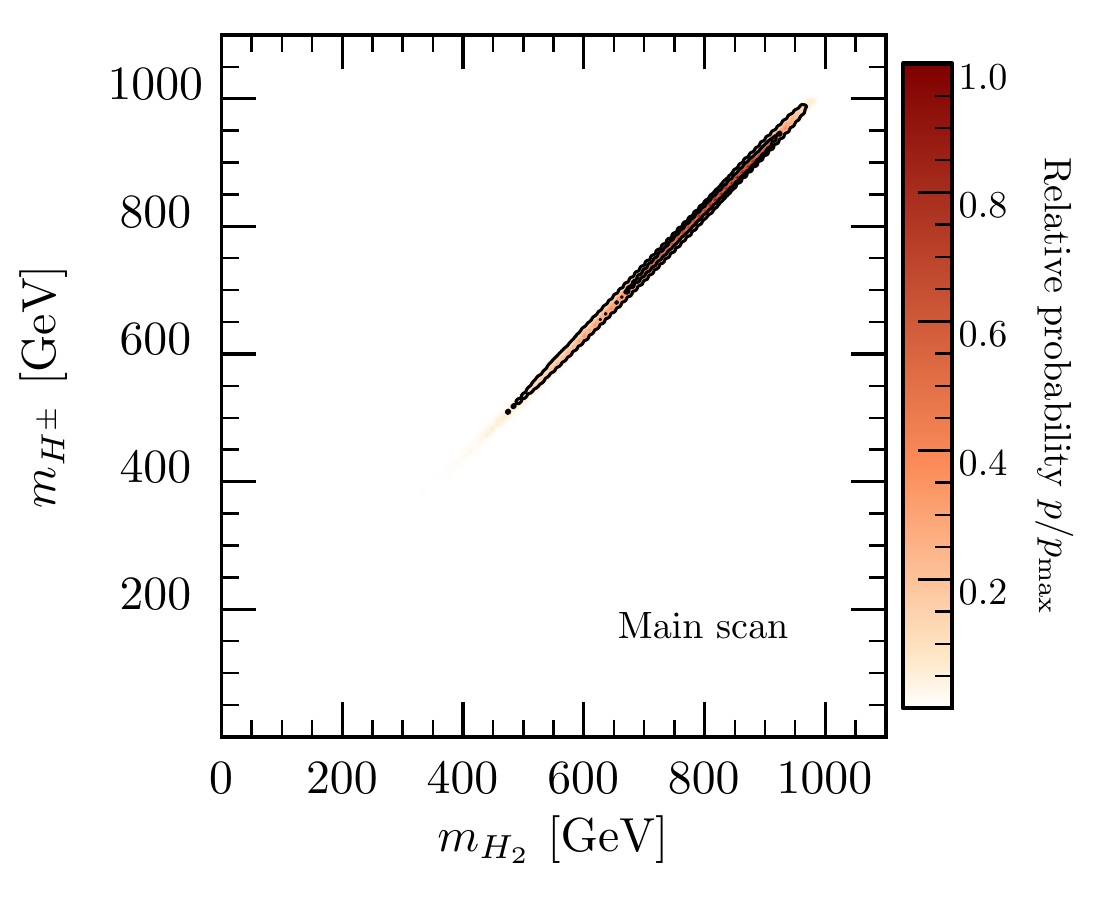}
\hspace{0.02\textwidth}
\includegraphics[width=0.48\textwidth]{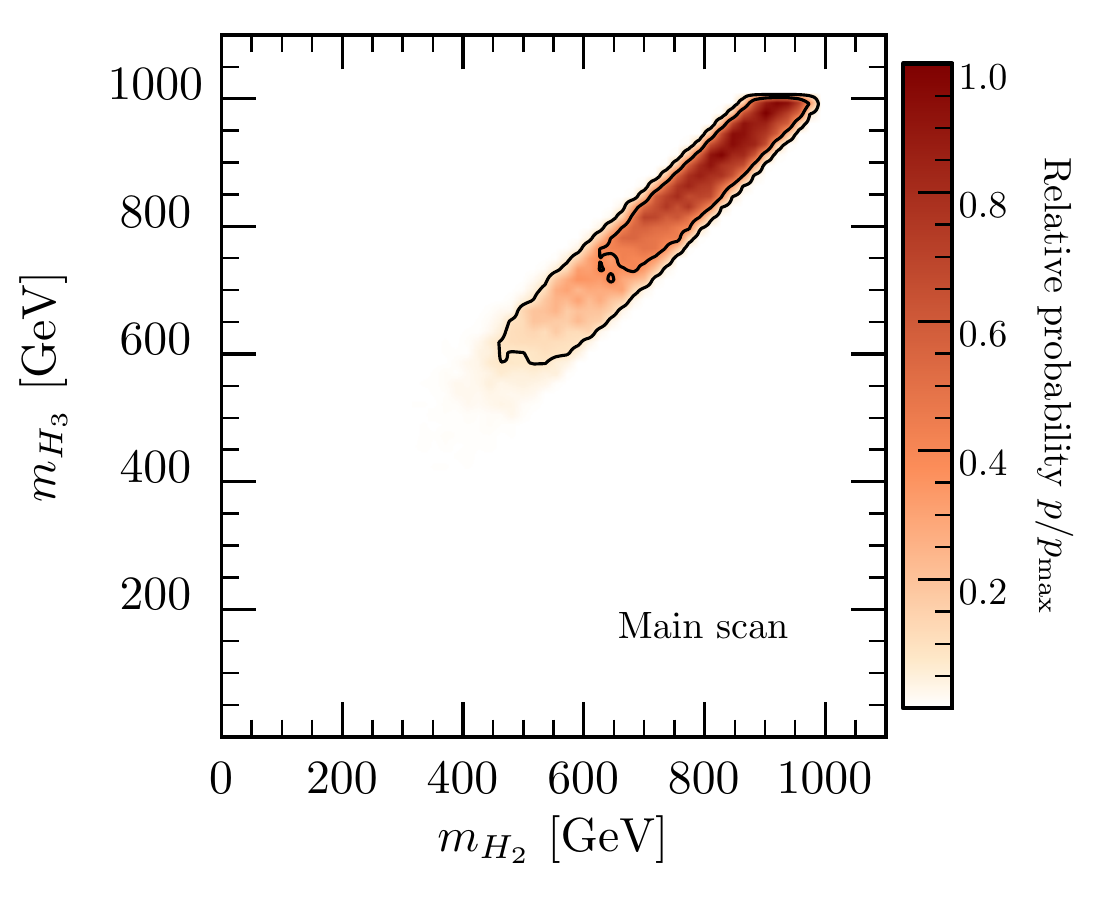}\\
\includegraphics[width=0.48\textwidth]{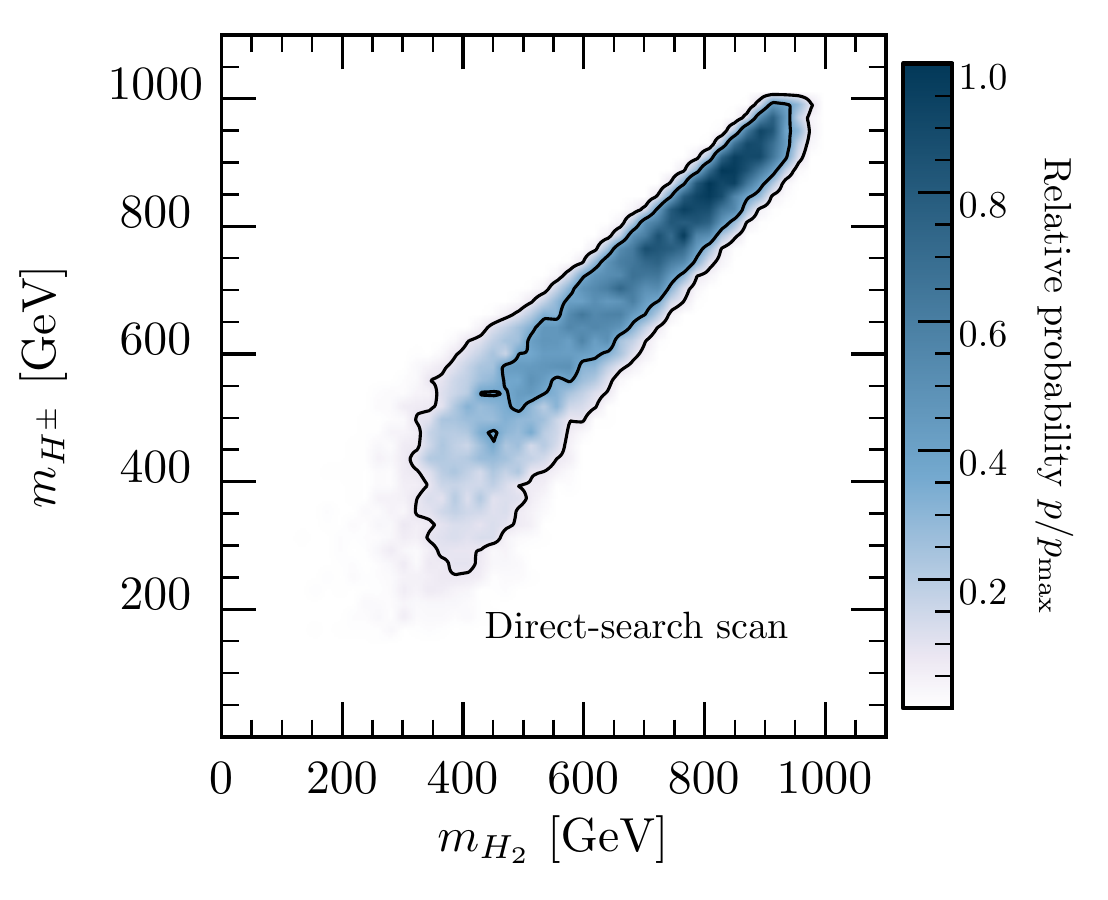}
\hspace{0.02\textwidth}
\includegraphics[width=0.48\textwidth]{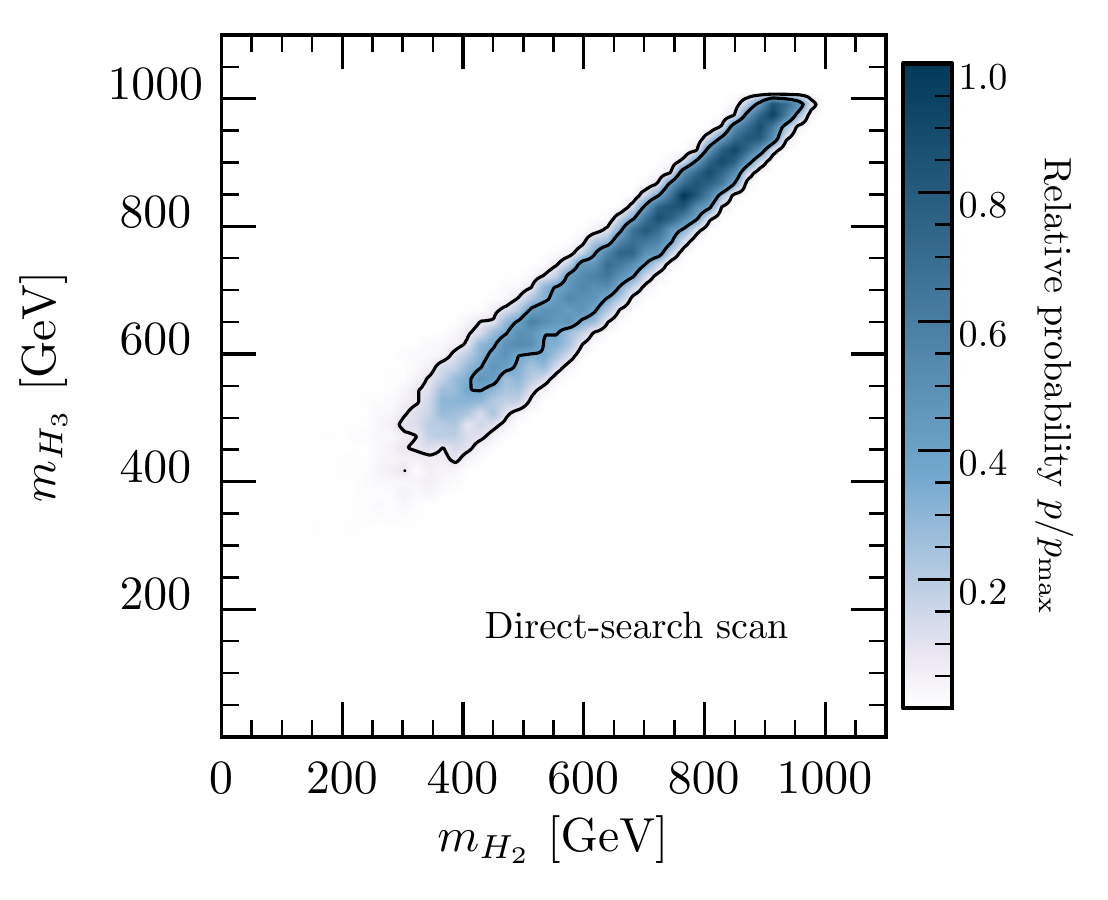}
\caption{Two-dimensional marginalized posterior distributions in the planes of $m_{H_2}$ vs.\ $m_{H^\pm}$ (left) and $m_{H_2}$ vs.\ $m_{H_3}$ (right). \textit{Top row:} Results from the main scan with all likelihood contributions included. The contours depict the $68\%$ and $95\%$ credible regions. \textit{Bottom row:} Results from the direct-search scan, which includes likelihood contributions from direct Higgs searches and measurements.}
\label{fig:posteriors_scalar_masses_2D}
\end{figure*}
Moving on to the posterior distributions from the main scan in red, we see that the preferred ranges for $\lambda_{3}$, $\lambda_{4}$ and $\lambda_{5}^R$ are further narrowed and low values of $\tan\beta$ and $\Re(m_{12}^2)$ are more strongly disfavoured. The most important additional effect come from the observables $\Delta\rho$ and $BR(b \rightarrow s \gamma)$. While $\Delta\rho$ limits the mass difference between $m_{H^\pm}$ and $m_{H_2}$, $BR(b \rightarrow s \gamma)$ disfavours low values of $m_{H^\pm}$. As a consequence, the preferred ranges for the scalar masses are pushed above $\sim 400$~GeV. This is illustrated in Figs.\ \ref{fig:posteriors_scalar_masses} and \ref{fig:posteriors_scalar_masses_2D}, which show one- and two-dimensional posterior distributions for the heavy scalars, for both the direct-search scan and the main scan. The inner and outer contours in Fig.\ \ref{fig:posteriors_scalar_masses_2D} are the $68\%$ and $95\%$ Bayesian credible regions, respectively. We note that the constraints coming from the direct Higgs searches and measurements limit the preferred $m_{H_3}-m_{H_2}$ mass difference to $m_{H_3}-m_{H_2} \lesssim 200$~GeV, as can be seen in the bottom right plot of Fig.\ \ref{fig:posteriors_scalar_masses_2D}.

\begin{figure*}[t]
\centering
\includegraphics[width=0.48\textwidth]{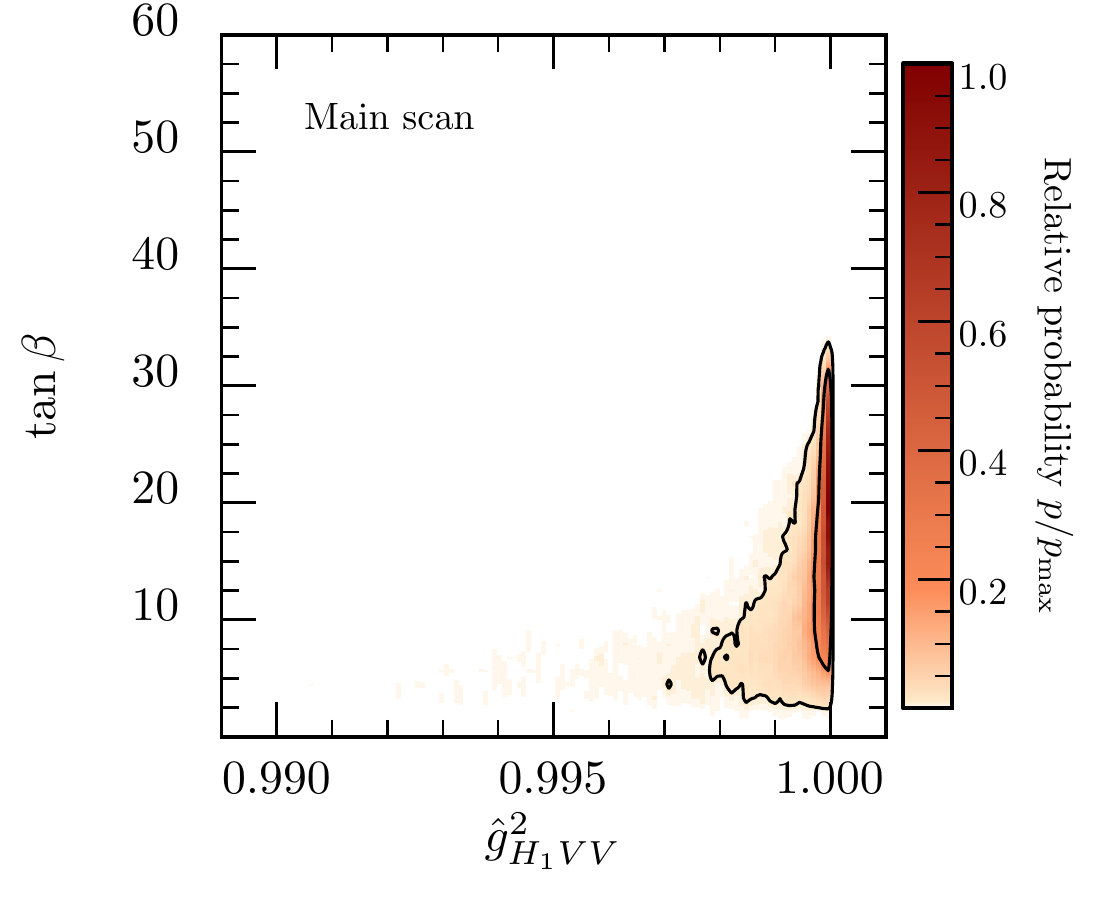}
\hspace{0.02\textwidth}
\includegraphics[width=0.48\textwidth]{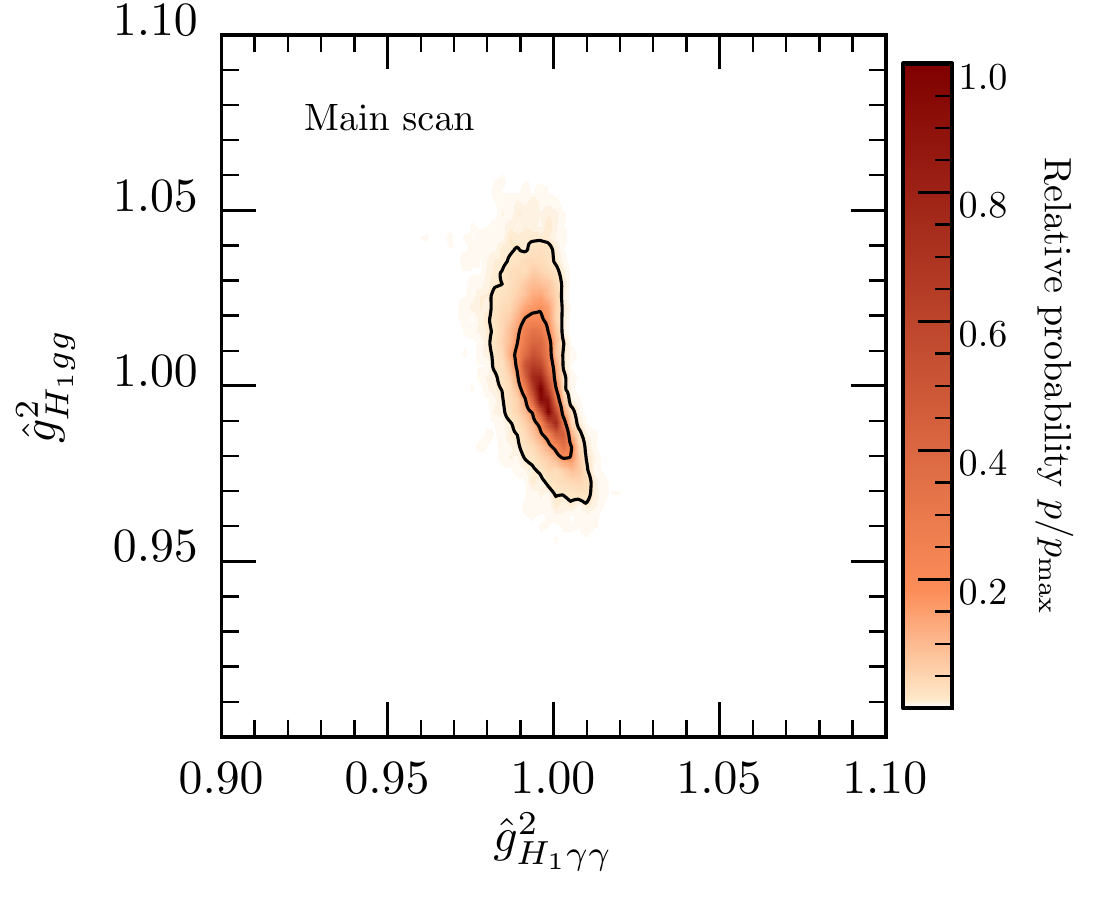}
\caption{Two-dimensional marginalized posterior distributions in the planes of $\hat{g}^2_{H_{1}VV}$ vs.\ $\tan\beta$ (left) and $\hat{g}^2_{H_{1}\gamma\gamma}$ vs.\ $\hat{g}^2_{H_{1}gg}$ (right). The contours show the $68\%$ and $95\%$ credible regions.}
\label{fig:posteriors_effective_couplings_2D}
\end{figure*}
The preferred parameter region exhibits strong SM alignment, with the couplings of $H_{2,3}$ to gauge boson pairs vanishing while $H_1$ adopts SM-like couplings. To illustrate this we plot the posterior distributions of the squared couplings of $H_1$ to gauge boson pairs normalized to the corresponding SM couplings,
\ba
  \hat{g}^2_{H_{1}ii} \equiv \left(\frac{g_{H_{1}ii}}{g_{H_{\text{SM}}ii}}\right)^2, \quad i = Z,W,\gamma,g,
\ea
where the couplings to gluons and photons are understood as the effective loop-induced couplings. The left-hand plot of Fig.\ \ref{fig:posteriors_effective_couplings_2D} shows the marginalized posterior in the plane of $\hat{g}^2_{H_{1}VV}$ ($V = Z,W$) and $\tan\beta$. The $95\%$ credible region prefers $\hat{g}^2_{H_{1}VV}$ values within a few permille of the SM value. The right-hand plot of Fig.\ \ref{fig:posteriors_effective_couplings_2D} depicts the marginalized posterior in the plane of the two loop-induced couplings $\hat{g}^2_{H_{1}\gamma\gamma}$ and $\hat{g}^2_{H_{1}gg}$. In this case, the fit prefers deviations from alignment of no more than a few percent.

The high degree of SM alignment preferred by the fit is in part driven by the volume effect discussed in Sec.\ \ref{sec:Parameter scan}. Even when all scalars are kept below $1$~TeV there is a large parameter volume available that produces a highly SM-like Higgs at $125$~GeV and avoids any tension with the other observables.\footnote{This volume effect would be even more pronounced had we used a flat prior for $\Re(m_{12}^2)$.} Thus, when considering the ``full'' $\text{2HDM}_5$ parameter space as scanned here, the most probable scenario for fitting all the experimental data is one in which the properties of $H_1$ agree closely with SM predictions. One manifestation of the strong SM alignment is the absence of a second ``branch'' extending towards smaller $\hat{g}^2_{H_{1}VV}$ values in the left-hand plot of Fig.\ \ref{fig:posteriors_effective_couplings_2D}. As shown for the $CP$-conserving type-II 2HDM in Refs.\ \cite{Ferreira:2014naa,Dumont:2014wha}, the Higgs signal rates predicted by that parameter region only agree with SM predictions to within $\sim10\%$.

\begin{figure*}[t]
\centering
\includegraphics[width=0.7\textwidth]{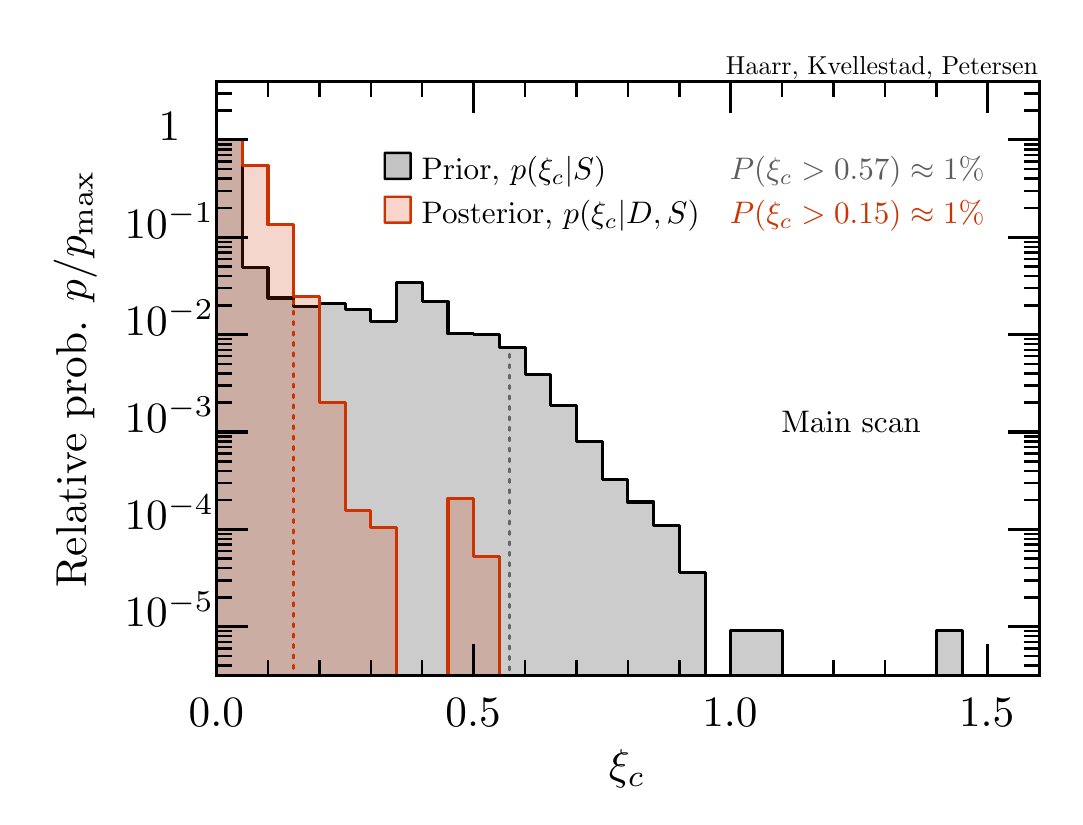}
\caption{Comparison of the prior predictive (grey) and posterior predictive (red) distributions for the phase transition strength $\xi_c$, given that the stability condition $S$ is satisfied. The dotted lines at $\xi_c = 0.15$ and $\xi_c = 0.57$ delimit the 1\% upper tails of the distributions. There are no posterior samples with $\xi_c > 0.6$, indicating that a FOPT is highly unlikely.}
\label{fig:prior_post_PT}
\end{figure*}
We now go on to investigate the probability for having a first-order phase transition. First we note that given our choice of prior distribution for the input parameters, the \textit{prior} probability for a strong phase transition is already quite small. This can be seen in the grey distribution in Fig.\ \ref{fig:prior_post_PT}, which shows the prior predictive distribution for $\xi_c$, given that the stability condition ($S$) is satisfied, $p(\xi_c|S)$. The distribution is based on around $1.4\times10^5$ samples satisfying the stability condition, out of a total of $2\times10^5$ prior samples. The $\sim 1\%$ upper tail of the distribution corresponds to $\xi_c > 0.57$.

Now the interesting question is whether fitting the model to the observed data has made large $\xi_c$ values a more probable prediction. The answer is contained in the posterior predictive distribution, $p(\xi_c|\bfD,S)$, shown as the red distribution in Fig.\ \ref{fig:prior_post_PT}. This distribution is based on the $3.3\times10^4$ posterior samples satisfying the stability condition. Despite the more limited statistics, we clearly see that the posterior distribution for $\xi_c$ drops off much more quickly compared to the prior distribution. In particular, the 1\% upper tail of the distribution starts already at $\xi_c > 0.15$ and there are no posterior samples with $\xi_c > 0.6$. We note that the few posterior samples with $\xi_c \approx 0.5$ have $\tan\beta \in (2,5)$ and scalar masses in the ranges  $m_{H_2} \in (550,650)$~GeV, $m_{H_3} \in (630,730)$~GeV and $m_{H^\pm} \in (580,700)$~GeV, which are all towards the lower ends of the respective posterior distributions.

\begin{figure*}[t]
\centering
\includegraphics[width=0.55\textwidth]{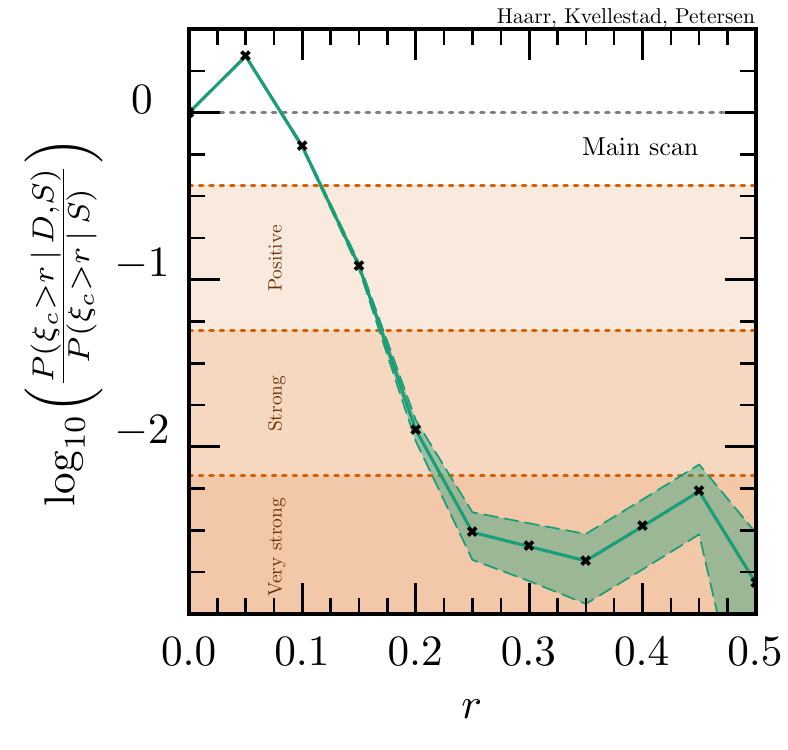}
\caption{The ratio of the posterior and prior probabilities for a phase transition strength $\xi_c > r$, given that the stability condition (S) for the effective potential is satisfied. The green band is the estimated $1\sigma$ uncertainty due to limited statistics. The plotted ratio corresponds to the reciprocal Bayes factor for the two hypotheses $H_a:S$ and $H_b:S,\,\xi_c>r$. The orange lines depict the Jeffreys scale limiting values for ``positive'', ``strong'' and ``very strong'' evidence in favour of $H_a$ \cite{kass1995bayes}.}
\label{fig:PT_ratio}
\end{figure*}
As an alternative way of illustrating the impact of the data, we can construct the ratio of the posterior and prior probabilities for $\xi_c > r$,  
\ba
  \frac{P(\xi_c>r|\bfD,S)}{P(\xi_c>r|S)} = \frac{P(\bfD|\xi_c>r,S)}{P(\bfD|S)}, 
  \label{eq:posterior_prior_ratio}
\ea
where the equality follows from rewriting $P(\xi_c>r|\bfD,S)$ using Bayes' theorem. Thus, the ratio describes how probable the observed data is across the region of parameter space satisfying both stability and $\xi_c > r$, compared to the wider region satisfying only the stability constraint. In Fig.\ \ref{fig:PT_ratio} we plot this ratio as a function of $r$ from $0$ to $0.5$. We see that when constraining the model to the region $\xi_c > 0.25$, the relative probability for the observed data $\bfD$ is reduced by two orders of magnitude. The right-hand side in Eq.\ (\ref{eq:posterior_prior_ratio}) can also be interpreted as the reciprocal Bayes factor for the two competing hypotheses $H_a: S$ and $H_b:S,\,\xi_c > r$. According to Jeffreys scale \cite{kass1995bayes}, the data constitutes ``very strong'' evidence in favor of $H_a$ for $\xi_c > 0.2$. This is indicated by the shaded bands in Fig.\ \ref{fig:PT_ratio}. 

Even though we divide out the prior probability for $\xi_c > r$ in Eq.\ (\ref{eq:posterior_prior_ratio}), the result in Fig.\ \ref{fig:PT_ratio} still has a dependence on the prior for the input parameters, as any Bayesian analysis will. This can be seen explicitly by expressing the right-hand side of Eq.\ (\ref{eq:posterior_prior_ratio}) as 
\ba
  \frac{P(\bfD|H_b)}{P(\bfD|H_a)} = \frac{\int p(\bfD|\bfTheta,H_b) p(\bfTheta|H_b) d\bfTheta}{\int p(\bfD|\bfTheta,H_a) p(\bfTheta|H_a) d\bfTheta}. 
  \label{eq:post_prior_ratio_prior_dep}
\ea
The factors $p(\bfTheta|H_a)$ and $p(\bfTheta|H_b)$ are nothing but the input prior distributions for the parameters $\bfTheta$ under the additional constraint of $H_a$ and $H_b$, respectively. In the limit where the chosen input prior $p(\bfTheta)$ reflects certainty in hypothesis $H_b$, i.e.\ only points satisfying stability and $\xi_c > r$ are assigned a non-zero prior probability, we would have $P(H_b) = 1$, and since $P(H_a|H_b) = 1$ the ratio in Eq.\ (\ref{eq:post_prior_ratio_prior_dep}) would be unity for any data $\bfD$. Thus, the value in our result lies in the fact that we use a fairly non-informative prior distribution and then explore in what direction the data take us.

%%%%%%%%%%%%%%%%%%%%
\subsection{Interplay of a strong phase transition and experimental constraints}
\label{subsec:Interplay}
%%%%%%%%%%%%%%%%%%%%
%
In the previous section we found that the $\text{2HDM}_5$ does not predict a first-order phase transition when fitted to the experimental data. However, we also saw that the sheer volume of the weakly constrained parameter regions had an important impact on the fit. One might therefore worry that our conclusion regarding the probability of a FOPT comes with a strong prior dependence. To check our conclusion we therefore perform a new parameter scan with \texttt{MultiNest}. The likelihood function is the same as in the main scan, except that we now also include the requirement $\xi_c > 1$ as a $95\%$~CL lower limit. Thus, the scan is forced to search for parameter regions with a strong phase transition while at the same time staying in as close agreement with other constraints as possible. We refer to this scan as the ``strong-PT scan''.

Due to the computational expense of calculating $\xi_c$ for every point in the scan, the scope of the scan must be restricted: First, in light of the preference for small $\lambda_{345}$ in the main scan, we now take $\lambda_{345}$ as an input parameter with a flat prior on $[-5,5]$. $\lambda_5^R$ is then determined from $\lambda_{5}^R = \lambda_{345} - \lambda_3 - \lambda_4$, with the constraint that $|\lambda_5^R| < 4\pi$. Further, we restrict the $\lambda_2$ prior to the range $[0,1]$. Our choice of priors, \texttt{MultiNest} settings and likelihood contributions are summarized in Table~\ref{tab:priorsStrongPT}. We here run two identical \texttt{MultiNest} scans with only 500 live points (\texttt{nlive = 500}) and combine the posterior samples. It should be noted that such a low number of live points is not enough to guarantee a complete exploration of a large parameter space. However, here we are mainly interested in using the scan to further understand and cross-check the results of the larger scan in the previous section. 

\begin{table*}[tb]
\centering
\textbf{Priors and settings for the strong-PT scan} \\
\begin{tabular}{lll|ll}
\hline
 Parameter & Range & Type & \texttt{MultiNest} setting & Value  \\
\hline
$\lambda_1$  & $[0,\,4\pi]$  & flat &\texttt{nlive} & 500  \\ 
$\lambda_2$  & $[0,\,1]$  & flat & \texttt{tol} & 0.5  \\ 
$\lambda_3$  & $[-4\pi,\,4\pi]$  & flat & \texttt{efr} & 0.8 \\ \cline{4-5}  
$\lambda_4$  & $[-4\pi,\,4\pi]$  & flat & Likelihood & Included \\ \cline{4-5}  
$\lambda_{345}$  & $[-5,\,5]$  & flat & Stability & Yes  \\ 
$\lambda_5^I$  & $[-4\pi,\,4\pi]$  & flat & Direct searches & Yes \\ 
$\tan\beta$  & $[0.1,\,60]$  & flat  & $\Delta \rho$ + B physics & Yes  \\ 
$\Re(m_{12}^2)$  & $[-1500^2,\, 1500^2]$ $\text{GeV}^2$  & two-sided log & Strong PT & Yes \\ \cline{4-5}
 & & & Sampled points & Posterior  \\ \cline{4-5} 
 & & & $5.2 \times 10^6$ & $4.7 \times 10^3$ \\
\hline
\end{tabular}
\caption{Prior distributions, likelihood contributions and \texttt{MultiNest} settings for the strong-PT scan.}
\label{tab:priorsStrongPT}
\end{table*}

\begin{figure*}[t]
\centering
\includegraphics[width=0.24\textwidth]{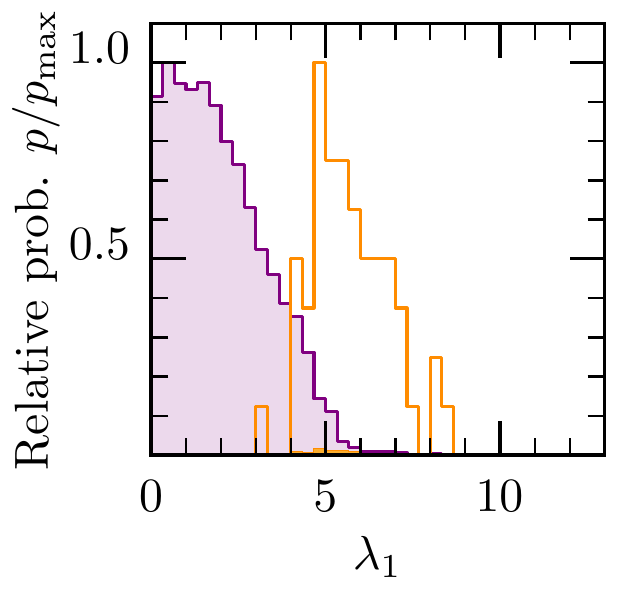}
\includegraphics[width=0.24\textwidth]{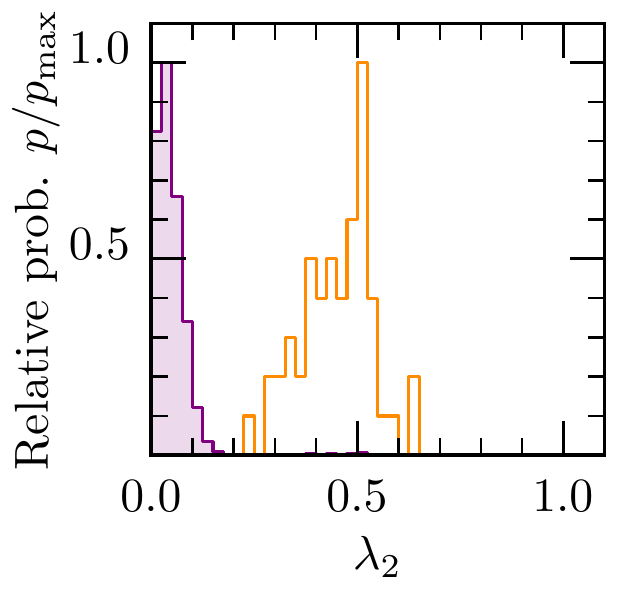}
\includegraphics[width=0.24\textwidth]{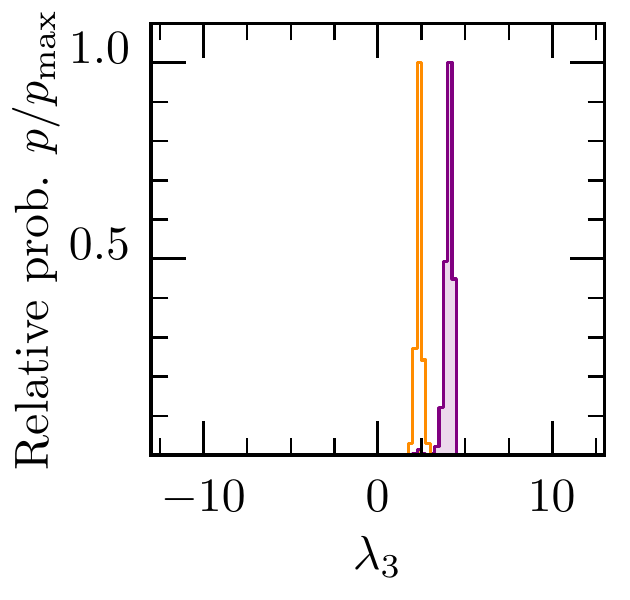}
\includegraphics[width=0.24\textwidth]{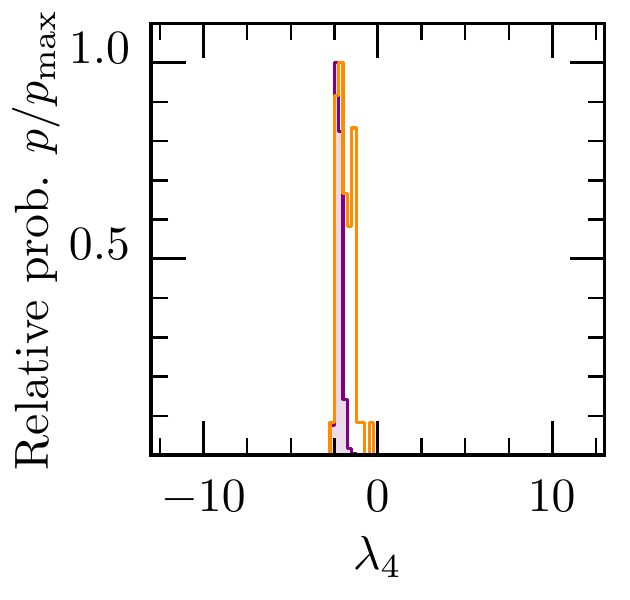}\\
\includegraphics[width=0.24\textwidth]{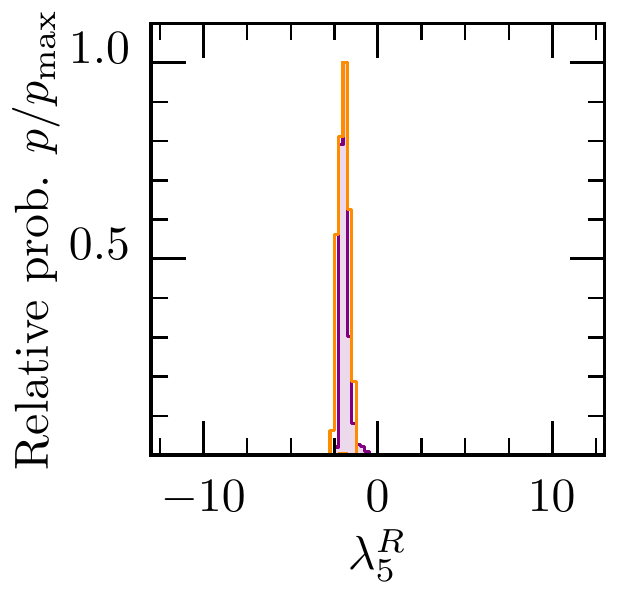}
\includegraphics[width=0.24\textwidth]{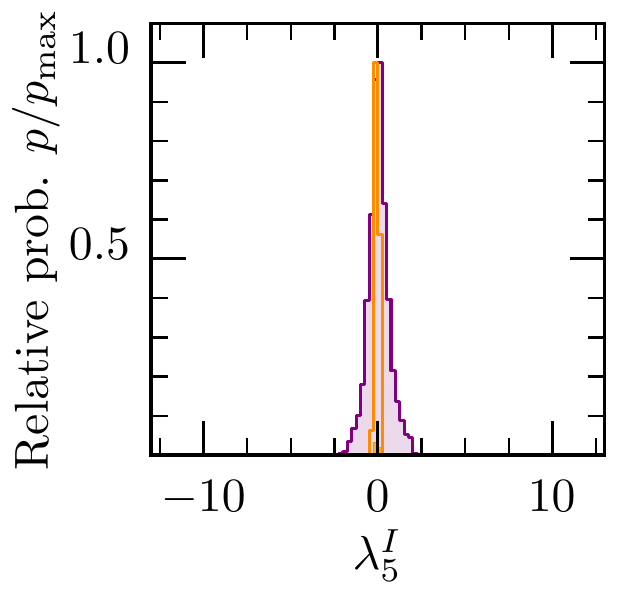}
\includegraphics[width=0.24\textwidth]{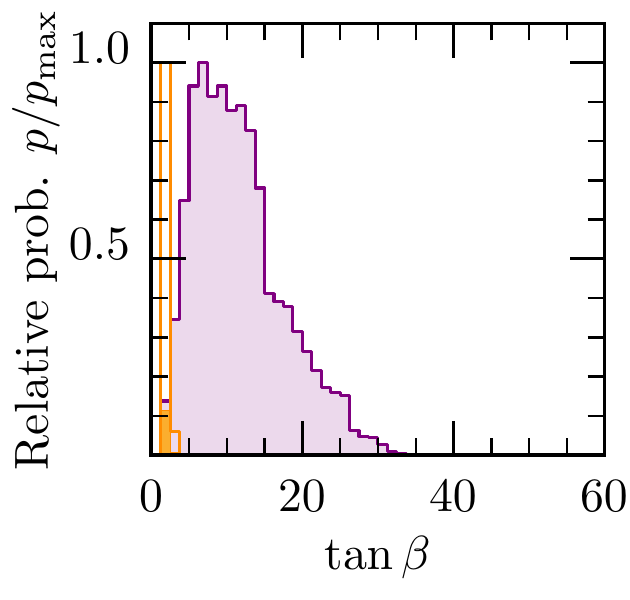}
\includegraphics[width=0.24\textwidth]{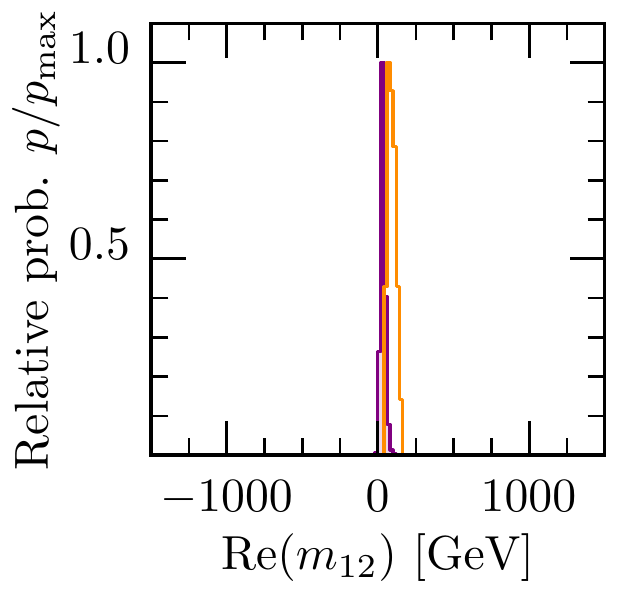}
\caption{Marginalized posterior distributions for the input parameters in the strong-PT scan. The orange distributions, scaled up for visibility, show the contribution from the scenario where $H_1$ is the 125 GeV Higgs. To ease comparison with the results in Fig.\ \ref{fig:posteriors} we choose to still plot $\lambda_5$ even though it was replaced by $\lambda_{345}$ as the input parameter.}
\label{fig:posteriors_strongPT}
\end{figure*}
In Fig.\ \ref{fig:posteriors_strongPT} we show the obtained posterior distributions for the parameters (purple). As a result of demanding a strong phase transition, all input parameters except for $\lambda_1$ and $\tan \beta$ are rather tightly constrained. The model has for the most part been pushed into the hidden-Higgs scenario, where the next-to-lightest Higgs, $H_2$, is identified with the $125$~GeV resonance. The ``standard scenario'' with $H_1$ as the $125$~GeV Higgs only accounts for around $1\%$ of the posterior probability. In Fig.\ \ref{fig:posteriors_strongPT} the contribution from the standard scenario to the overall posterior is shown as orange histograms, scaled up for visibility.

Before discusssing why the preferred parameter region in this scan is in tension with observations, we note a couple of characteristics that agree well with the findings of previous studies: 
Firstly, allowing for a non-zero $CP$-violating phase does not seem to be important for having a strong phase transition \cite{Dorsch:2013wja,Fromme:2006cm}. This is seen in the preference for small $\lambda_5^I$. When $\lambda_5^I = 0$ the Jarlskog-like $CP$-invariants vanish and we have no $CP$ violation and therefore no $CP$-violating phase. Secondly, when $H_1$ is the $125$~GeV Higgs there is a preference for $\tan \beta$ close to one \cite{Dorsch:2013wja}. In Table \ref{tab:strong_PT_points} we list two example points, one with $m_{H_1} \approx 125$~GeV and one with $m_{H_2} \approx 125$~GeV.
\begin{table*}[tb]
\centering
\begin{tabular}{lccccccc|cccc}
\hline
$\lambda_1$ & $\lambda_2$ & $\lambda_3$ & $\lambda_4$ & $\lambda_5^R$ & $\lambda_5^I$ & $\tan\beta$ & $\Re(m_{12}^2)$ & $m_{H_1}$ & $m_{H_2}$ & $m_{H_3}$ & $m_{H^\pm}$ \\
\hline
$6.40$ & $0.51$ & $2.44$ & $-2.47$ & $-1.73$ & $0.06$ & $2.14$ & $120^2$ & $126$ & $341$ & $344$ & $376$  \\
$0.64$ & $0.01$ & $4.33$ & $-2.31$ & $-1.95$ & $0.66$ & $7.69$ & $55^2$ & $95$ & $126$ & $352$ & $362$  \\
%6.3968e-01      7.9541e-03      4.3274e+00      -2.3144e+00     -1.9493e+00     6.6367e-01      7.6938e+00      5.5366e+01
%9.5065e+01      1.2576e+02      3.5170e+02      3.6260e+02
\hline
\end{tabular}
\caption{Two example points with $\xi_c > 1$ taken from the strong-PT scan. The first point is an example of the standard scenario ($m_{H_1} \approx 125$~GeV), while the second point illustrates the hidden-Higgs scenario ($m_{H_2} \approx 125$~GeV). All masses are given in GeV.}
\label{tab:strong_PT_points}
\end{table*}

\begin{figure*}[t]
\centering
\includegraphics[width=0.24\textwidth]{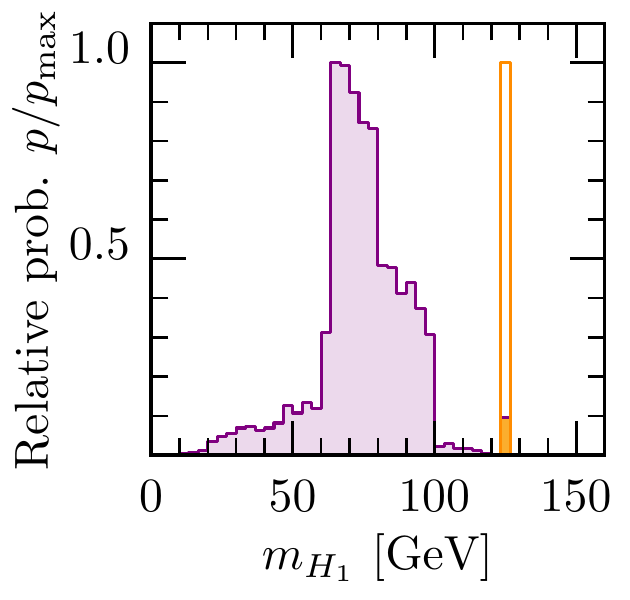}
\includegraphics[width=0.24\textwidth]{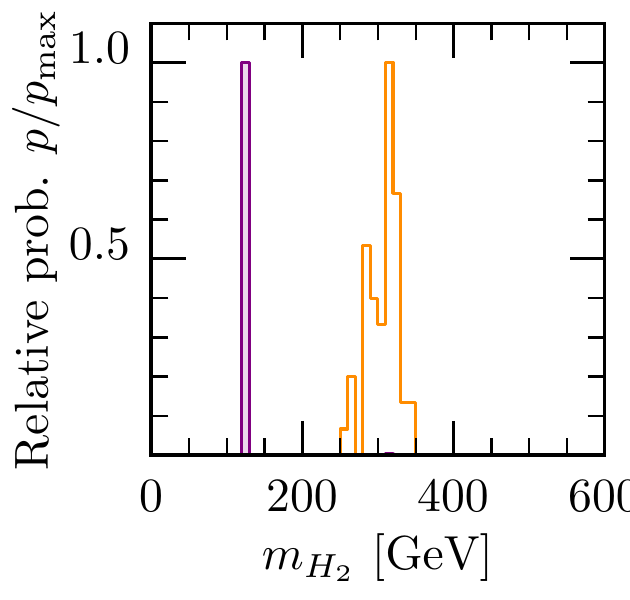}
\includegraphics[width=0.24\textwidth]{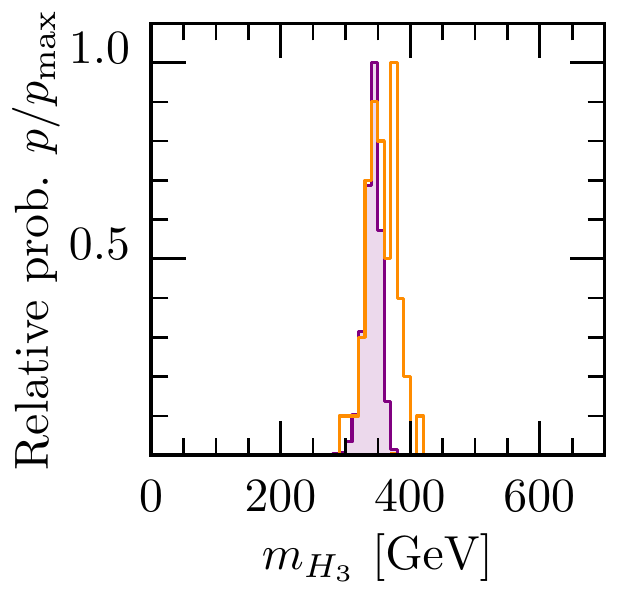}
\includegraphics[width=0.24\textwidth]{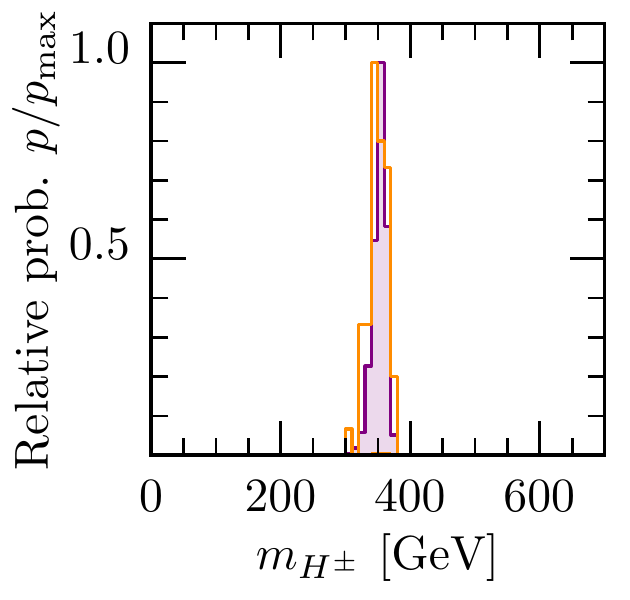}\\
\caption{Marginalized posterior distributions for the scalar masses in the strong-PT scan. The posterior contribution from the scenario with $H_1$ being the $125$~GeV Higgs is shown in orange, scaled up for visibility.}
\label{fig:posteriors_strongPT_masses}
\end{figure*}

\begin{figure*}[t]
\centering
\includegraphics[width=0.32\textwidth]{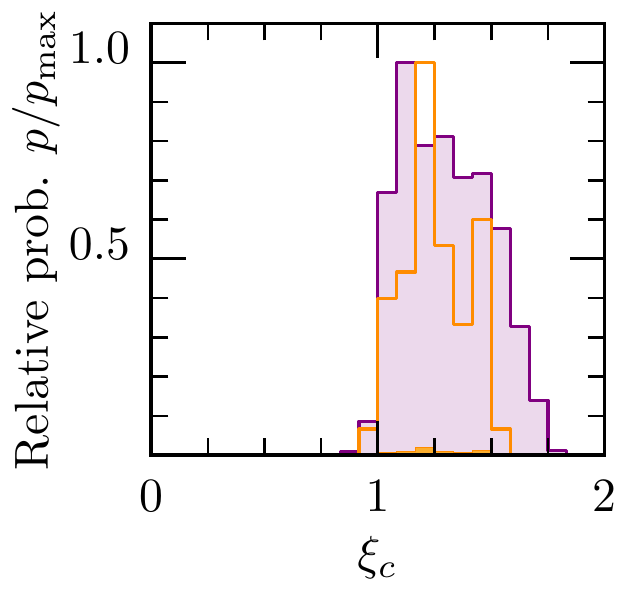}
\includegraphics[width=0.32\textwidth]{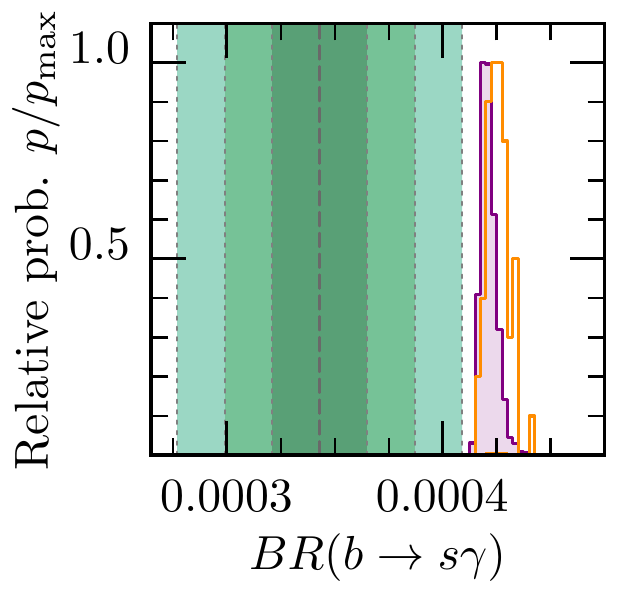}\\
\caption{Marginalized posterior distributions for the phase transition strength $\xi_c$ (left) and $BR(b\rightarrow s \gamma)$ (right) in the strong-PT scan. The green bands depict the $1\sigma$, $2\sigma$ and $3\sigma$ regions around the observed value. The contribution from the standard scenario with $H_1$ as the $125$~GeV Higgs is shown in orange, scaled up for visibility.}
\label{fig:posteriors_strongPT_observables}
\end{figure*}

Requiring a strong phase transition has the effect of forcing down the mass scale for the scalars. As seen in Fig.\ \ref{fig:posteriors_strongPT_masses}, the preferred ranges for all the scalar masses are below $400$~GeV. In particular, the predicted range for $m_{H^\pm}$ is below the $490$~GeV lower bound for avoiding large contributions to $BR(b \rightarrow s \gamma)$. This is illustrated in Fig.\ \ref{fig:posteriors_strongPT_observables}. The left plot is the posterior distribution for $\xi_c$, showing that the scan has indeed identified a parameter region giving $\xi_c > 1$. The right plot shows the posterior distribution for $BR(b \rightarrow s \gamma)$. The entire range of predicted $BR(b \rightarrow s \gamma)$ values is more than $3\sigma$ above the observed value, as indicated by the vertical green bands. This demonstrates that the scan was not able to find a parameter region with strong phase transition without accepting significant tension with other observables, thus confirming the result in Fig.\ \ref{fig:PT_ratio}.

To further illustrate the interplay between the $\xi_c > 1$ requirement and the other constraints we perform two grid scans in the best-fit parameter regions of the strong-PT scan.\footnote{These grid scans are not listed in Table \ref{tab:Scans} as they only serve to highlight some aspects of the strong-PT scan results.} The first of these two sub-scans is focused on the standard scenario with $H_1$ as the 125 GeV resonance, while the second sub-scan covers a parameter region in the hidden-Higgs scenario. The scanned parameter planes are defined in Table~\ref{tab:gridscan}. For the standard scenario we fix the parameters $\lambda_1$, $\lambda_3$, $\lambda_5^I$ and $\tan \beta$, take $\lambda_5^R = 0.7\lambda_4$ and scan the plane of $\lambda_4$ vs $\Re(m_{12}^2)$. At each point in the plane we choose $\lambda_2$ such as to minimize the difference $|m_{H_1} - 125\;\text{GeV}|$. For the hidden-Higgs grid scan we fix $\lambda_1$, $\lambda_3$, $\lambda_4$, $\lambda_5^R$ and $\lambda_5^I$, scan the $\tan\beta$ vs $\Re(m_{12}^2)$ plane and choose $\lambda_2$ by minimizing $|m_{H_2} - 125\;\text{GeV}|$. It should be noted that while we always choose $\lambda_2$ to minimize $|m_{H_{1,2}} - 125\;\text{GeV}|$, in large regions of the scanned parameter planes this freedom is not enough to actually obtain $m_{H_{1,2}} \approx 125$~GeV.

\begin{figure*}[t]
\centering
\includegraphics[width=0.485\textwidth]{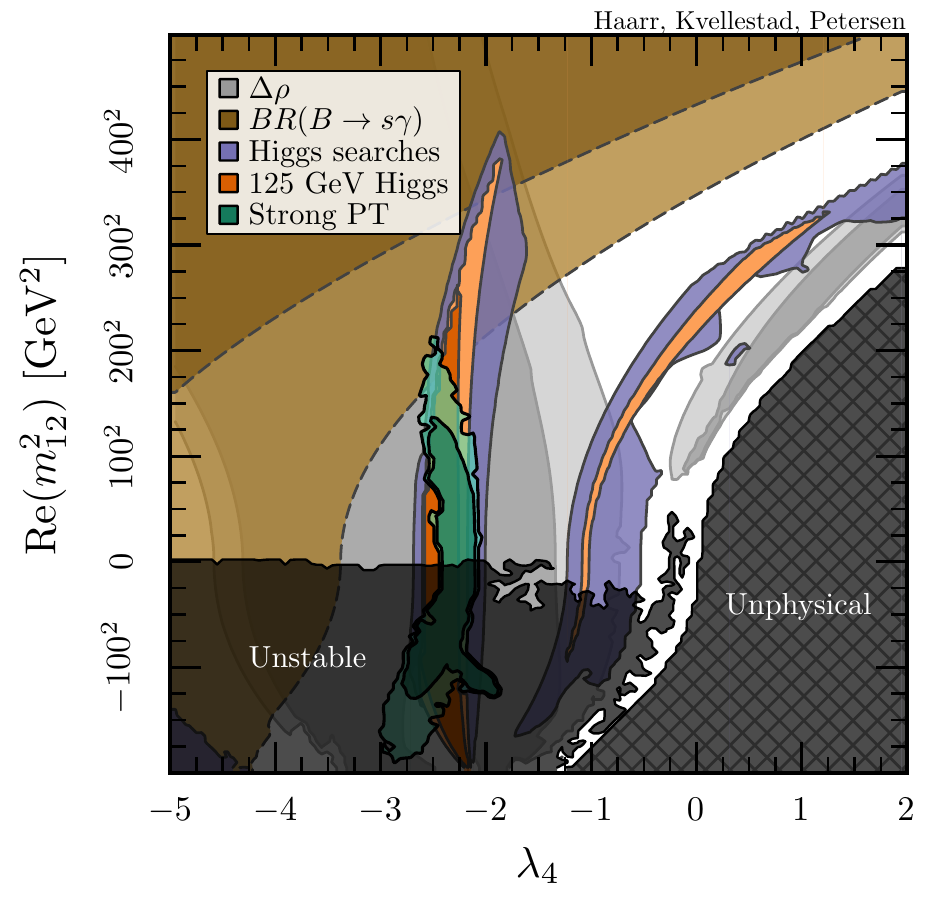}
\hspace{0.01\textwidth}
\includegraphics[width=0.485\textwidth]{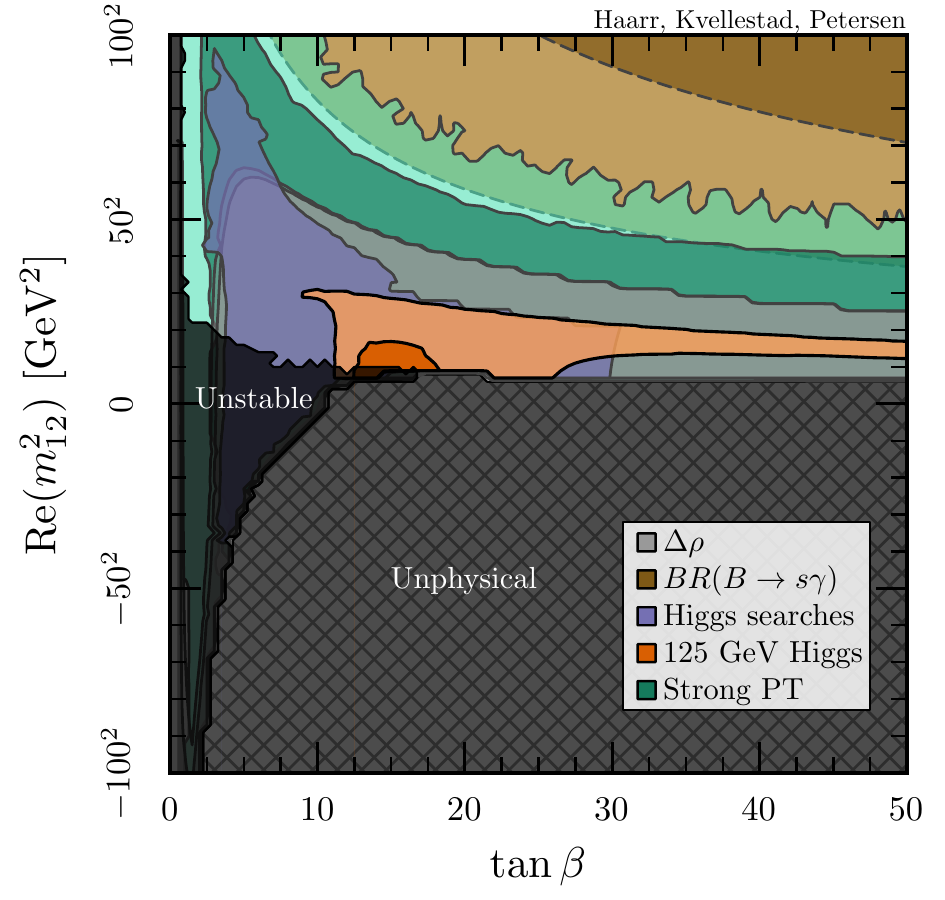}
\caption{Lower-dimensional grid scans exhibiting regions of strong electroweak phase transition. \textit{Left:} The plane of $\lambda_4$ vs $\Re(m_{12}^2)$, with the remaining parameters chosen according to the ``standard scenario'' in Table \ref{tab:gridscan}. \textit{Right:} The plane of $\tan\beta$ vs $\Re(m_{12}^2)$, with the other parameters set to the hidden-Higgs scenario in Table \ref{tab:gridscan}. Dark regions marked ``unstable'' and ``unphysical'' fail the stability condition and have negative squared masses, respectively. 
The other coloured regions depict the following: \textit{Teal (cyan):} Phase transition strength satisfying $\xi_c >1$ ($\xi_c>0.5$). \textit{Dark (light) orange:} Within $2\sigma$ ($3\sigma$) of the best-fit point for the $125$ GeV Higgs data. \textit{Purple:} Allowed at the $95\%$ CL by collider searches for additional Higgs bosons. \textit{Dark (light) brown:} Predicted $BR(b \rightarrow s \gamma)$ within $2\sigma$ ($3\sigma$) of the observed value. \textit{Dark (light) grey:} Predicted $\Delta\rho$ within $2\sigma$ ($3\sigma$) of the observed value.}
\label{fig:gridscan}
\end{figure*}

\begin{table*}[tb]
\centering
\begin{tabular}{lll}
\hline
Parameter \quad\quad\quad& Standard scenario \quad\quad\quad\quad&  Hidden-Higgs scenario \\
\hline
$\lambda_1$  & $6.5$  & $1.3$\\
$\lambda_2$  & $\min\,(|m_{H_1} - 125\;\text{GeV}|)$ & $\min\,(|m_{H_2} - 125\;\text{GeV}|)$ \\
$\lambda_3$  & $2.5$  & $4.3$ \\
$\lambda_4$  & $[-5,2]$ & $-2.3$  \\
$\lambda_5^R$  & $0.7\lambda_4$  & $-2$ \\
$\lambda_5^I$  & $0$ & $0.3$  \\
$\tan\beta$  & $2.2$  & $[0.5,50]$ \\
$\Re(m_{12}^2)$  & $[-200^2,\, 500^2]$ $\text{GeV}^2$  & $[-100^2,\, 100^2]$ $\text{GeV}^2$ \\
\hline
\end{tabular}
\caption{Choice of parameter planes for the grid scans. The $125$~GeV Higgs is identified with $H_1$ in the standard scenario and with $H_2$ in the hidden-Higgs scenario.}
\label{tab:gridscan}
\end{table*}
In Fig.\ \ref{fig:gridscan} we show the results of the two grid scans, with the scan of the standard scenario in the left-hand plot and the hidden-Higgs scan in the right-hand plot. In both plots there are dark regions marked ``unstable'' and ``unphysical''. The ``unstable'' regions fail the stability requirement detailed in section \ref{subsec:determining PT strength}, while in the ``unphysical'' regions spectrum generation fails due to negative squared masses. The other coloured regions depict parameter regions favoured by different constraints: The dark (light) grey regions satisfy $\Delta \rho$ at $2\sigma$ ($3\sigma$), the dark (light) brown regions satisfy $BR(b \rightarrow s \gamma)$ at $2\sigma$ ($3\sigma$) and the purple regions are regions that are allowed at the $95\%$ CL by all the Higgs searches checked by \texttt{HiggsBounds}. In the dark (light) orange regions the $\chi^2$ from \texttt{HiggsSignals} for the $125$ GeV Higgs measurements satisfy $\Delta \chi^2 = \chi^2 - \chi^2_{\textrm{min}} \leq 6.18\;(11.83)$, corresponding to 2$\sigma$ (3$\sigma$), where we take $\chi^2_{\textrm{min}}$ to be the minimum $\chi^2$ value observed across all the \texttt{MultiNest} scans $(\chi^2_{\textrm{min}} = 75.6,\;\textrm{ndf}=81)$. Finally, the teal (cyan) regions have a phase transition strength satisfying $\xi_c >1$ $(\xi_c>0.5)$.

The grid scans illustrate some of the important ways in which the $\xi_c > 1$ constraint determines the preferred parameter regions in the strong-PT scan. First, increasing the mass scale by raising $\Re(m_{12}^2)$ decreases $\xi_c$. From the right-hand plot we see that this decrease in $\xi_c$ can be counteracted by decreasing $\tan\beta$. This explains the clear preference for low $\tan\beta$ in the standard scenario, as this scenario prefers higher $\Re(m_{12}^2)$ compared to the hidden-Higgs case. Further, in the left-hand plot we see that for a fixed $\Re(m_{12}^2)$ the fit to $BR(b\rightarrow s \gamma)$ can be improved by going to large and negative $\lambda_{4}$ and $\lambda_5^R$ (recall that we have taken $\lambda_5^R = 0.7\lambda_4$), as this increases $m_{H^\pm}$. However, increasing $|\lambda_4|$ and $|\lambda_5^R|$ eventually also affects the masses of $H_1$ and $H_2$, causing conflict with both LHC Higgs data and the strong PT requirement. Thus, we end up with a $> 2\sigma$ tension between $\xi_c>0.5$ and $BR(b \rightarrow s \gamma)$. For $\xi_c>1$ this tension is greater than 3$\sigma$. We also note that negative $\Re(m_{12}^2)$ typically gives an unstable potential. 

In conclusion we find that the strong-PT scan confirms the result of the main scan, namely that current data does not favour a first-order phase transition in the $\text{2HDM}_5$, and that this result is largely due to the conflict between the low mass scale required for a first-order phase transition and the high $m_{H^\pm}$ required to avoid tension with $BR(b\rightarrow s \gamma)$. On the other hand, recent studies of the impact of LHC data on $CP$-conserving 2HDMs indicate that the hidden-Higgs scenario should be able accomodate charged Higgs masses up to about $600$~GeV \cite{Dumont:2014wha,Chang:2015goa}, at which level there would be no significant tension with $BR(b \rightarrow s \gamma)$, while still keeping the two lightest scalars around or below $125$~GeV. In the next subsection we will investigate the origin of the stricter upper bound on $m_{H^\pm}$ observed in our strong-PT scan. We will find that a $H^\pm$ much lighter than $600$~GeV is in fact a general consequence of the hidden-Higgs scenario and the loop-level spectrum calculation we employ.

Finally, we remind the reader that any constraints coming from processes involving Yukawas are type dependent. In particular, the constraint from $BR(b \rightarrow s \gamma)$ on the type-I model is less severe than for the type-II case that we study here. In the $\text{2HDM}_5$ with type-I couplings, the effect of this constraint is to tune the charged Higgs mass to $m_{H^{\pm}} \sim v\approx246$~GeV \cite{Dorsch:2013wja}. Furthermore, some parameter regions in the type-I model seem to provide interesting possibilities for detection at the LHC \cite{Dorsch:2014qja}.

%
%%%%%%%%%%%%%%%%%%%%
\subsection{A closer look at the hidden-Higgs scenario}
\label{subsec:Hidden Higgs}
%%%%%%%%%%%%%%%%%%%%
%
Even if no requirement is put on the phase transition strength, an upper bound on the heavy scalar masses arises in the hidden-Higgs scenario due to the limited parametric freedom: After requiring EWSB with $v \approx 246\,\text{GeV}$, the only free mass scale in the theory is $\sqrt{|\Re(m_{12}^2)|}$. In the standard scenario, where $H_1$ is SM-like, $v$ sets the correct mass scale for $m_{H_1} \approx 125$~GeV. The free parameter $\Re(m_{12}^2)$ can then be used to push the remaining scalars up to a high mass scale, as seen in the main scan in Sec.\ \ref{sec:Results and discussion}. This scale separation is no longer possible in the hidden-Higgs scenario due to the stronger requirement of having $m_{H_1} < m_{H_2} \approx 125$~GeV. A large mass gap up to $H_3$ and $H^{\pm}$ then relies on having large quartic couplings, meaning that constraints on these couplings translate into upper bounds on $m_{H_3}$ and $m_{H^{\pm}}$. We note that increasing $\tan\beta$ will also help raise $m_{H^\pm}$ through the first term in Eq.\ (\ref{ChargedHiggsMass}), but this effect is sub-dominant when $\Re(m_{12}^2) < v^2$.

In the $CP$-conserving type-II model studied in Ref.\ \cite{Dumont:2014wha}, the light $CP$-even $h^0$ is the hidden state and the heavy $CP$-even $H^0$ plays the role of the $125$~GeV Higgs. 
The study identifies constraints on the scalar masses and the mixing of the $CP$-even states coming from the 7 and 8 TeV LHC data. A similar analysis, focused on deriving constraints on the scalar potential parameters, is presented in Ref.\ \cite{Chang:2015goa}. The authors of \cite{Chang:2015goa} analyse a set of parameter points drawn from flat probability distributions in the real parameters $\lambda_{1,2,3,4,5}$, $\tan\beta$ and $m_{12}^2$, discarding points that are in significant tension with any of the applied experimental constraints. While no statistical fit is performed, the surviving set of parameter points indicate that the SM alignment required by the LHC Higgs results impose an upper bound on the masses of $H^\pm$ and $A^0$ of $m_{H^\pm,\,A^0} \lesssim 600$~GeV, similar to what is found in \cite{Dumont:2014wha}.
\begin{table*}[tb]
\centering
\textbf{Priors and settings for the hidden-Higgs scans} \\
\begin{tabular}{lll|ll}
\hline
 Parameter & Range & Type & \texttt{MultiNest} setting & Value  \\
\hline
$\lambda_1$  & $[0,\,4\pi]$  & flat &\texttt{nlive} & 2000  \\ 
$\lambda_2$  & $[0,\,1]$  & flat & \texttt{tol} & 0.5  \\ 
$\lambda_3$  & $[-4\pi,\,4\pi]$  & flat & \texttt{efr} & 0.8 \\ \cline{4-5}  
$\lambda_4$  & $[-4\pi,\,4\pi]$  & flat & Likelihood & Included \\ \cline{4-5}  
$\lambda_{345}$  & $[-5,\,5]$  & flat & Stability & Yes  \\ 
$\lambda_5^I$  & $[-4\pi,\,4\pi]$  & flat & Direct searches & Yes \\ 
$\tan\beta$  & $[0.1,\,60]$  & flat  & $\Delta \rho$ + B physics & Yes  \\  
$\Re(m_{12}^2)$  & $[-500^2,\, 500^2]$ $\text{GeV}^2$  & two-sided log & Strong PT & No \\ \cline{4-5}
 & & & Sampled points & Posterior  \\ \cline{4-5}
 & & & $1.6 \times 10^7$ & $1.1 \times 10^4$ \\   % loop-level scan
 \hline
\end{tabular}
\caption{Prior distributions, likelihood contributions and \texttt{MultiNest} settings for the tree-level and loop-level hidden-Higgs scans. For the tree-level scan the stability requirement in the likelihood refers to the check of global minimum and positivity of the tree-level scalar potential as in \cite{Ivanov:2015nea,ElKaffas:2006gdt}. The number of visited points and posterior samples refer to the loop-level scan. The tree-level scan visited around $1.1 \times 10^7$ points, producing $9.5 \times 10^3$ posterior samples.}
\label{tab:priorsHidden}
\end{table*}

For our investigation of the hidden-Higgs scenario we scan the relevant subspace of the $\text{2HDM}_5$ parameter space with \texttt{MultiNest}, using the likelihood function described in Sec.\ \ref{subsec:likelihood}. Thus, in the following we do not include any constraint on the phase transition strength. In \cite{Chang:2015goa} the scalar spectrum is calculated at tree level from the input parameters. To compare our results against this analysis we run two separate scans, one in which we switch off the loop corrections in \texttt{SPheno} and one where the loop corrections are included. In the following we refer to these scans individually as the ``tree-level scan'' and ``loop-level scan'', and collectively as the ``hidden-Higgs scans''. 

The scan setup is summarized in Table~\ref{tab:priorsHidden}. Note that the stability check for the tree-level scan involves checking for global minimum and positivity using inequalities from \cite{Ivanov:2015nea} and \cite{ElKaffas:2006gdt}. As for the strong-PT scan, we restrict the prior for $\lambda_2$ to the range $[0,1.0]$ and take $\lambda_{345}$ as an input parameter instead of $\lambda_5^R$. Further, the prior range for $\Re(m_{12}^2)$ is limited to $[-500^2,\, 500^2]$~$\text{GeV}^2$. To ensure that we only scan the hidden-Higgs scenario we impose the additional hard constraint that $m_{H_1} < 120$~GeV. Thus, our overall effective prior is   
\ba
  \pi(\bfTheta) = H\left(1 - \dfrac{m_{H_1}(\bfTheta)}{120\,\text{GeV}}\right) \prod\limits_j \pi_j(\theta_j),
  \label{eq:prior}
\ea
where $\pi_j$ are the priors for the individual input parameters $\theta_j$.
 
As previously discussed, we have not imposed the tree-level perturbative unitarity constraint \cite{Ginzburg:2005dt} in our scans. While relatively unimportant for the previous scans, this constraint significantly impacts the hidden-Higgs scans presented here. To illustrate this we present two sets of posterior results, one with the posteriors resulting from our scans and one where we have discarded posterior samples that violate perturbative unitarity. In the tree-level analysis in \cite{Chang:2015goa} the perturbative unitarity constraint is included in the scan. Also, in contrast to our analysis, the requirement that the vacuum is a global minimum of the potential is not imposed.

\begin{figure*}[t]
\centering
\includegraphics[width=0.24\textwidth]{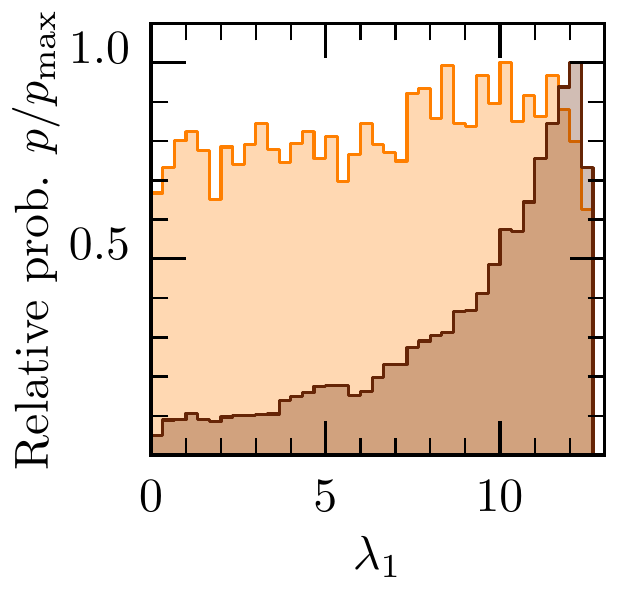}
\includegraphics[width=0.24\textwidth]{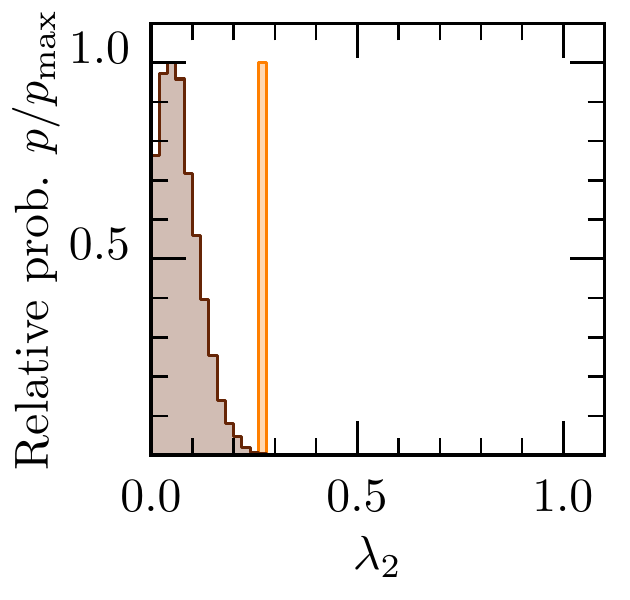}
\includegraphics[width=0.24\textwidth]{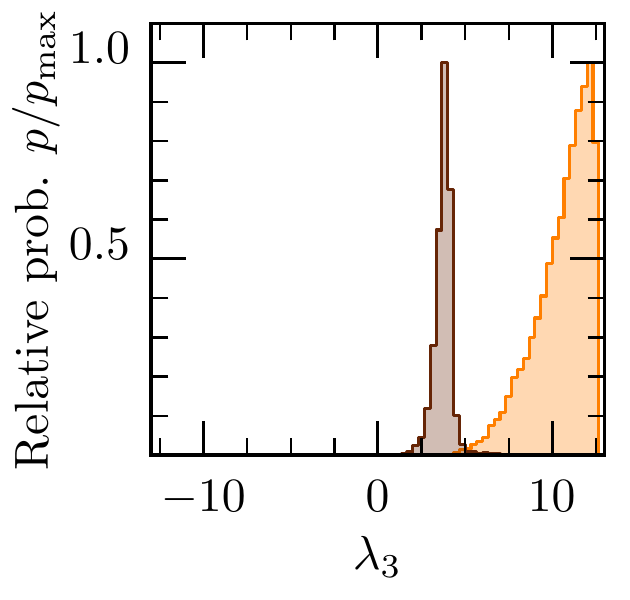}
\includegraphics[width=0.24\textwidth]{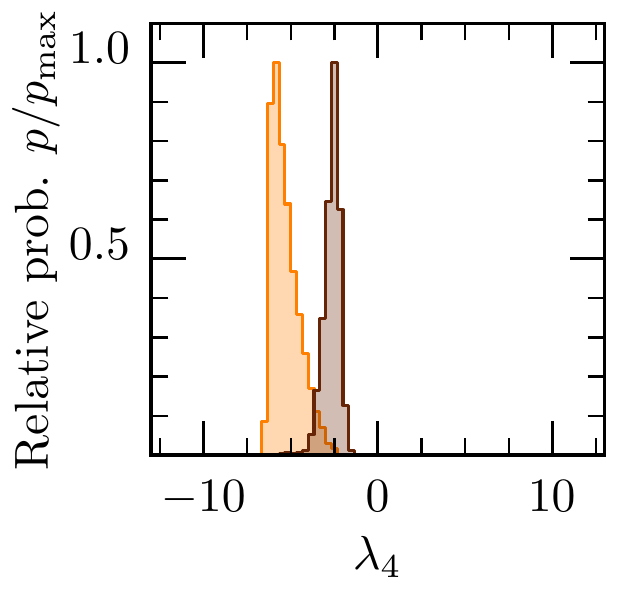}\\
\includegraphics[width=0.24\textwidth]{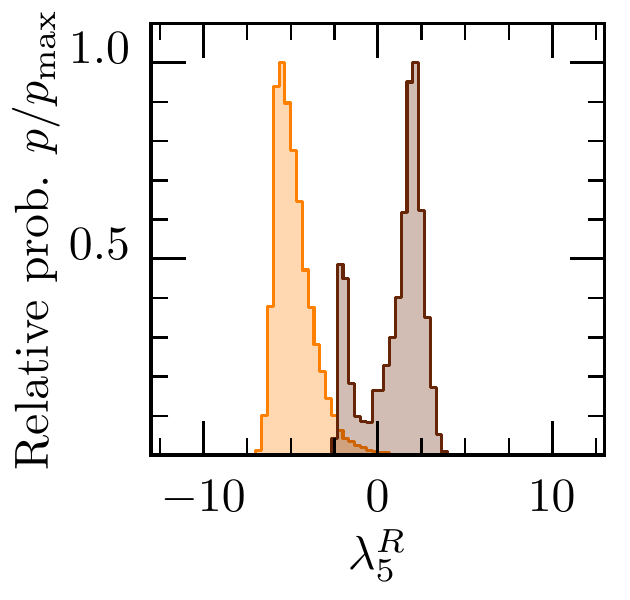}
\includegraphics[width=0.24\textwidth]{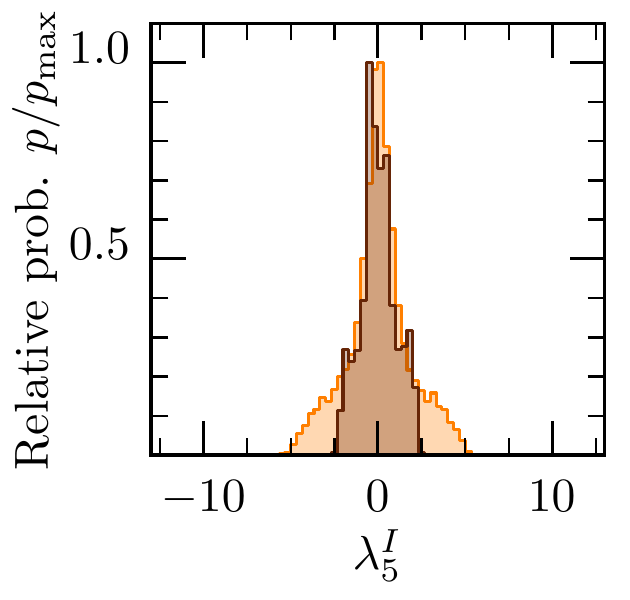}
\includegraphics[width=0.24\textwidth]{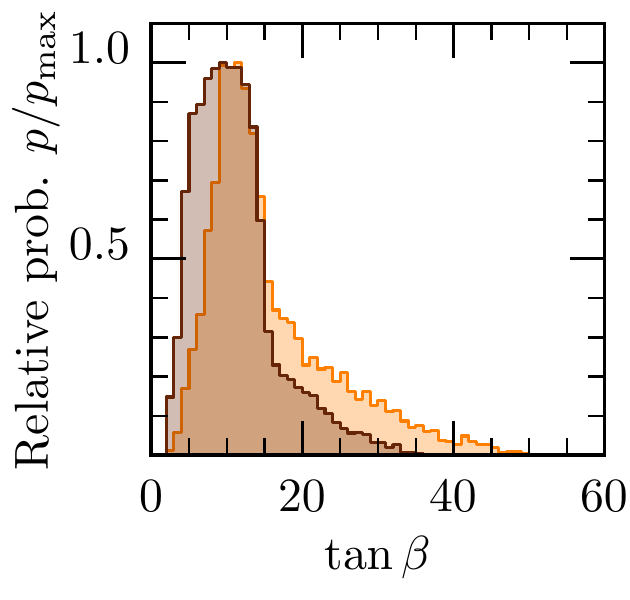}
\includegraphics[width=0.24\textwidth]{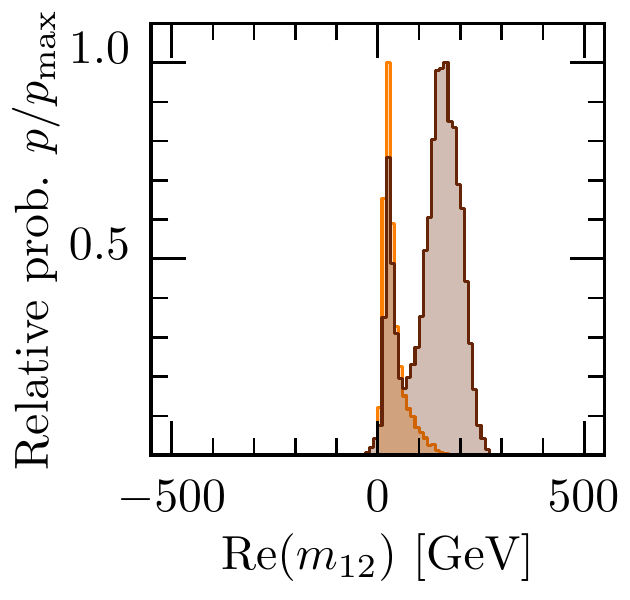}\\
\includegraphics[width=0.24\textwidth]{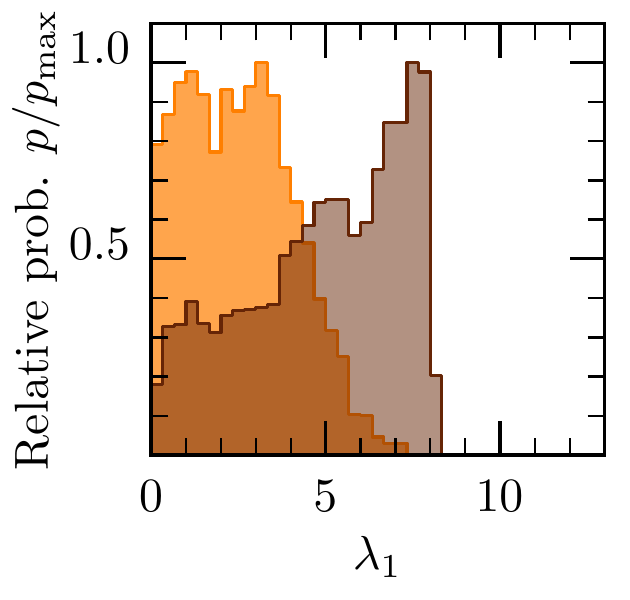}
\includegraphics[width=0.24\textwidth]{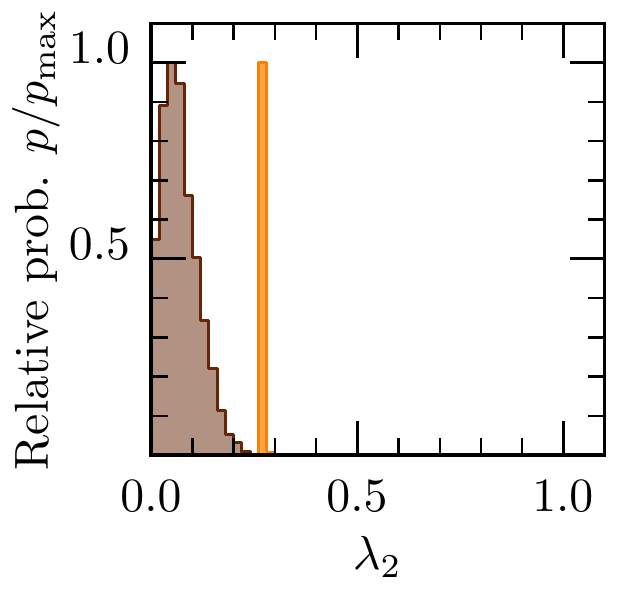}
\includegraphics[width=0.24\textwidth]{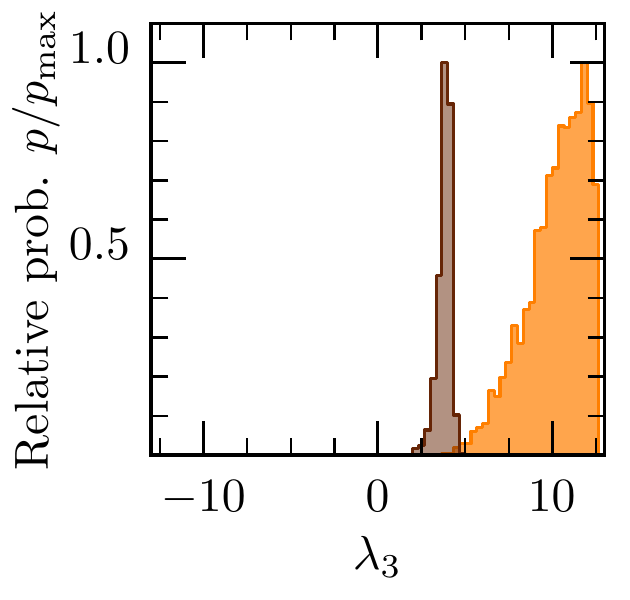}
\includegraphics[width=0.24\textwidth]{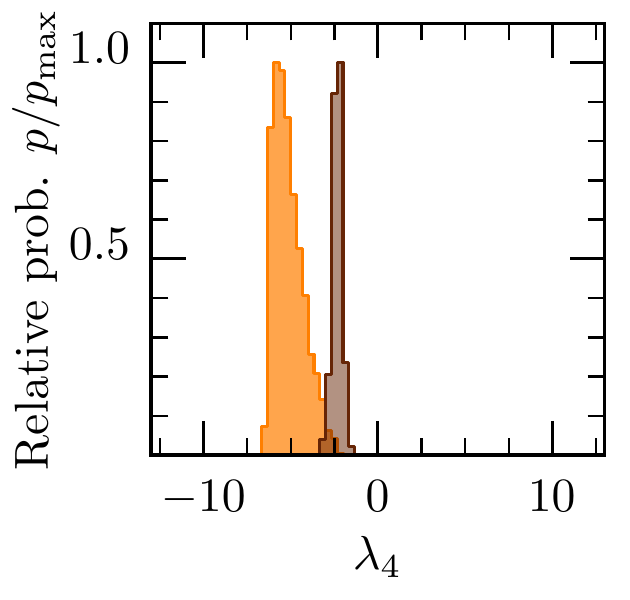}\\
\includegraphics[width=0.24\textwidth]{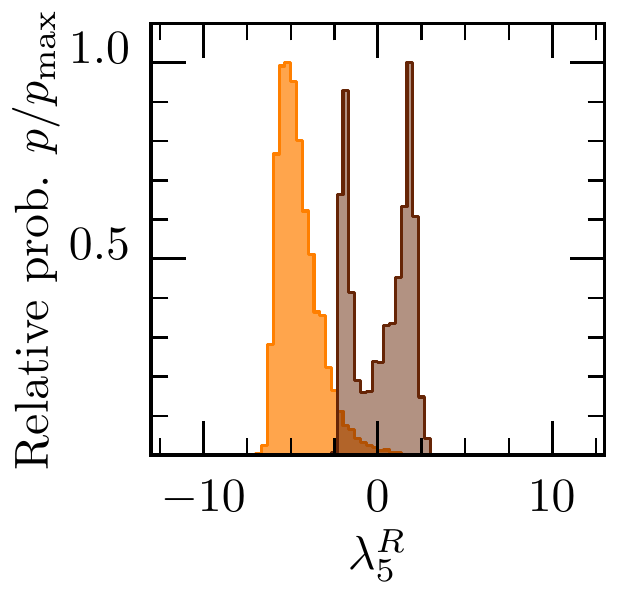}
\includegraphics[width=0.24\textwidth]{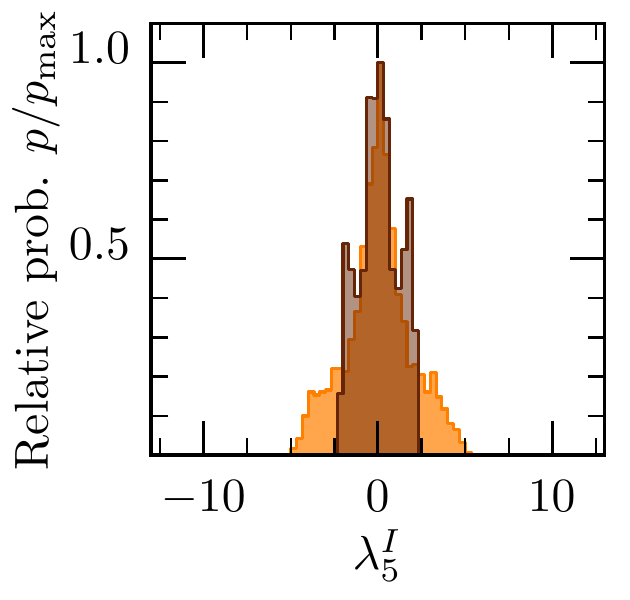}
\includegraphics[width=0.24\textwidth]{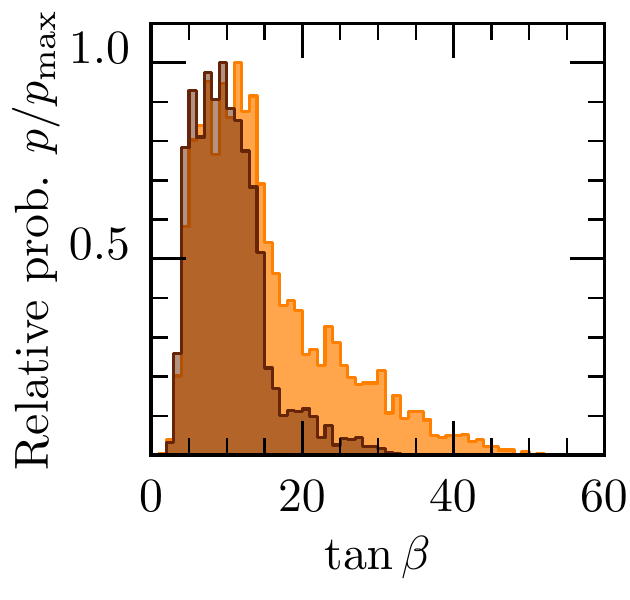}
\includegraphics[width=0.24\textwidth]{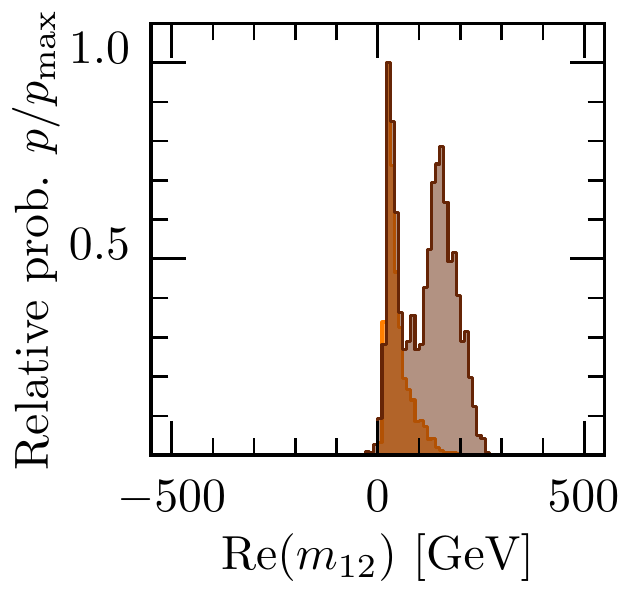}
\caption{Marginalized posterior distributions for the parameters in the hidden-Higgs scans. \textit{Top two rows:} The parameter posteriors resulting from the tree-level scan (orange) and loop-level scan (brown). \textit{Bottom two rows:} The same posterior distributions after discarding samples that fail the tree-level perturbative unitarity constraint.}  
\label{fig:posteriors_hidden_higgs}
\end{figure*}

In the upper two rows of Fig.\ \ref{fig:posteriors_hidden_higgs} we show the posterior results of the tree-level (orange) and loop-level (brown) scans. The bottom two rows (darker colours) contain the same posterior distributions after requiring tree-level perturbative unitarity. Focusing first on the tree-level scan, we find that the preferred parameter ranges show good agreement with the parameter bounds for the type-II $CP$-conserving model in \cite{Chang:2015goa}, in particular after taking into account the perturbative unitarity constraint. The main difference is that our posterior distributions are somewhat broader compared to the results in \cite{Chang:2015goa}. This is to be expected from the difference in methodology and the additional freedom in our model coming from the extra $\lambda_5^I$ parameter.

\begin{figure*}[t]
\centering
\includegraphics[width=0.32\textwidth]{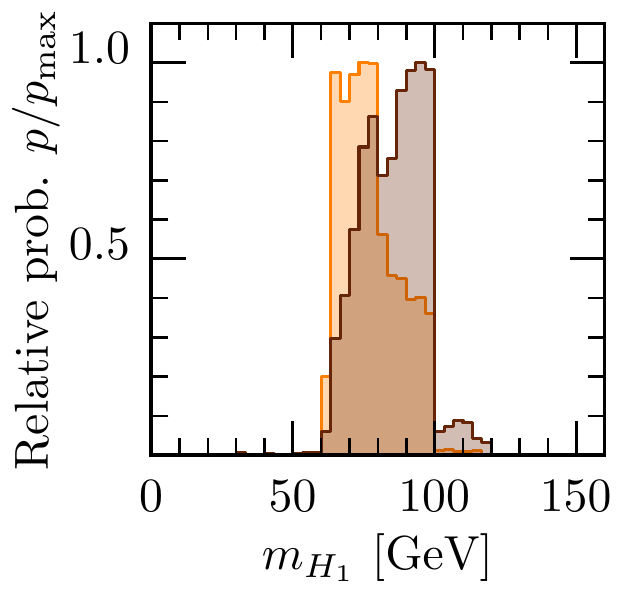}
\includegraphics[width=0.32\textwidth]{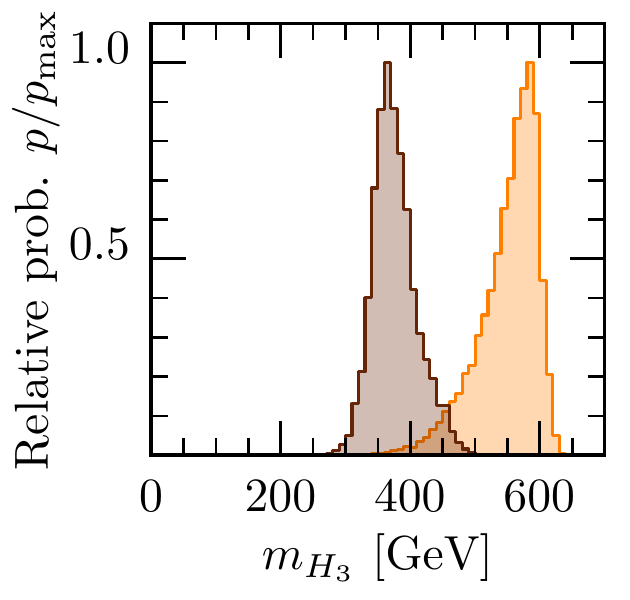}
\includegraphics[width=0.32\textwidth]{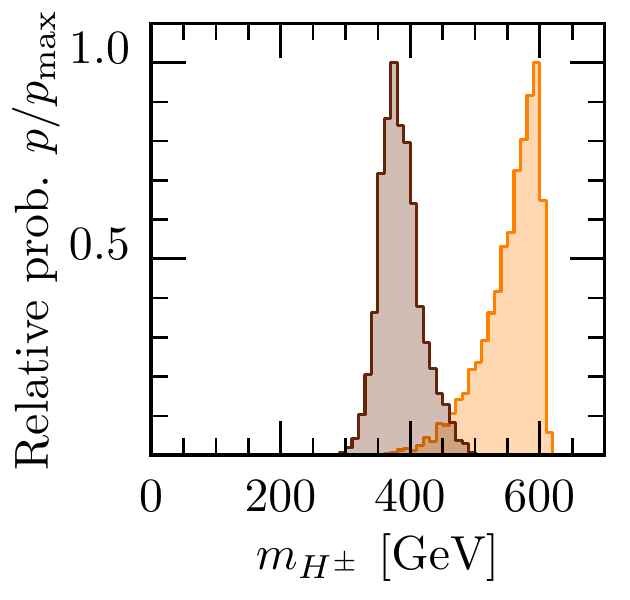}\\
\includegraphics[width=0.32\textwidth]{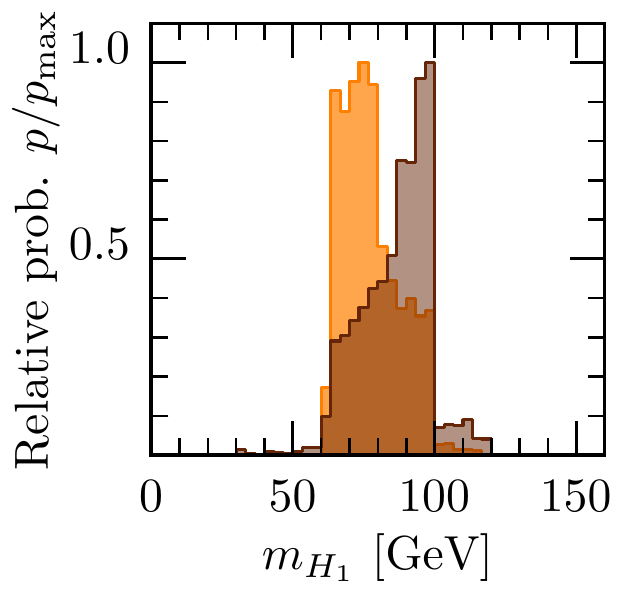}
\includegraphics[width=0.32\textwidth]{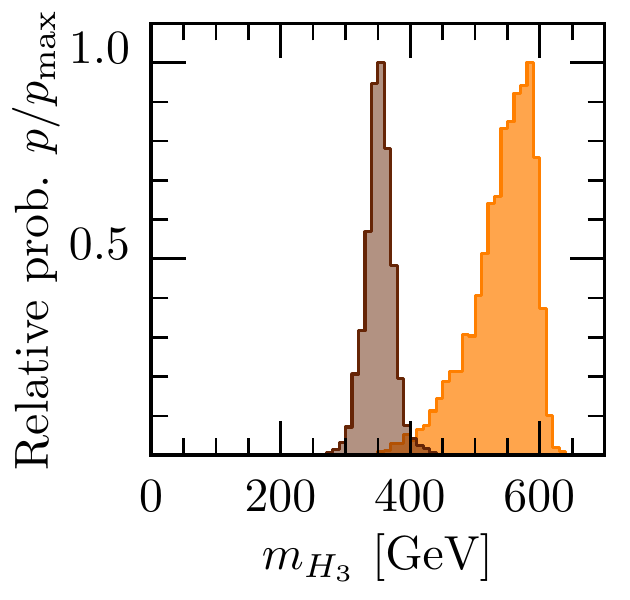}
\includegraphics[width=0.32\textwidth]{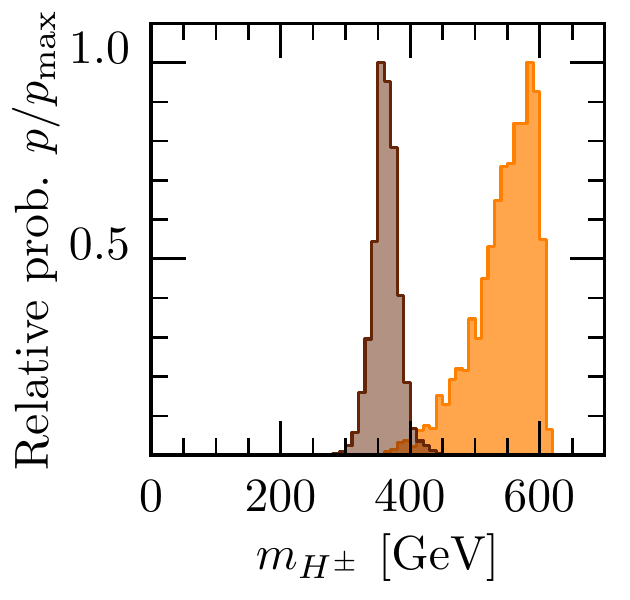}
\caption{Marginalized posterior distributions for $m_{H_1}$, $m_{H_3}$ and $m_{H^\pm}$. \textit{Top row:} The posteriors resulting from the tree-level scan (orange) and loop-level scan (brown). \textit{Bottom row:} The same distributions after discarding samples that fail the tree-level perturbative unitarity constraint.}
\label{fig:posteriors_scalar_masses_hidden_higgs}
\end{figure*}

\begin{figure*}
\centering
\includegraphics[width=0.32\textwidth]{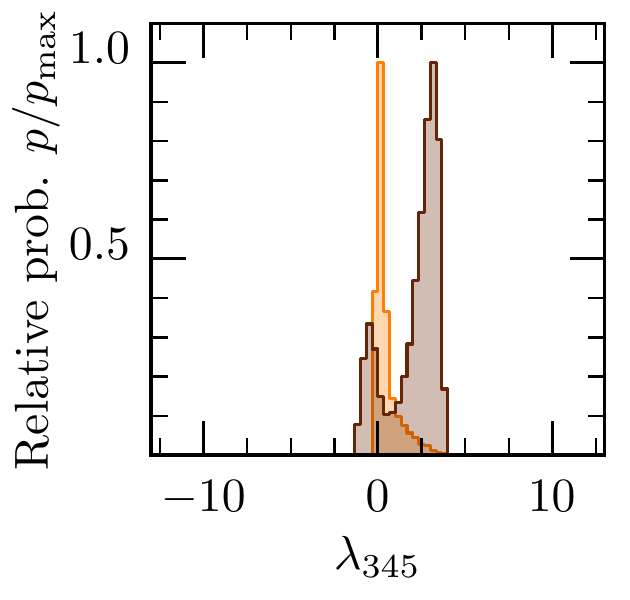}
\includegraphics[width=0.32\textwidth]{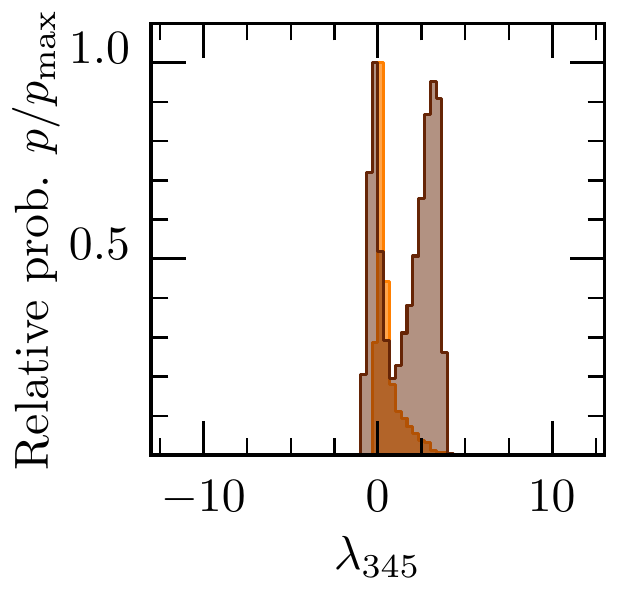}
\caption{\textit{Left:} Marginalized posterior distributions for $\lambda_{345} \equiv \lambda_3 + \lambda_4 + \lambda_5^R$ from the tree-level scan (orange) and the loop-level scan (brown). \textit{Right:} The same distribution after requiring tree-level perturbative unitarity.}
\label{fig:posteriors_la345_hidden_higgs}
\end{figure*}

The posterior distributions for the tree-level $m_{H_1}$, $m_{H_3}$ and $m_{H^\pm}$ are shown in Fig.\ \ref{fig:posteriors_scalar_masses_hidden_higgs} (orange). Our preferred ranges for the scalar masses agree well with the results that \cite{Chang:2015goa} find for $m_{h^0}$, $m_{A^0}$ and $m_{H^\pm}$. Two bounds on the preferred mass ranges are worth commenting: First, we see the same upper bound on $m_{H^\pm}$ around $600$~GeV as in \cite{Dumont:2014wha,Chang:2015goa}. This is a consequence of an interplay between the LHC Higgs data and the perturbativity bound of $\lambda_3 < 4\pi$: At tree level the LHC Higgs data prefers small values of $\lambda_{345}$. This can be seen in Fig.\ \ref{fig:posteriors_la345_hidden_higgs} and is also found in \cite{Chang:2015goa}. At the same time, $BR(b \rightarrow s \gamma)$ requires a large $m_{H^\pm}$. With $\Re(m_{12}^2)$ constrained by the requirement of having both $H_1$ and $H_2$ light, the tree-level $m_{H^\pm}$ can only be pushed up by large and negative $(\lambda_4 + \lambda_5)$, see Eq.\ (\ref{ChargedHiggsMass}). The combined effect of the $\lambda_{345} \sim 0$ preference and the perturbativity bound is then $\lambda_3 \sim -(\lambda_4 + \lambda_5^R) < 4\pi$, which gives the upper bound on $m_{H^\pm}$. Second, we see a clear preference for $m_{H_1} > m_{H_2}/2 \approx 62.5$~GeV. Below this mass the decay channel $H_2 \rightarrow H_1 H_1$ opens up, which quickly leads to conflict with the observed Higgs data.\footnote{See Ref.\ \cite{Bernon:2014nxa} for a dedicated study of the viability of parameter regions with $m_{H_1} < m_{H_2}/2$.} This effect is also observed in \cite{Dumont:2014wha,Chang:2015goa}.


\begin{figure*}
\centering
\includegraphics[width=0.7\textwidth]{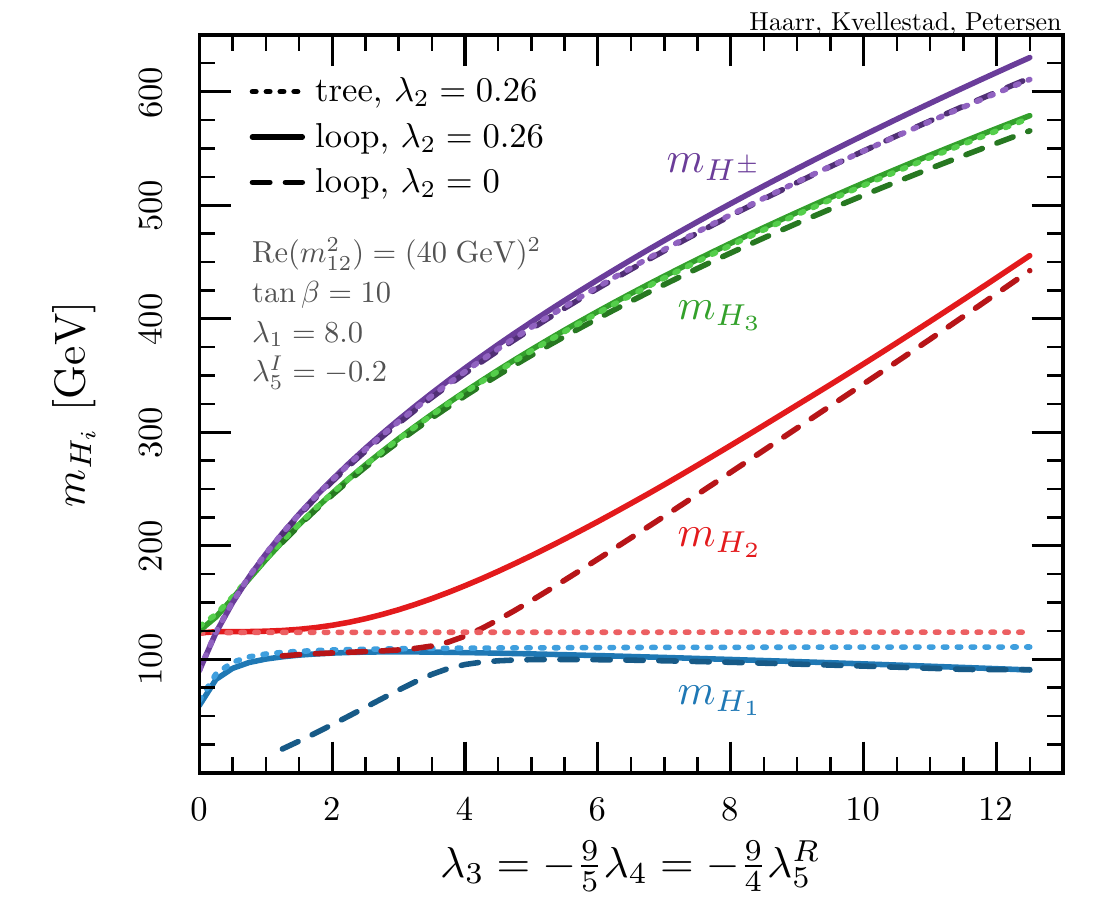}
\caption{A comparison of the scalar masses as functions of $\lambda_3 = -\frac{9}{5}\lambda_4 = -\frac{9}{4}\lambda_5^R$. The dotted and solid lines depict tree-level and one-loop masses respectively, for $\lambda_2=0.26$. The dashed lines show one-loop masses for $\lambda_2=0$. The remaining parameters are set as follows: $\lambda_1 = 8.0$, $\lambda_5^I = -0.2$, $\tan\beta = 10$, $\Re(m_{12}^2) = (40\;\text{GeV})^2$.}
\label{fig:la3_mH}
\end{figure*}

We now move on to the result of the loop-level scan. As seen from Fig.\ \ref{fig:posteriors_hidden_higgs} (brown distributions), calculating the spectrum at one-loop order leads to a preference for smaller magnitudes of $\lambda_3$, $\lambda_4$ and $\lambda_5^R$ compared to the result of the tree-level scan. This can be understood as follows: In the tree-level case, the dependence of $m_{H_1}$, $m_{H_2}$ on $\lambda_4$ and $\lambda_5^R$ can be cancelled by adjusting $\lambda_3$ such that $\lambda_{345}$ remains constant. Thus, $m_{H^\pm}$ can be pushed up without affecting $m_{H_1}$ and $m_{H_2}$. This is no longer true for the one-loop masses, with the consequence that the hidden-Higgs scenario $m_{H_1} < m_{H_2} \approx 125$~GeV  is confined to a smaller range in the quartic couplings. We illustrate this in Fig.\ \ref{fig:la3_mH}. Here we plot the scalar masses as functions of $\lambda_3 = -\frac{9}{5}\lambda_4 = -\frac{9}{4}\lambda_5^R$ while fixing all other parameters. This choice ensures a constant $\lambda_{345} = 0$, meaning that $\lambda_5^R$ is the only non-constant contribution to the tree-level masses through the term $(\mathcal{M}^2)_{33} = \mu^2 - v^2 \lambda_5^R$ in Eq.\ (\ref{eq:neutral_mass_matrix}). Thus, the two lightest tree-level masses (blue and red dotted lines) remain constant with increasing $|\lambda_5^R|$. The tree-level masses are plotted for $\lambda_2 = 0.26$ which gives $m_{H_2} \approx 125$~GeV. The one-loop masses for the same $\lambda_2$ value are shown as solid lines. In this case $m_{H_2}$ exceeds $125$~GeV already for $\lambda_3 \approx 2$, corresponding to $\lambda_4 \approx -1.1$ and $\lambda_5^R \approx -0.9$. By taking $\lambda_2 \rightarrow 0$ the departure of the one-loop $m_{H_2}$ can be delayed somewhat. For instance, setting $\lambda_2 = 0$ in the scenario plotted in Fig.\ \ref{fig:la3_mH} (dashed lines) extends the $m_{H_2} \lesssim 125$~GeV region up to $\lambda_3 \approx 4$, corresponding to $\lambda_4 \approx -2.2$ and $\lambda_5^R \approx -1.8$. (Note that in this case a slightly larger $\Re(m_{12}^2)$ would be needed to obtain $m_{H_2} \approx 125$~GeV in the region of small $\lambda_3$.)

An additional hidden-Higgs parameter region, with $\lambda_5^R > 0$, opens up at loop level. A positive $\lambda_5^R$ reduces $m_{H^\pm}$, but this reduction is compensated by an increase in $\Re(m_{12}^2)$. This parameter region manifests as distinct peaks centred at $\lambda_5^R \approx 2$ and $\Re(m_{12}^2) \approx (150\;\text{GeV})^2$ in the loop-level scan posteriors in Fig.\ \ref{fig:posteriors_hidden_higgs}, and a corresponding peak at $\lambda_{345} \approx 3$ in Fig.\ \ref{fig:posteriors_la345_hidden_higgs}. With $\lambda_5^R > 0$ the hidden-Higgs scenario can extend towards larger $|\lambda_4|$, evident as a tail towards $\lambda_4 \approx -5$ in the upper right-hand plot of Fig.\ \ref{fig:posteriors_hidden_higgs}. These parameter choices produce an unstable tree-level scalar potential, which explains why this region is not present in the tree-level scan results. In the loop-level scan this parameter region can survive the stability check for the one-loop potential described in Sec.\ \ref{subsec:determining PT strength}, especially for large $\lambda_1$.\footnote{Note that the numerical stability check of $V(\eta_1,\eta_2,\eta_3)$ in Sec.\ \ref{subsec:determining PT strength} is limited in field space to the region $|\eta_i| < 20m_t$, whereas the inequalities used to check positivity of $V_{\text{tree}}(\eta_1,\eta_2,\eta_3)$ are derived from the limit $|\eta_i| \rightarrow \infty$.} However, large $\lambda_1$ values conflicts with the tree-level perturbative unitarity constraint, as can be seen in the bottom two rows of Fig.\ \ref{fig:posteriors_hidden_higgs}. Thus, the relative posterior weight for this parameter region is reduced when perturbative unitarity is required.

In summary, the loop-level dependence of $m_{H_1}$,$m_{H_2}$ on $\lambda_4$ and $\lambda_5^R$, along with the stability and perturbative unitarity constraints, restrict the hidden-Higgs scenario to parameter regions with $m_{H^\pm}$ much below the $600$~GeV bound observed at tree-level. This can be seen in the posterior for the one-loop $m_{H^\pm}$ in the upper right-hand plot of Fig.\ \ref{fig:posteriors_scalar_masses_hidden_higgs} (brown). The preferred $m_{H^\pm}$ range resulting from the loop-level scan extends up to $m_{H^\pm}$ just below $500$~GeV. The upper tail of the $m_{H^\pm}$ distribution is associated with the $\lambda_5^R > 0$, $\Re(m_{12}^2) \gtrsim (100\;\text{GeV})^2$ parameter region described above. Imposing tree-level perturbative unitarity cuts into this parameter region, with the result that the preferred $m_{H^\pm}$ range is restricted to $m_{H^\pm} < 440$~GeV (lower right-hand plot). For the $\lambda_5^R < 0$, $\Re(m_{12}^2) \lesssim (100\;\text{GeV})^2$ region also found in the tree-level scan, the probable range for the loop-level $m_{H^\pm}$ is limited to $m_{H^\pm} < 400$~GeV.

\begin{figure*}[t]
\centering
\includegraphics[width=0.32\textwidth]{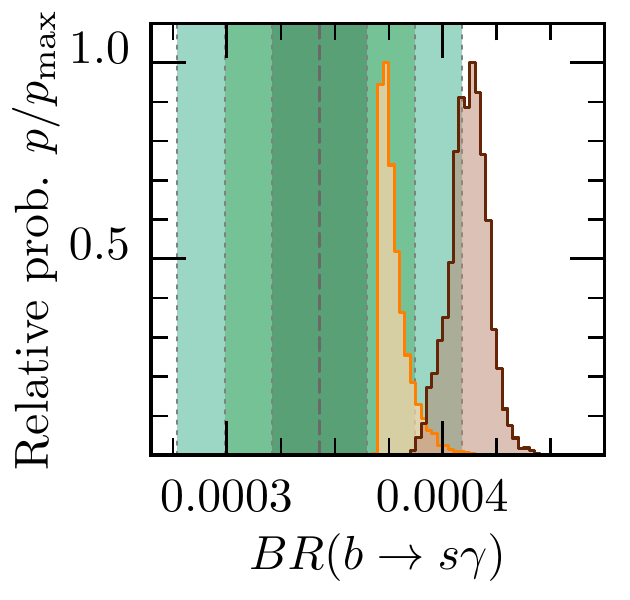}
\includegraphics[width=0.32\textwidth]{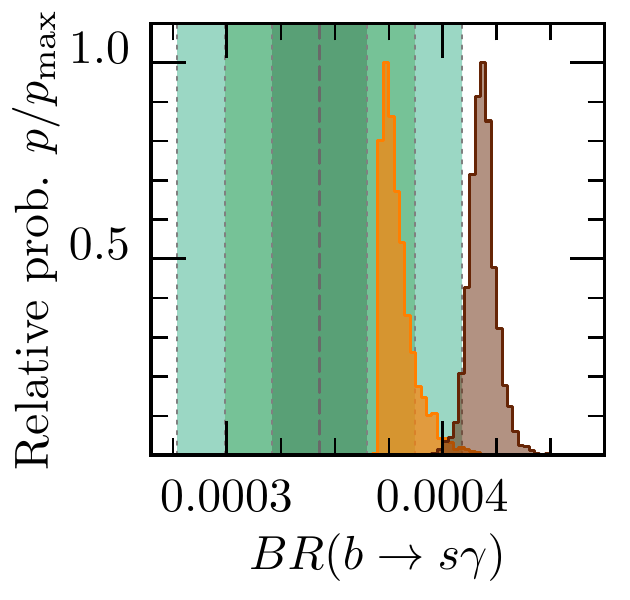}
\caption{\textit{Left:} Marginalized posterior distributions for $BR(b\rightarrow s \gamma)$ from the tree-level (orange) and loop-level (brown) scans. \textit{Right:} The same distributions after discarding samples that fail the tree-level perturbative unitarity constraint. The green bands show the $1\sigma$, $2\sigma$ and $3\sigma$ regions around the observed value.}
\label{fig:posteriors_observables_hidden_higgs}
\end{figure*}

So while at tree level the best-fit region for the hidden-Higgs scenario stays clear of strong tension with the observed $BR(b \rightarrow s \gamma)$, this is no longer true at loop level. We illustrate this in Fig.\ \ref{fig:posteriors_observables_hidden_higgs} by comparing the posteriors for $BR(b \rightarrow s \gamma)$ from the tree-level (orange) and loop-level (brown) scans to the $3\sigma$ range for the observed value. The preferred range for $BR(b \rightarrow s \gamma)$ in the loop-level scan see a $\gtrsim 2\sigma$ tension with the observed value (left-hand plot). This tension is further strengthened when perturbative unitarity is required (right-hand plot).
%

%
%%%%%%%%%%%%%%%%%%%%
\section{Conclusions}
\label{sec:Conclusions}
%%%%%%%%%%%%%%%%%%%%
%
In this study we have performed a series of Bayesian fits of the $\text{2HDM}_5$ with type-II Yukawa couplings. Our main aim has been to investigate how current experimental constraints affect the probability for a strong first-order phase transition. We find that fitting the model to the 125 GeV Higgs data, collider searches for additional Higgs bosons, the electroweak precision observable $\Delta\rho$ and a selection of $B$-physics observables severly lowers this probability. 

The above conclusion has been checked by means of an additional fit where we constrain the model to regions of parameter space with a first-order phase transition. This is done by including the requirement $\xi_c > 1$ as a $95\%$ CL lower limit in the likelihood function. This requirement introduces a strong preference for light scalar masses, with the consequence that the best-fit region is dominated by the ``hidden-Higgs'' scenario. In this scenario the next-to-lightest scalar, $H_2$, is the $125$~GeV Higgs and the lighter state $H_1$ has remained undetected. In comparison, the standard Higgs scenario with $H_1$ as the $125$~GeV state only accounts for about $1\%$ of the posterior probability in this fit. The two scenarios have characteristics that agree with previously studied models of baryogenesis in 2HDMs: First, we see no indication that a non-zero $CP$-violating phase is important to obtain a first-order transition. Second, we find a preference for small $\tan\beta$ when $H_1$ is the $125$~GeV state.

The insistence of $\xi_c > 1$ in the likelihood forces the model into a part of parameter space where there is significant tension with experimental constraints: Whereas a first-order phase transition prefers a low mass scale for the additional scalars, a $m_{H^\pm}$ heavier than about $490$~GeV is required for the predicted $BR(b \rightarrow s \gamma)$ to stay within $2\sigma$ of the observed value. With only one free mass parameter in the theory, $\Re(m_{12}^2)$, a large mass splitting between $H^\pm$ and the lighter $H_{1,2}$ can only be achieved by going to large values in the quartic couplings, but this brings the model into conflict data on the $125$~GeV Higgs. This tension between the experimental constraints and the $\xi_c > 1$ requirement confirms the conclusion that a first-order phase transition is not a likely prediction of the type-II $\text{2HDM}_5$.

For similar reasons we find that the hidden-Higgs scenario in the type-II $\text{2HDM}_5$ is disfavoured by current data, regardless of the phase transition strength: The requirement that $m_{H_1} < m_{H_2} \approx 125$~GeV forces down the mass scale $\Re(m_{12}^2)$, with the consequence that the masses of the heavier scalars get bounded from above by constraints on the quartic couplings. With tree-level spectrum calculation this upper bound on $m_{H_3}$ and $m_{H^\pm}$ is around $600$~GeV and is a consequence of the perturbativity bound $\lambda_3 < 4\pi$. However, when the mass spectrum is calculated at one loop we find that the preferred parameter region has $m_{H_3},m_{H^\pm} < 500$~GeV. This stronger upper bound is a consequence of the stability condition for the one-loop effective potential, and the fact that at one-loop order, increasing $-(\lambda_4 + \lambda_5^R)$ to raise $m_{H^\pm}$ eventually also causes $m_{H_2} > 125$~GeV, in conflict with the LHC data. Further, if tree-level perturbative unitarity is required we find that $m_{H_3},m_{H^\pm} < 440$~GeV is preferred. As a result of these loop-level bounds on $m_{H^\pm}$ there is significant tension between the predictions of the hidden-Higgs scenario and the experimental data, most notably $BR(b \rightarrow s \gamma)$.

We end by mentioning a few ways in which the current study can be further improved and expanded. First, the likelihood function can be extended to include contributions from more observables, most importantly LHC Higgs results at $13$~TeV, a more complete set of flavour observables and electric dipole moments. Further, the check for stability of the scalar potential can be made more sophisticated by also accepting parameter points for which the broken minimum is metastable. Also, using \texttt{SARAH} and \texttt{SPheno}, we can implement the RG-improved effective potential as in \cite{Blinov:2015vma}. This could in turn enable an extension of the stability check by requiring the potential to remain stable also at scales higher than the electroweak scale, similar to what was done in \cite{Chowdhury:2015yja}. Finally, the accuracy with which the posterior distributions have been determined should ideally be improved by running \texttt{MultiNest} with a higher number of live points and stricter convergence criteria, in particular for the fit with $\xi_c > 1$ included in the likelihood. Given enough CPU time this would also enable a frequentist analysis in terms of profile likelihoods, to serve as a useful comparison to the Bayesian analysis presented here.

In future work we plan to extend the current analysis to 2HDMs with other Yukawa structures.

%
%%%%%%%%%%%%%%%%%%%%
\section*{Acknowledgements}\label{sec:Acknowledgements} 
%%%%%%%%%%%%%%%%%%%%
% 
We thank Anders Tranberg for useful discussions and valuable input. Also, we thank Florian Staub for help with using the \texttt{SARAH}/\texttt{SPheno} interface. In this work we made extensive use of the \texttt{PySLHA} \cite{Buckley:2013jua} package to ease communication between the different numerical tools, and the \texttt{pippi} \cite{Scott:2012qh} package for plotting posterior distributions. The CPU intensive parameter scans in this work was carried out using the Abel Cluster, owned by the Norwegian metacenter for High Performance Computing (NOTUR) and the University of Oslo. The computing time was made available through NOTUR allocation NN9284K, financed through the Research Council of Norway. AH and AK thank the Niels Bohr Institute for generous hospitality while part of this work was completed.

\appendix
%
%%%%%%%%%%%
\section{Field- and temperature-dependent masses}
\label{app: Field dependent masses}
%%%%%%%%%%
%
% 
%%%%%%%%%%%
\subsection{Neutral scalars}
\label{app: Neutral scalars}
%%%%%%%%%%
%
The tree-level Higgs mass matrix gets a debye-corrected mass matrix \cite{Cline:2011mm}:
\ba
\delta m^2_{ij} = \frac{T^2}{24} \sum_k N_k \frac{\partial^2 m^2_k(\phi_n)}{\partial \phi_i \partial \phi_j}
\ea
where in our case $\phi_i=\eta_i$, for $i=1,2,3$ and $\phi_4=\text{G}_0$. The $m_k^2$ are the zero-temperature, field-dependent masses of the relevant particles, see Eq.\ (\ref{particle degeneracy numbers}). For the Higgs-sector the calculation is significantly simplified by utilizing that the sum of neutral-squared masses equals the trace of the tree-level mass matrix.\footnote{The trace is basis independent. This can be seen using its cyclic property, which cancels the rotation matrices.} The total four-by-four, field- and temperature-dependent mass matrix is:
\ba
\mathcal{N}^2_{ij} = \mathcal{M}^2_{ij}+\delta m^2_{ij}
\ea
with components given by:
\ba
%%%%%%%%%11
\mathcal{N}^2_{\eta_1,\eta_1}&=&\frac{1}{2} \bigg(\eta _3^2 \lambda _1 \sin ^2\beta +2 \mu ^2 \sin ^2\beta + 3 \eta _1^2 \lambda _1+\eta _2^2 \lambda _3+\eta _2^2 \lambda _4
\\ \nonumber
&-&2 \cos \beta \left(\eta _3 \lambda ^{I}_5 \left(\eta _2+v \sin \beta \right)
-3 \eta _1 \lambda _1 v\right)+\eta _2^2 \lambda ^{R}_5
\\ \nonumber
&+&\cos ^2\beta  \left(\eta _3^2 \left(\lambda _3+\lambda _4-\lambda ^{R}_5\right)+2 \lambda _1 v^2\right)+2 \eta _2 v \sin \beta  \left(\lambda _3+\lambda _4+\lambda ^{R}_5\right)\bigg)
\\ \nonumber
&+&\frac{1}{24} T^2 \bigg[\frac{9 g^2}{2}+\frac{3 g\prime^2}{2}+6 \lambda _1+4 \lambda _3+2 \lambda _4-\frac{24 m_b^2  }{v^2\cos ^2\beta}\bigg]
\ea
%%%%%%%%%22
\ba
\mathcal{N}^2_{\eta_2,\eta_2} &=& \frac{1}{2} \bigg(\eta _3^2 \lambda _3 \sin ^2\beta +\eta _3^2 \lambda _4 \sin ^2\beta +2 \mu ^2 \cos ^2\beta 
\\ \nonumber
&+&2 \eta _1 \left(-\eta _3 \sin \beta  \lambda ^{I}_5+v \cos \beta  \lambda ^{R}_5+\lambda _3 v \cos \beta +\lambda _4 v \cos \beta \right)
\\ \nonumber
&-&2 \eta _3 v \sin \beta  \cos \beta  \lambda ^{I}_5-\eta _3^2 \sin ^2\beta  \lambda ^{R}_5+\eta _1^2 \left(\lambda _3+\lambda _4+\lambda ^{R}_5\right)
\\ \nonumber
&+&\lambda _2 \left(\eta _3^2 \cos ^2\beta +3 \eta _2^2+2 v^2 \sin ^2\beta +6 \eta _2 v \sin \beta \right)\bigg)
\\ \nonumber
&+&
\frac{1}{24} T^2 \bigg[\frac{9 g^2}{2}+\frac{3 g\prime^2}{2}+6 \lambda _2+4 \lambda _3+2 \lambda _4-\frac{24 m_t^2 }{v^2\sin ^2\beta }\bigg]
\ea
%%%%%%%%%%%33
\ba
\mathcal{N}^2_{\eta_3,\eta_3} &=& \bigg(\frac{\eta _1}{2} \bigg[\cos \beta  \bigg(3 \eta _3 \sin 2 \beta  \lambda ^{I}_5-4 \eta _2 \sin \beta  \lambda ^{R}_5
\\
&+&v \cos 2 \beta  \lambda ^{R}_5-3 v \lambda ^{R}_5+2 \lambda _3 v \cos ^2\beta +2 \lambda _4 v \cos ^2\beta \bigg)
\\ \nonumber
&+&\lambda _1 v \sin \beta  \sin 2 \beta \bigg]+ \frac{3}{2}\eta _3 \sin 2 \beta  \lambda ^{I}_5 \left(\eta _2 \sin \beta +v\right)
\\ \nonumber
&+&\frac{3}{4} \eta _3^2 \left(2 \lambda _1 \sin ^4\beta +2 \lambda _2 \cos ^4\beta +\sin ^2 2 \beta  \left(\lambda _3+\lambda _4+\lambda ^{R}_5\right)\right)
\\ \nonumber
&+&\frac{1}{2} \eta _1^2 \left(\lambda _1 \sin ^2 \beta +\cos ^2\beta  \left(\lambda _3+\lambda _4-\lambda ^{R}_5\right)\right)
\\ \nonumber
&+&\frac{1}{2} \bigg[\eta _2^2 \left(\lambda _2 \cos ^2\beta +\sin ^2\beta  \left(\lambda _3+\lambda _4-\lambda ^{R}_5\right)\right)+2 \left(\mu ^2-v^2 \lambda ^{R}_5\right)
\\ \nonumber
&+& 2 \eta _2 v \sin \beta  \left(\lambda _3 \sin ^2\beta +\lambda _4 \sin ^2\beta +\lambda _2 \cos ^2\beta -\sin ^2\beta  \lambda ^{R}_5-2 \cos ^2\beta  \lambda ^{R}_5\right)\bigg]\bigg)
\\ \nonumber
&+&\frac{T^2}{24}
\\ \nonumber
&\times & \bigg(\bigg[ \left(6 \lambda _1 \sin ^2\beta +6 \lambda _2 \cos ^2\beta +\frac{9}{2} g^2+ \frac{3}{2} g\prime^2 + 4 \lambda _3+2 \lambda _4\right)
\\ \nonumber
&-&24 \frac{m_b^2}{v^2} \tan^2 \beta - 24 \frac{m_t^2}{v^2} \cot ^2\beta \bigg] \bigg)
\ea
%%%%%%%%%%%44
\ba
\mathcal{N}^2_{\text{G}_0,\text{G}_0} &=& \frac{1}{2} \sin ^2\beta  \bigg(2 \eta _1 \bigg[\eta _3 \sin \beta  \lambda ^{I}_5-2 \eta _3 \cos \beta  \cot \beta  \lambda ^{I}_5+2 \eta _2 \cot \beta  \lambda ^{R}_5
\\ \nonumber
&+&v \cos \beta  \lambda ^{R}_5+\lambda _3 v \cos \beta +\lambda _4 v \cos \beta +\lambda _1 v \cos \beta  \cot ^2\beta \bigg]
\\ \nonumber
&+&\eta _3 \cot \beta  \csc \beta  \lambda ^{I}_5 \left(\eta _2 (3 \cos 2 \beta -1)-2 v \sin \beta \right)+\eta _1^2 \left(\lambda _1 \cot ^2\beta +\lambda _3+\lambda _4-\lambda ^{R}_5\right)
\\ \nonumber
&+&\frac{1}{4} \eta _3^2 \csc ^2\beta  \left(3 \lambda _1 \sin ^2 2 \beta +3 \lambda _2 \sin ^2 2 \beta +(3 \cos 4 \beta +1) \left(\lambda _3+\lambda _4+\lambda ^{R}_5\right)\right)
\\ \nonumber
&+&\eta _2 \bigg[\cot ^2\beta  \left(\lambda ^{R}_5 \left(2 v \sin \beta -\eta _2\right)+\lambda _3 \left(\eta _2+2 v \sin \beta \right)+\lambda _4 \left(\eta _2+2 v \sin \beta \right)\right)
\\ \nonumber
&+&\lambda _2 \left(\eta _2+2 v \sin \beta \right)\bigg]\bigg)
\\ \nonumber
&+&\frac{T^2}{24} \left[6 \lambda _2 \sin ^2\beta +6 \lambda _1 \cos ^2\beta +\frac{9 g^2}{2}+\frac{3 g\prime^2}{2}+4 \lambda _3+2 \lambda _4-\frac{24 m_b^2}{v^2}-\frac{24 m_t^2}{v^2}\right]
\ea
%%%%%%%12
\ba
\mathcal{N}^2_{\eta_1,\eta_2} &=& \frac{1}{2} \bigg(-\mu ^2 \sin 2 \beta -2 \eta _2 \eta _3 \sin \beta  \lambda ^{I}_5-2 \eta _1 \eta _3 \cos \beta  \lambda ^{I}_5-2 \eta _3 v \lambda ^{I}_5
\\ \nonumber
&-&\eta _3^2 \sin 2 \beta  \lambda ^{R}_5+2 \eta _1 \eta _2 \lambda ^{R}_5+v^2 \sin 2 \beta  \lambda ^{R}_5+2 \eta _1 v \sin \beta  \lambda ^r{}_5+2 \eta _2 v \cos \beta  \lambda ^{R}_5
\\ \nonumber
&+&2 \lambda _3 \left(\eta _2+v \sin \beta \right) \left(\eta _1+v \cos \beta \right)+2 \lambda _4 \left(\eta _2+v \sin \beta \right) \left(\eta _1+v \cos \beta \right)\bigg)
\ea
%%%%%%%13
\ba
\mathcal{N}^2_{\eta_1,\eta_3} &=& \frac{1}{2} \bigg(3 \eta _3^2 \sin \beta  \cos ^2\beta  \lambda ^{I}_5-\lambda ^{I}_5 \bigg[\eta _2^2 \sin \beta +v^2 \sin \beta
\\ \nonumber
&+&2 \eta _1 \cos \beta  \left(\eta _2+v \sin \beta \right)+2 \eta _2 v\bigg]
\\ \nonumber
&+&\eta _3 \bigg[\cos \beta \bigg(\lambda ^{R}_5 \left(-4 \eta _2 \sin \beta -2 \eta _1 \cos \beta +v (\cos 2 \beta -3)\right)
\\ \nonumber
&+&2 \lambda _3 \cos \beta  \left(\eta _1+v \cos \beta \right)
\\ \nonumber
&+&2 \lambda _4 \cos \beta  \left(\eta _1+v \cos \beta \right)\bigg)+2 \lambda _1 \sin ^2\beta  \left(\eta _1+v \cos \beta \right)\bigg]\bigg)
\ea
%%%%%%%14
\ba
\mathcal{N}^2_{\eta_1,\text{G}_0} &=& \frac{1}{8} \bigg(\eta _3^2 (-(\cos \beta +3 \cos 3 \beta )) \lambda ^{I}_5+4 \lambda ^{I}_5 \left(\eta _2^2 \cos \beta -2 \eta _1 \sin \beta  \left(\eta _2+v \sin \beta \right)\right)
\\ \nonumber
&+&2 \eta _3 \bigg[2 \eta _1 \lambda _4 \sin 2 \beta -2 \eta _1 \sin 2 \beta  \lambda ^{R}_5+4 \eta _2 \cos 2 \beta \lambda ^{R}_5-3 v \sin \beta  \lambda ^{R}_5+v \sin 3 \beta  \lambda ^{R}_5
\\ \nonumber
&-&2 \lambda _1 \sin 2 \beta  \left(\eta _1+v \cos \beta \right)+2 \lambda _3 \sin 2 \beta  \left(\eta _1+v \cos \beta \right)+\lambda _4 v \sin \beta +\lambda _4 v \sin 3 \beta \bigg]\bigg)
\ea
%%%%%%%23
\ba
\mathcal{N}^2_{\eta_2,\eta_3} &=& \frac{1}{2} \bigg(3 \eta _3^2 \sin ^2\beta  \cos \beta  \lambda ^{I}_5
\\ \nonumber
&-&\lambda ^{I}_5 \bigg[\eta _1^2 \cos \beta +2 \eta _1 \left(\eta _2 \sin \beta +v\right)+v \cos \beta  \left(2 \eta _2 \sin \beta +v\right)\bigg]
\\ \nonumber
&+&\eta _3 \bigg[\sin \beta  \bigg(-\lambda ^{R}_5 \left(2 \eta _2 \sin \beta +4 \eta _1 \cos \beta +v (\cos 2 \beta +3)\right)+2 \lambda _3 \sin \beta  \left(\eta _2+v \sin \beta \right)
\\ \nonumber
&+&2 \lambda _4 \sin \beta  \left(\eta _2+v \sin \beta \right)\bigg)+2 \lambda _2 \cos ^2\beta  \left(\eta _2+v \sin \beta \right)\bigg]\bigg)
\ea
%%%%%%%24
\ba
\mathcal{N}^2_{\eta_2,\text{G}_0} &=& \frac{1}{8} \bigg(\eta _3^2 (\sin \beta -3 \sin 3 \beta ) \lambda ^{I}_5+4 \lambda ^{I}_5 \left(-\eta _1^2 \sin \beta +2 \eta _1 \eta _2 \cos \beta +2 \eta _2 v \cos ^2\beta \right)
\\ \nonumber
&-&4 \eta _3 \bigg[\eta _2 \lambda _4 \sin 2 \beta +2 \eta _1 \sin ^2\beta  \lambda ^{R}_5-\eta _2 \sin 2 \beta  \lambda ^{R}_5-2 \eta _1 \cos ^2\beta  \lambda ^{R}_5-2 v \cos ^3\beta  \lambda ^{R}_5
\\ \nonumber
&-&\lambda _2 \sin 2 \beta  \left(\eta _2+v \sin \beta \right)+\lambda _3 \sin 2 \beta  \left(\eta _2+v \sin \beta \right)+\lambda _4 v \sin \beta  \sin 2 \beta \bigg]\bigg)
\ea
%%%%%%%34
\ba
\mathcal{N}^2_{\eta_3,\text{G}_0} &=& \frac{1}{8} \bigg(2 \eta _1 \bigg[-\eta _3 \cos \beta  \lambda ^{I}_5-3 \eta _3 \cos 3 \beta  \lambda ^{I}_5+4 \eta _2 \cos 2 \beta  \lambda ^{R}_5-3 v \sin \beta  \lambda ^{R}_5
\\ \nonumber
&+&v \sin 3 \beta  \lambda ^{R}_5+\lambda _4 v \sin \beta +\lambda _4 v \sin 3 \beta -4 \lambda _1 v \sin \beta  \cos ^2\beta +4 \lambda _3 v \sin \beta  \cos ^2\beta \bigg]
\\ \nonumber
&-&2 \eta _3 \lambda ^{I}_5 \left(\eta _2 (3 \sin 3 \beta -\sin \beta )+4 v \cos 2 \beta \right)-2 \eta _1^2 \sin 2 \beta  \left(\lambda _1-\lambda _3-\lambda _4+\lambda ^{R}_5\right)
\\ \nonumber
&+&3 \eta _3^2 \left(4 \lambda _2 \sin \beta  \cos ^3\beta -4 \lambda _1 \sin ^3\beta  \cos \beta -\sin 4 \beta  \left(\lambda _3+\lambda _4+\lambda ^{R}_5\right)\right)
\\ \nonumber
&+&4 \eta _2 \cos \beta  \bigg[-\eta _2 \lambda _4 \sin \beta +\eta _2 \sin \beta  \lambda ^{R}_5+v \cos 2 \beta  \lambda ^{R}_5+v \lambda ^{R}_5+\lambda _2 \sin \beta  \left(\eta _2+2 v \sin \beta \right)
\\ \nonumber
&-&\lambda _3 \sin \beta  \left(\eta _2+2 v \sin \beta \right)+\lambda _4 v \cos 2 \beta -\lambda _4 v\bigg]\bigg)
\\ \nonumber
&-&\frac{T^2}{16 v^2\cos \beta  \sin \beta } \bigg[8 \left(\cos 2 \beta  \left(m_b^2+m_t^2\right)-m_b^2+m_t^2\right)+\lambda _1 v^2 \sin ^2 2 \beta -\lambda _2 v^2 \sin ^2 2 \beta  \bigg]
\ea
The temperature- and field-dependent masses $m^2_i(\eta_j,T)$ are obtained by diagonalizing this matrix.
%
%
%
%%%%%%%%%%%
\subsection{Charged scalars}
\label{app: Charged scalars}
%%%%%%%%%%
%
Debye corrections are added like for the neutral sector. We denote the two-by-two mass matrix of the charged sector by $\mathcal{P}^2_{ij}$. It is field- and temperature-dependent.
\ba
\mathcal{P}^2_{\text{H}^{+},\text{H}^{-}} &=& \frac{1}{2} \sin ^2\beta \tan \beta \bigg(\eta _1^2 \cot \beta  \left(\lambda _3 \cot ^2\beta +\lambda _1\right)
\\ \nonumber
&+&2 \eta _1 \cot \beta  \bigg[\cot \beta  \left(\eta _3 \cos \beta  \lambda ^{I}_5-\eta _2 \lambda ^{R}_5-v \sin \beta  \lambda ^{R}_5-\lambda _4 \left(\eta _2+v \sin \beta \right)+\lambda _3 v \cos \beta  \cot \beta \right)
\\ \nonumber
&+&\lambda _1 v \cos \beta \bigg]+2 \eta _3 \cot ^2 \beta  \lambda ^{I}_5 \left(\eta _2 \sin \beta +v\right)
\\ \nonumber
&+&\eta _3^2 \cos \beta  \left(\lambda _1 \sin \beta +\cos \beta  \cot \beta  \left(\lambda _2 \cot ^2 \beta +2 \left(\lambda _3+\lambda _4+\lambda ^{R}_5\right)\right)\right)
\\ \nonumber
&+&\frac{1}{2} \cot \beta  \csc ^2 \beta  \bigg[2 \eta _2^2 \left(\lambda _3 \sin ^2 \beta +\lambda _2 \cos ^2 \beta \right)+4 \mu ^2-2 v^2 \lambda ^{R}_5
\\ \nonumber
&+&2 \eta _2 v \sin \beta  \left(2 \lambda _3 \sin ^2 \beta +2 \lambda _2 \cos ^2 \beta -2 \cos ^2 \beta  \left(\lambda _4+\lambda ^{R}_5\right)\right)-2 \lambda _4 v^2\bigg]\bigg)
\\ \nonumber
&+&\frac{T^2}{24} \bigg[\left(6 \lambda _1 \sin ^2\beta +6 \lambda _2 \cos ^2\beta +\frac{9}{2} g^2+\frac{3}{2} g\prime^2 +4 \lambda _3+2 \lambda _4\right)-24 \frac{m_b^2}{v^2} \tan^2 \beta - 24 \frac{m_t^2}{v^2} \cot ^2\beta \bigg]
\ea
\ba
\mathcal{P}^2_{\text{H}^{+},\text{G}^{-}} &=& \frac{1}{16} \bigg(-4 \eta _1^2 \left(\lambda _1-\lambda _3\right) \sin 2 \beta
\\ \nonumber
&-&4 \eta _3^2 \sin 2 \beta  \left(\lambda _1 \sin ^2\beta -\lambda _2 \cos ^2\beta +\lambda _3 \cos 2 \beta +\lambda _4 \cos 2 \beta +i \lambda ^{I}_5+\cos 2 \beta  \lambda ^{R}_5\right)
\\ \nonumber
&-&4 \eta _1 \bigg[2 i \eta _3 \lambda _4 \cos \beta -2 \eta _2 \lambda _4 \cos 2 \beta +\eta _3 \cos \beta  \lambda ^{I}_5+\eta _3 \cos 3 \beta  \lambda ^{I}_5-2 i \eta _2 \lambda ^{I}_5-2 i v \sin \beta \lambda ^{I}_5
\\ \nonumber
&-&2 i \eta _3 \cos \beta  \lambda ^{R}_5-2 \eta _2 \cos 2 \beta  \lambda ^{R}_5+v \sin \beta  \lambda ^{R}_5-v \sin 3 \beta  \lambda ^{R}_5+\lambda _4 v \sin \beta -\lambda _4 v \sin 3 \beta
\\ \nonumber
&+&4 \lambda _1 v \sin \beta  \cos ^2\beta -4 \lambda _3 v \sin \beta  \cos ^2\beta \bigg]-8 \eta _3 \left(\eta _2 \sin \beta +v\right) \left(\cos 2 \beta  \lambda ^{I}_5+i \lambda _4-i \lambda ^{R}_5\right)
\\ \nonumber
&+&8 \eta _2 \cos \beta  \left(v \left(\lambda _4 \cos 2 \beta +i \lambda ^{I}_5+\cos 2 \beta  \lambda ^{R}_5\right)+\lambda _2 \sin \beta  \left(\eta _2+2 v \sin \beta \right)-\lambda _3 \sin \beta  \left(\eta _2+2 v \sin \beta \right)\right)\bigg) 
\\ \nonumber
&-&\frac{T^2}{16 v^2\cos \beta  \sin \beta } \bigg[8 \left(\cos 2 \beta  \left(m_b^2+m_t^2\right)-m_b^2+m_t^2\right)+\lambda _1 v^2 \sin ^2 2 \beta -\lambda _2 v^2 \sin ^2 2 \beta \bigg]
\ea
\ba
\mathcal{P}^2_{\text{G}^{+},\text{H}^{-}} &=& (\mathcal{P}^2_{\text{H}^{+},\text{G}^{-}})^* 
\ea
\ba
\mathcal{P}^2_{\text{G}^{+},\text{G}^{-}} &=& \frac{1}{2} \sin ^2\beta  \bigg(\eta _1^2 \left(\lambda _1 \cot ^2\beta +\lambda _3\right)
\\ \nonumber
&+&2 \eta _1 \cot \beta  \left(\eta _2 \lambda _4-\eta _3 \cos \beta  \lambda ^{I}_5+\eta _2 \lambda ^{R}_5+v \sin \beta  \lambda ^{R}_5+\lambda _3 v \sin \beta +\lambda _4 v \sin \beta +\lambda _1 v \cos \beta  \cot \beta \right)
\\ \nonumber
&-&2 \eta _3 \cot \beta  \lambda ^{I}_5 \left(\eta _2 \sin \beta +v\right)
\\ \nonumber
&+&\eta _3^2 \left(\lambda _3 \sin ^2 \beta +\lambda _1 \cos ^2\beta +\lambda _2 \cos ^2\beta -2 \lambda _4 \cos ^2\beta +\lambda _3 \cos ^2\beta  \cot ^2\beta -2 \cos ^2\beta  \lambda ^{R}_5\right)
\\ \nonumber
&+&\eta _2 \left(\cot ^2\beta  \left(2 v \sin \beta  \left(\lambda _4+\lambda ^{R}_5\right)+\lambda _3 \left(\eta _2+2 v \sin \beta \right)\right)+\lambda _2 \left(\eta _2+2 v \sin \beta \right)\right)\bigg)
\\ \nonumber
&+&\frac{T^2}{24}\bigg[6 \lambda _2 \sin ^2\beta +6 \lambda _1 \cos ^2\beta +\frac{9 g^2}{2}+\frac{3 g\prime^2}{2}+4 \lambda _3+2 \lambda _4-\frac{24 m_b^2}{v^2}-\frac{24 m_t^2}{v^2}\bigg]
\ea
%
%
%
%%%%%%%%%%%
\subsection{Gauge bosons}
\label{app: Gauge bosons}
%%%%%%%%%%
%
For the gauge bosons we revert to the original gauge field basis and write the mass matrix \
\beq
M^2(\phi,T)& =& M^2(\phi) + M_T^2(T)=\nonumber\\
&&\left(\begin{array}{cccc}
g^2\phi^2/4&0&0&0\\
0&g^2\phi^2/4&0&0\\
0&0&g^2\phi^2/4&-gg'\phi^2/4\\
0&0&-gg'\phi^2/4&g'^{2}\phi^2/4\end{array}\right)
+\left(\begin{array}{cccc}
2g^2T^2&0&0&0\\
0&2g^2T^2&0&0\\
0&0&2g^2T^2&0\\
0&0&0&2g'^2T^2\\
\end{array} \right)\delta_{T},\nonumber\\
\eeq
where $\phi =\eta_1 v \cos \beta+\eta_2 v \sin \beta+\frac{1}{2} \left(\eta_1^2+\eta_2^2+\eta_3^2+v^2\right)$ and the $\delta_T$ indicates that only transverse degrees of freedom get a temperature correction. 
%
%
%
%%%%%%%%%%%
\subsection{Fermions}
\label{app:Fermions}
%%%%%%%%%%
%
Fermions do not get a thermal correction. The top quark gets it mass from $\Phi_2$ and so does not depend on $\eta_1$. In a similar manner the bottom quark is granted its mass by $\Phi_1$ and does not depend on $\eta_2$.
\ba
M_t^2(\eta_j) &=& \frac{m_t^2 \left(\eta _3^2 \cot ^2\beta +\left(\eta _2 \csc \beta +v\right)^2\right)}{v^2}
\\
M_b^2(\eta_j) &=& \frac{m_b^2 \left(\eta _3^2 \tan ^2\beta +\left(\eta _1 \sec \beta +v\right)^2\right)}{v^2} 
\ea
%
%
%
%%%%%%%%%%%%%
% References
%%%%%%%%%%%%%
%%%%%%%%%%%%%%%%%%%%%%%%%%%%%%%%%%%%%%%%
 
%
%\input{biblio_sorted.tex}
%
%
\end{document}